\documentclass[twocolumn,letterpaper,superscriptaddress,amsmath,amssymb,prx,floatfix,showkeys]{revtex4-2}

\usepackage{graphicx}
\usepackage{booktabs}
\usepackage{dcolumn}
\usepackage{siunitx}
\usepackage{bm}
\usepackage{float}
\usepackage[T1]{fontenc}
\usepackage{xcolor}
\usepackage{textcomp,gensymb} 
\usepackage{lineno}
\usepackage{soul}
\usepackage{hyperref} 
\hypersetup{colorlinks=true,linkcolor=blue,filecolor=magenta,urlcolor=cyan}
\newcounter{para}

\begin{document}

\title{Optimizing laser coupling, matter heating, and particle acceleration from solids using multiplexed ultraintense lasers}

\author{Weipeng Yao} 
\email{yao.weipeng@polytechnique.edu}
\affiliation{LULI - CNRS, CEA, Sorbonne Universit\'e, Ecole Polytechnique, Institut Polytechnique de Paris - F-91128 Palaiseau cedex, France}
\affiliation{Sorbonne Universit\'e, Observatoire de Paris, Universit\'e PSL, CNRS, LERMA, F-75005, Paris, France}

\author{Motoaki Nakatsutsumi} 
\affiliation{LULI - CNRS, CEA, Sorbonne Universit\'e, Ecole Polytechnique, Institut Polytechnique de Paris - F-91128 Palaiseau cedex, France}

\author{S\'ebastien Buffechoux} 
\affiliation{LULI - CNRS, CEA, Sorbonne Universit\'e, Ecole Polytechnique, Institut Polytechnique de Paris - F-91128 Palaiseau cedex, France}

\author{Patrizio Antici} 
\affiliation{INRS-EMT, 1650 boul, Lionel-Boulet, Varennes, QC, J3X 1S2, Canada}

\author{Marco Borghesi} 
\affiliation{Center for Plasma Physics, School of Mathematics and Physics, Queen's University Belfast, Belfast BT7 1NN, United Kingdom}

\author{Andrea Ciardi} 
\affiliation{Sorbonne Universit\'e, Observatoire de Paris, Universit\'e PSL, CNRS, LERMA, F-75005, Paris, France}

\author{Sophia N. Chen} 
\affiliation{``Horia Hulubei'' National Institute for Physics and Nuclear Engineering, 30 Reactorului Street, RO-077125, Bucharest-Magurele, Romania}

\author{Emmanuel d'Humi\`eres} 
\affiliation{University of Bordeaux, Centre Lasers Intenses et Applications, CNRS, CEA, UMR 5107, F-33405 Talence, France}

\author{Laurent Gremillet} 
\affiliation{CEA, DAM, DIF, F-91297 Arpajon, France}
\affiliation{Universit\'{e} Paris-Saclay, CEA, LMCE, 91680 Bruy\`{e}res-le-Ch\^{a}tel, France}

\author{Robert Heathcote} 
\affiliation{Central Laser Facility, STFC Rutherford Appleton Laboratory, Didcot, UK}

\author{Vojt\v ech Horn\'y} 
\affiliation{LULI - CNRS, CEA, Sorbonne Universit\'e, Ecole Polytechnique, Institut Polytechnique de Paris - F-91128 Palaiseau cedex, France}
\affiliation{CEA, DAM, DIF, F-91297 Arpajon, France}
\affiliation{Universit\'{e} Paris-Saclay, CEA, LMCE, 91680 Bruy\`{e}res-le-Ch\^{a}tel, France}

\author{Paul McKenna} 
\affiliation{SUPA, Department of Physics, University of Strathclyde, Glasgow, G4 0NG, UK}

\author{Mark N. Quinn} 
\affiliation{SUPA, Department of Physics, University of Strathclyde, Glasgow, G4 0NG, UK}

\author{Lorenzo Romagnani} 
\affiliation{LULI - CNRS, CEA, Sorbonne Universit\'e, Ecole Polytechnique, Institut Polytechnique de Paris - F-91128 Palaiseau cedex, France}

\author{Ryan Royle} 
\affiliation{Department of Physics, University of Nevada, Reno, Nevada 89557, USA}

\author{Gianluca Sarri} 
\affiliation{Center for Plasma Physics, School of Mathematics and Physics, Queen's University Belfast, Belfast BT7 1NN, United Kingdom}

\author{Yasuhiko Sentoku} 
\affiliation{Institute of Laser Engineering, Osaka University, 2-6 Yamadaoka, Suita, Osaka 565-0871, Japan}

\author{Hans-Peter Schlenvoigt} 
\affiliation{LULI - CNRS, CEA, Sorbonne Universit\'e, Ecole Polytechnique, Institut Polytechnique de Paris - F-91128 Palaiseau cedex, France}

\author{Toma Toncian} 
\affiliation{Institut f\"ur Laser und Plasmaphysik, Heinrich Heine Universit\"at D\"usseldorf, D\"usseldorf, Germany}

\author{Olivier Tresca} 
\affiliation{SUPA, Department of Physics, University of Strathclyde, Glasgow, G4 0NG, UK}

\author{Laura Vassura} 
\affiliation{LULI - CNRS, CEA, Sorbonne Universit\'e, Ecole Polytechnique, Institut Polytechnique de Paris - F-91128 Palaiseau cedex, France}

\author{Oswald Willi} 
\affiliation{Institut f\"ur Laser und Plasmaphysik, Heinrich Heine Universit\"at D\"usseldorf, D\"usseldorf, Germany}

\author{Julien Fuchs} 
\email{julien.fuchsg@polytechnique.edu}
\affiliation{LULI - CNRS, CEA, Sorbonne Universit\'e, Ecole Polytechnique, Institut Polytechnique de Paris - F-91128 Palaiseau cedex, France}

\date{\today}


\begin{abstract}

Realizing the full potential of ultrahigh-intensity lasers for particle and radiation generation will require multi-beam arrangements due to technology limitations. Here, we investigate how to optimize their coupling with solid targets. Experimentally, we show that overlapping two intense lasers in a mirror-like configuration onto a solid with a large preplasma can greatly improve the generation of hot electrons at the target front and ion acceleration at the target backside. The underlying mechanisms are analyzed through multidimensional particle-in-cell simulations, revealing that the self-induced magnetic fields driven by the two laser beams at the target front are susceptible to reconnection, which is one possible mechanism to boost electron energization. In addition, the resistive magnetic field generated during the transport of the hot electrons in the target bulk tends to improve their collimation. Our simulations also indicate that such effects can be further enhanced by overlapping more than two laser beams.


\end{abstract}


\maketitle

\bigskip




\section{Introduction}
\label{INTRO}

The advent of multi-petawatt (PW) laser systems \cite{danson2019petawatt} opens new perspectives in many research areas, including compact particle and radiation sources \cite{fuchs2006laser, roth2013bright, higginson2018near}, condensed matter physics \cite{barberio2018laser}, probing of dense matter \cite{glinec2005high, manvcic2010picosecond, mahieu2018probing}, laboratory astrophysics \cite{chen2015scaling, higginson2019laboratory}, chemistry \cite{prasselsperger2021real}.
Yet, the feasibility of large-area gratings \cite{nguyen2006gratings} and mirrors with both broadband reflectivity and high-fluence-resistant coating \cite{chorel2018robust} currently limits the maximum power that can be delivered by a single laser beam to about 10~PW. Therefore, the quest for ever-increasing laser power will necessarily involve the combination of multiple independent beamlets, each being at the limit of the technology. 

Such a strategy is already being pursued in several projects, such as the Laser M\'egaJoule's PETAL system in France \cite{batani2014development}, the National Ignition Facility's Advanced Radiographic Capability (NIF-ARC) in the US \cite{di2015commissioning}, the Laser for Fast Ignition Experiment (LFEX) in Japan \cite{arikawa2016ultrahigh}, the Superintense Ultrafast Laser Facility (SULF) in China \cite{liang2020recent}. 
This approach, however, raises the question of how the individual beamlets should be arranged in order to maximize their overall coupling with the target. The present study will focus on opaque, solid targets, since the interaction of intense lasers with transparent, dilute plasmas, such as those suitable for wakefield acceleration of electrons, presents different challenges
\cite{steinke2016multistage, debus2019circumventing}. 
In addition, our investigations show that it is actually interesting to not only increase the energy of a single laser beam, as the multi-beam scheme benefits the quality of the produced particles.

The optimization of the coupling of a single, intense laser beam with a dense (overcritical) plasma has been the subject of extensive research \cite{santala2000effect, mackinnon2002enhancement, green2008effect, chawla2013effect, chen2013comparisons, fujioka2015heating, ziegler2021proton}. When using several beams, it was shown experimentally that temporally stacked laser pulses could enhance the guiding of hot electrons within the target \cite{scott2012controlling, malko2019enhanced}, or the target normal sheath acceleration (TNSA) \cite{wilks2001energetic, mora2003plasma} of ions  at the target rear side. The improvement of the latter process  was achieved either by laser shaping the target \cite{markey2010spectral, scott2012multi, morace2019enhancing} or by lengthening the effective ion acceleration time \cite{yogo2017boosting}.
Recently, an alternative scheme \cite{raymond2018relativistic, palmer2019field} employed two synchronized, but this time spatially separated, intense laser pulses, so that the antiparallel magnetic fields produced around the target surface by the laser-driven electron currents \cite{sarri2012dynamics, schumaker2013ultrafast} could reconnect. Magnetic reconnection (MR) is a process that converts magnetic field energy into kinetic particle energy \cite{nilson2006magnetic, rosenberg2015slowing}. As shown in Refs.~\cite{raymond2018relativistic, palmer2019field}, MR can boost the generation of nonthermal electrons in relativistic laser-plasma interactions. Another study using a similar beam arrangement conjectured that MR could also arise at the target back side \cite{golovin2020enhancement}, and hence impact ion acceleration. Other configurations have been tested numerically \cite{kim2022efficient}. For example, relativistic MR could be triggered by laser pulses propagating side by side in undercritical plasmas \cite{gu2019electromagnetic}.
It was also predicted that, compared to a single pulse, two laser pulses with halved intensity/energy and focused at opposite incidence angles onto sharp-gradient, thin solid foils could favor fast electron generation via vacuum heating, and thus also ion acceleration \cite{ferri2019enhanced, ferri2020effects, rahman2021particle}.

Complementing these previous works, we here examine, both experimentally and numerically, the processes of electron and ion acceleration in solids irradiated by two transversely separated, synchronized laser beams. In particular, we show that an optimum can be achieved that improves the particle yields and beam qualities. Note that, contrary to what is usually considered in numerical studies, the configuration addressed here involves a large-scale preplasma in front of the dense target, a common situation in realistic petawatt-level laser interactions \cite{burdonov2021characterization, raffestin2021enhanced}.
Our ultimate goal is to develop a testbed for future 10-PW-scale multi-beam laser facilities at the leading edge of technology, where the energy of each laser beam cannot be increased.


\begin{figure}
    \centering
    \includegraphics[width=0.45\textwidth]{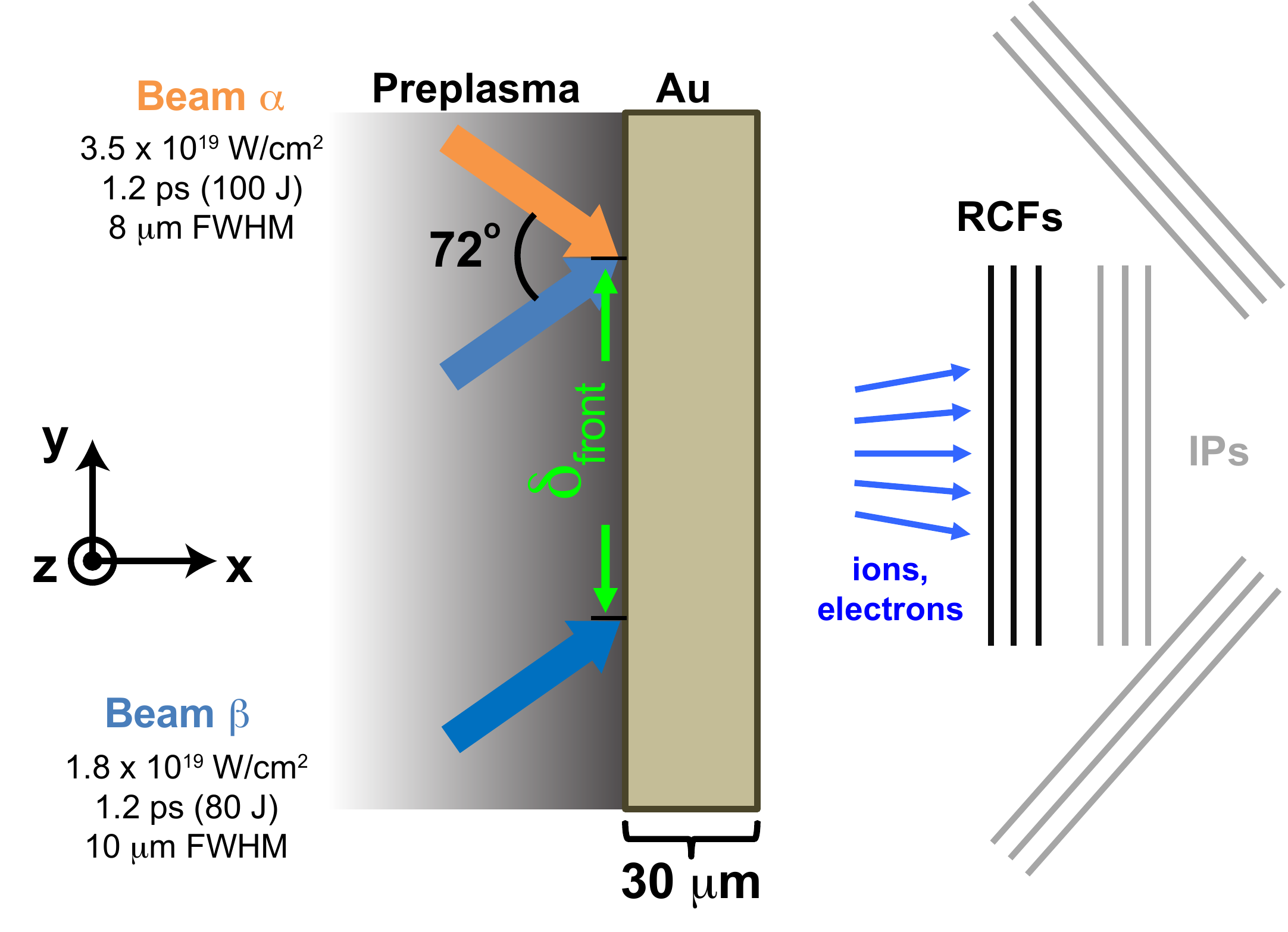}
    \caption{\textbf{Schematic of the experiment} using two intense laser beams (denoted as $\alpha$ and $\beta$) irradiating a solid Au target (with a large-scale preplasma at the target front), with opposite incidence angles and a variable separation distance ($\delta_{\rm front}$) between the laser spots on the target front surface. In all cases, the focus of the laser beams coincides with the target surface. The outgoing hot electrons are diagnosed by image plate (IP) stacks, located along each laser beam axis, as well as in the target normal direction. The accelerated ions are characterized by a radiochromic film (RCF) stack located in the target normal direction. 
    }
    \label{fig:setup}
\end{figure}

\section{Experimental results}
\label{EXP}

The experiment was carried out at the Vulcan Target Area West (TAW) laser facility at the Rutherford Appleton Laboratory (RAL). We will provide evidence that the use of two overlapping laser beams (denoted $\alpha$ and $\beta$ in the setup sketched in Fig.~\ref{fig:setup}) in a mirror-like geometry can substantially augment the hot-electron generation at the front side, the subsequent ion acceleration at the rear side, and the collimation of both outgoing electron and ion beams.

One beam ($\alpha$), with a pulse duration of approximately $1.2\,\rm ps$ and a pulse energy of around $100\,\rm J$, was focused onto the target \textcolor{black}{using an $f/3$ off-axis parabola} at an incidence angle of $36\degree$. This resulted in an $\sim 8\,\rm \mu m$ FWHM spot and an on-target intensity of $\sim 3.5\times 10^{19}\,\rm W\,cm^{-2}$. The other beam ($\beta$), with a pulse duration of $\sim 1.2\,\rm ps$ and a pulse energy of $\sim 80\,\rm J$, irradiated the target symmetrically, i.e., at an incidence angle of $-36\degree$. \textcolor{black}{Using also an $f/3$ off-axis parabola,} it was focused to a $\sim 10\,\rm \mu m$ FWHM spot \textcolor{black}{(due to slightly less optimal wavefront correction)}, yielding an on-target intensity of $\sim 1.8\times 10^{19}\,\rm W\,cm^{-2}$. Both laser beams had a $1.053\,\rm \mu m$ central wavelength and impinged onto the target at $p$-polarization. As they originated from the same oscillator, the shot-to-shot jitter was within the pulse duration, with a verified temporal overlap precision of $25\,\%$, and thus negligible.
\textcolor{black}{The pointing stability was approximately of 1 focal spot, i.e., $10\,\rm \mu m$. The focus of each laser beam remained constant within a longitudinal range of $\sim 100\,\rm \mu m$.} 
The targets consisted of $30\,\rm \mu m$-thick gold foils, \textcolor{black}{placed anew in the chamber before every shot with a $\sim 15\,\rm \mu m$ precision, and thus lying within the region of highest laser intensity.}

The outgoing proton distribution was recorded using a stack of radiochromic films (RCFs) \cite{bolton2014instrumentation}, centered along the rear target normal direction. The fast electrons escaping the target \cite{link2011effects} were detected by several stacks \cite{rusby2015measurement} of five photostimulable, FUJIFILM TR type image plates (IPs), each coated with a 1.5~mm Aluminum layer to filter out low-energy electrons. The IP stacks were placed along both the laser axis and the target normal directions. The preplasma expansion \cite{wagner2014pre} and the optical self-emission at $526 \pm 5\,\rm nm$ from the plasma were monitored along a line of sight parallel to the target surface.
The laser prepulse was measured to have an average intensity of $\sim 4.0\times 10^{13}\,\rm W\,cm^{-2}$ over a $\sim 0.5\,\rm ns$ duration \cite{chen2007creation}. The resulting preplasma was simulated by the radiation-hydrodynamics code \textsc{multi} \cite{Ramis_1988}, predicting a density scale-length $L_n \simeq 100\,\rm \mu m$ (fitting the density profile as $e^{-x/L_n}$, where $x$ is the longitudinal spatial coordinate). This value is consistent with the location of the edge of the refracted probe beam \cite{michaelis1981refractive}, observed to be $\sim 200\,\rm \mu m$ away from the target front. 

The distance between the centers of the two laser spots ($\delta_{\rm front}$), as measured at the initial (before preplasma formation) front side of the target, was consistently varied from $0\,\rm \mu m$ to $120\,\rm \mu m$, while keeping all other parameters constant. The following three irradiation configurations were considered:

\begin{description}
    \item[Case 0] A single laser beam, either beam $\alpha$ or beam $\beta$,  irradiated the target.
    \item[Case 1] Two non-overlapping laser beams were used with $\delta_{\rm front}$ up to $120\,\rm \mu m$.
    \item[Case 2] Two laser beams are overlapped at the front side of the target with $\delta_{\rm front} = 0$.
\end{description}

\textcolor{black}{Since the transverse preplasma scale-length exceeds the lateral shift imposed on each beam ($\delta_{\rm front}/2$), we do not expect this shift to modify the interaction physics, as can indeed be seen in the proton acceleration (as detailed below). Moreover, as the large extent of the preplasma hampers the propagation of the reflected beams, their mirror-type arrangement at the target front does not endanger the laser system.}

\begin{figure}
    \centering
    \includegraphics[width=0.45\textwidth]{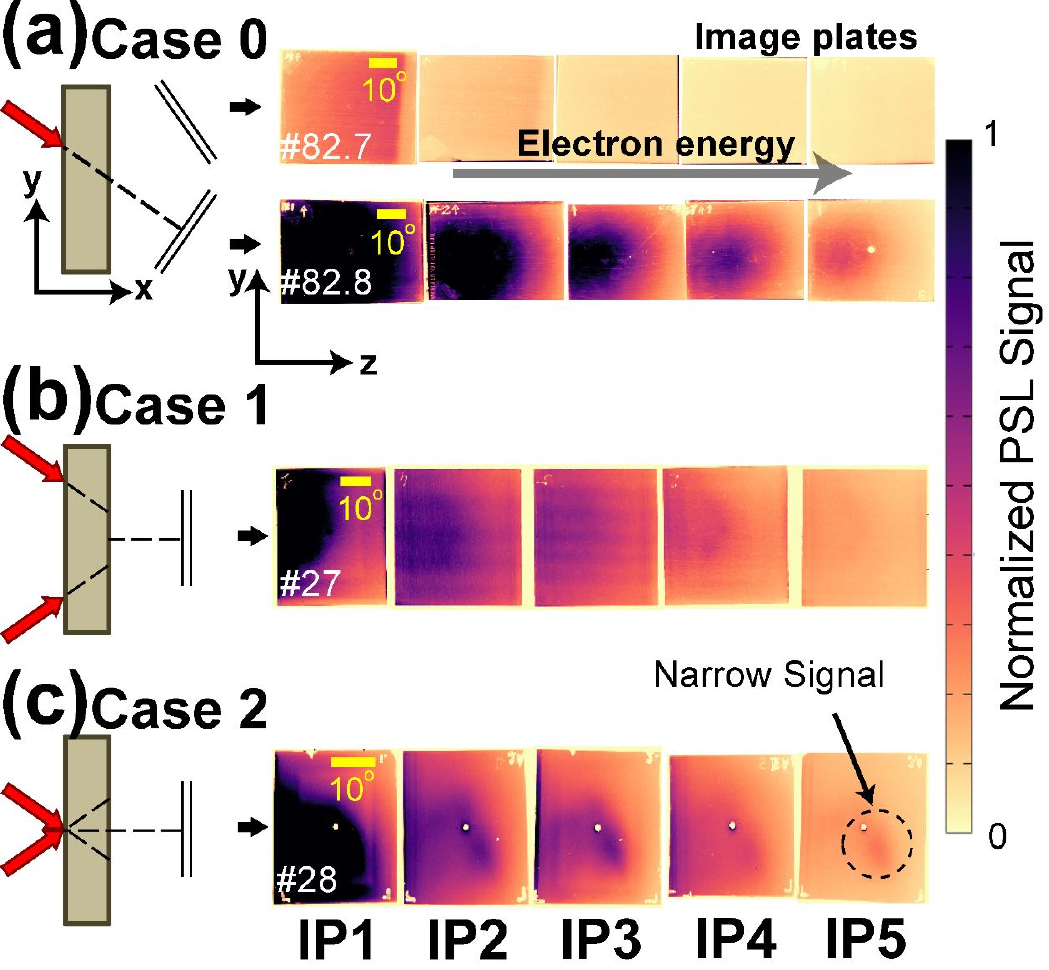}
    \caption{\textbf{Electron signals as recorded by the IP stacks} (a) along the laser direction for the single beam case (Case 0) and (b, c) along the target normal for the dual-beam cases, with either (b) $\delta_{\rm front} = 120\,\rm \mu m$ (Case 1) or (c) $\delta_{\rm front} = 0$ (Case 2). 
    Note that the stripped modulations observed for some shots are induced by a defective scanner readout, and hence are not physical.
    The associated laser and IP setup are sketched on the left. The IP were positioned with IP1 the closest to the target and IP5 the farthest from the target. Hence, the deepest IP5 can only be reached by the highest energy electrons.
    All IPs share the same colormap, as displayed on the right with normalized  photo-stimulated luminescence (PSL) number. The angular scales (yellow bars) vary between the IP images because these are located at different distances from the target. Note also that the laser beam axes do not intersect the IPs positioned along the target normal in panels (b) and (c).
    }
    \label{fig:IP_profile}
\end{figure}

The raw IP profiles recorded in Case 0 are displayed in Fig.~\ref{fig:IP_profile}(a). Notably, only the IPs positioned along the laser propagation direction detected a significant signal. This suggests that the energetic electrons responsible for this signal were mainly generated along the laser beam direction via the $\mathbf{j}\times \mathbf{B}$ mechanism in the preplasma at the target front \cite{wilks1992absorption, malka1996experimental}.

Figures~\ref{fig:IP_profile}(b) and (c) compare the raw IP data obtained along the target normal using two laser beams. For $\delta_{\rm front} = 120\,\rm \mu m$ [panel (b)], significant signals are observed only up to the 4th plate. By contrast, for $\delta_{\rm front} = 0$ [Fig.~\ref{fig:IP_profile}(c)], the signal shows a clear enhancement up to the 5th plate, implying a stronger emission of hot electrons along the target normal. Besides being intensified, the IP signals also appear to be narrower, suggesting that the hot electrons were more collimated, possibly due to enhanced resistive magnetic fields in the target bulk, as will be discussed later.

\begin{figure*}
    \centering
    \includegraphics[width=0.8\textwidth]{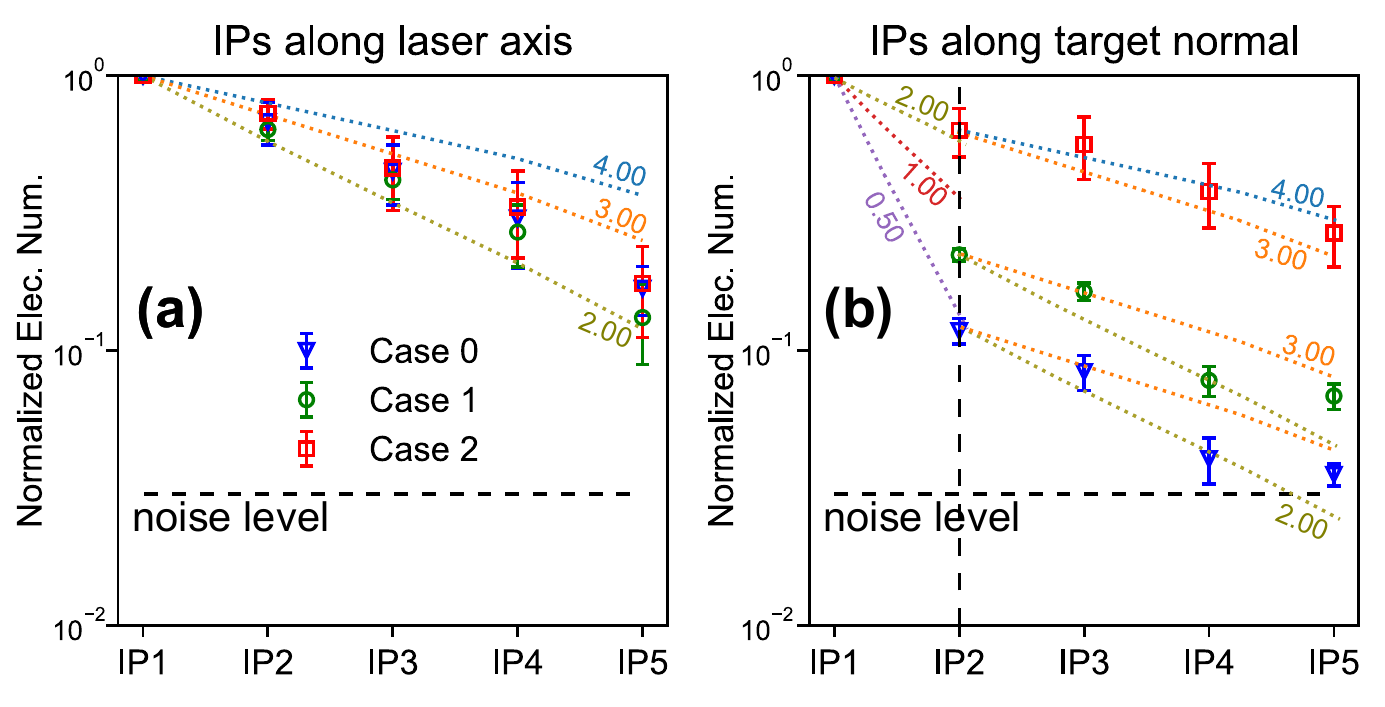}
    \caption{\textbf{Quantitative analysis of the electron signals as recorded by the IP stacks} (a) along the laser axis and (b) along the target normal. Note that the normalized electron number in each IP is retrieved from the variation in the IP signal using Monte Carlo \textsc{fluka} simulations and the calibration conducted in Ref.~\cite{boutoux2015study}. The data points represent the signal averaged over three shots performed under similar conditions, while the error bars correspond to the minimum and maximum values over those shots. The horizontal black dashed line represents the noise level at around 0.03. Simulation results are plotted as dotted lines and the associated numbers at the right end of the dotted lines indicate the injected electron temperatures (in MeV). Note that in (b), we use a two-temperature distribution, separated by the vertical black dashed line located at IP2.
    }
    \label{fig:IPs}
\end{figure*}

The spectrum of the hot electrons measured away from the target with the IPs is representative of the electrons inducing the ion-accelerating sheath field \cite{link2011effects}. 
To analyze the fast electron signal recorded by the IPs (see Fig.~\ref{fig:setup}), we followed the procedure detailed in Ref.~\cite{rusby2015measurement} and performed Monte Carlo \textsc{fluka} simulations \cite{fasso2005fluka, battistoni2007fluka, bohlen2014fluka}. 
Figure~\ref{fig:IPs} shows the experimental data (points with error bars) overlaid with the \textsc{fluka} simulation results (dotted lines as a ruler). The numbers at the right end of the dotted lines are the injected hot-electron temperatures ($T_h$, in MeV units) in the \textsc{fluka} simulations. By comparing the slope of the data points to the simulated dotted lines, we can retrieve the hot-electron temperature in the experiment for each case. Specifically, in Fig.~\ref{fig:IPs}(a), the IP data acquired along the laser axis suggest similar values of $T_h \simeq 3.0 \pm 1.0 \,\rm MeV$, consistent with the ponderomotive scaling \cite{kluge2011electron}. The contribution from Bremsstrahlung photons generated in the laser target is expected to be negligible \cite{rusby2015measurement}. \textcolor{black}{By contrast, the IP signals recorded along the target normal [Fig.~\ref{fig:IPs}(b)] can only be reproduced using two-temperature hot-electron distributions, with different temperatures values in the three cases.} 
In detail, for either well-separated laser beams (Case 1, green) or a single laser (Case 0, blue), the temperature retrieved after IP2 (i.e. the temperature fitting the slope of the data points from IP2 to IP5) is similar, i.e., around 2-3~MeV. However, when two overlapping laser beams are used (Case 2, red), it rises to 3-4~MeV. This enhanced temperature of the hot electrons is also accompanied by an order-of-magnitude increase in their number. 

\begin{figure*}
    \centering
    \includegraphics[width=0.8\textwidth]{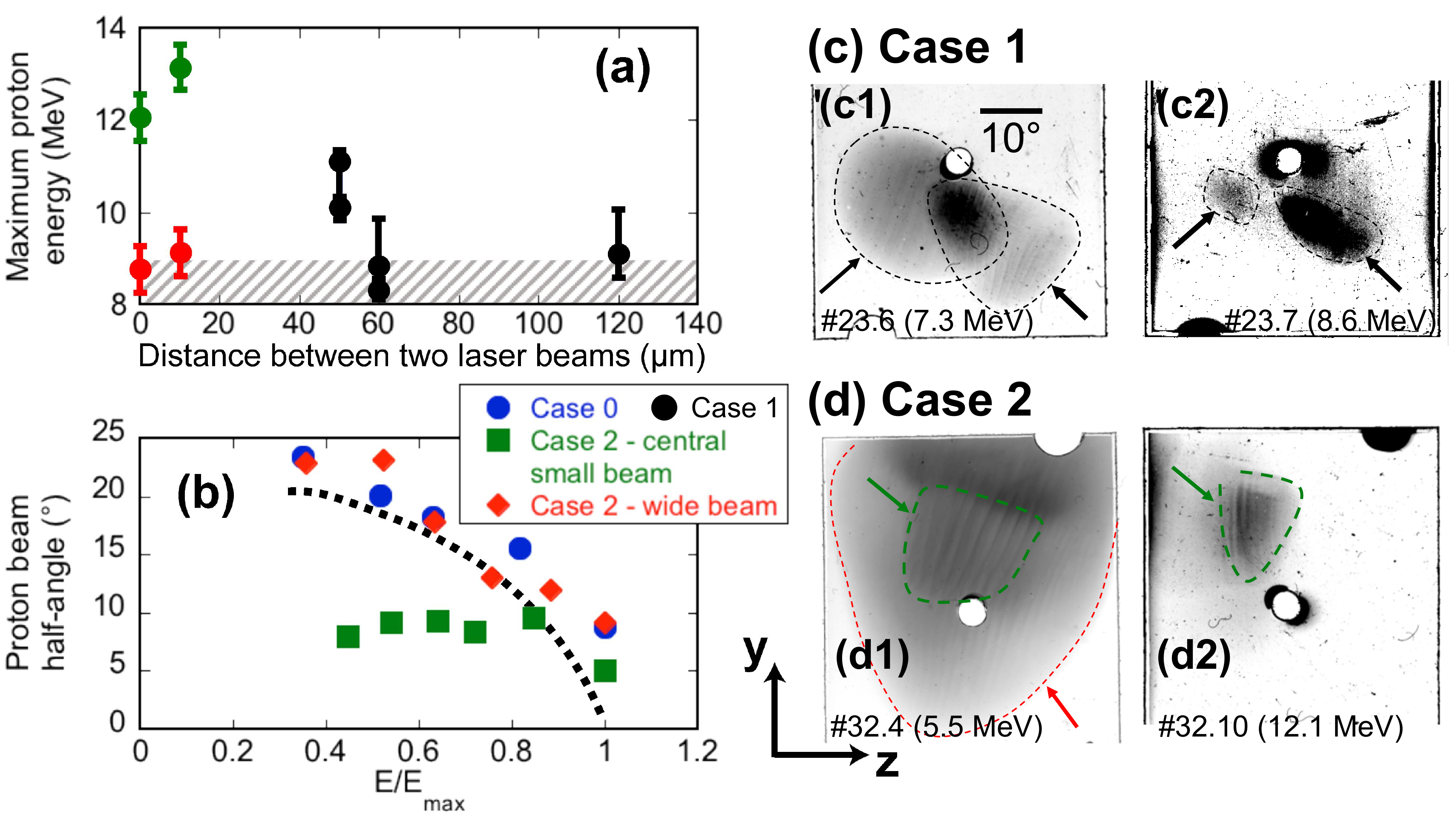}
    \caption{\textbf{Enhancement of the proton cutoff energy and collimation brought about by coupling two intense laser beams.} (a) Variation in the maximum proton energy (as inferred from the RCF data) when varying the spatial separation between the two beams at the front target surface, i.e., $\delta_{\rm front}$. The points represent the signal averaged over two to three shots performed in the same conditions, while the error bar represents the minimum and maximum values over those shots. The gray hashed area indicates the maximum proton energy obtained in the single-beam configuration (Case 0). The black data points correspond to two separated beams with varying interspacing \textcolor{black}{(Case 1)}, the green data points represent the central small beam in the overlapped configuration, i.e., Case 2, marked by the \textcolor{black}{green} dashed contour in panel (d); while the red data point represents the wide beam \textcolor{black}{(also in Case 2)}, having the standard divergence of TNSA proton beams, marked by the \textcolor{black}{red} dashed contour and blue arrows in panel (d). (b) Variation in the recorded half-angle subtended by the protons, as a function of their energy (normalized to the corresponding cutoff energy). The dashed line plots the energy-dependent angular distribution observed in many experiments to be characteristic of TNSA protons \cite{bolton2014instrumentation}. Note that the RCFs are positioned along the target normal, as shown in Fig.~\ref{fig:setup}. (c) Raw RCF data, corresponding to protons of (c1) 7.3 and (c2) 8.6~MeV mean energy, in Case 1 with well separated lasers ($\delta_{\rm front} = 120\,\rm \mu m$). Two distinct standard TNSA beams (driven simultaneously but independently) can be identified, as marked by the black dashed contours and arrows. (d) Raw RCF data, corresponding to protons of (d1) 5.5 and (d2) 12.1~MeV mean energy, in Case 2 with overlapping laser beams. Two different beam signals can be identified. One
    is identified by the \textcolor{black}{red dashed contour and} arrow. The other
    is identified by the \textcolor{black}{green} dashed contour and arrow. 
    Note that the hole in the RCF was managed for downstream spectrometry measurements (not shown).}
    \label{fig:RCF_proton}
\end{figure*}

The characteristics of the accelerated protons, as diagnosed by the RCFs, are summarized in Fig.~\ref{fig:RCF_proton}. The highest proton cutoff energy [Fig.~\ref{fig:RCF_proton}(a)] is obtained for a laser beam inter-spacing $\delta_{\rm front} \le 10\,\rm \mu m$ \textcolor{black}{(i.e., when the beams overlap within the pointing stability of the lasers)} whereas it quickly decreases to the value associated with a single beam (represented by the gray hashed area) when $\delta_{\rm front} \ge 60\,\rm \mu m$, (i.e., when the two beams no longer overlap). The slight deviation from zero for the optimal $\delta_{\rm front}$ value is ascribed to shot-to-shot fluctuations.
While the increase in proton energy when the laser beams are combined comes as no surprise (due to denser hot electrons in the sheath), there is clearly an unexpected benefit in terms of proton collimation. 

Figure~\ref{fig:RCF_proton}(b) depicts the variation in the proton angular divergence with energy (normalized to the cutoff energy) for the three cases considered. As a reference, the dashed curve plots the energy-dependent divergence obtained in Ref.~\cite{bolton2014instrumentation}. As expected, using a single laser beam (Case 0, blue dots) results in an angular distribution typical of TNSA.
In Case 1, each of the two observed proton beams follows the same trend, although this is not shown in Fig.~\ref{fig:RCF_proton}(b) for readability purposes. However, different results are obtained for the overlapping laser beams of Case 2. As evidenced by Fig.~\ref{fig:RCF_proton}(d1), the raw RCF signal then reveals a relatively wide proton beam (indicated by a black dashed contour and a blue arrow), characterized by a standard TNSA-type angular distribution [red diamonds in Fig.~\ref{fig:RCF_proton}(b)], inside which lies a narrower and denser beam (indicated by a yellow dashed contour and arrow) with a markedly smaller divergence [green squares in Fig.~\ref{fig:RCF_proton}(b)]. The energy cutoff of the wide beam is measured to be $\sim 9\,\rm MeV$ (see red points in Fig.~\ref{fig:RCF_proton}(a)), which is similar to that found in Cases 0 and 1 [see panel (c2)], but also smaller than the $\sim 12\,\rm MeV$ cutoff energy of the central, more collimated beam [see Figs.~\ref{fig:RCF_proton}(a) and (d2)]. The observation of the narrow central (and higher-energy) proton beam suggests that the enhancement in Case 2 does not merely result from the overlap of two TNSA beams.

If the laser beams do not overlap, as in Case 1 [Fig.~\ref{fig:RCF_proton}(c)], two TNSA-type proton beams are obtained without any enhancement, i.e., neither in energy nor in collimation. The two distinct proton beam envelopes that we observe are simply due to the large separation distance ($120\,\rm \mu m$) between the two laser beams. Thus, the centers of the sheaths produced by each beam are similarly separated. Knowing that each sheath has a diameter of the same order \cite{antici2008hot}, and that the proton beam pattern merely reflects the electron spatial distribution on the target rear \cite{cowan2004ultralow}, it is not surprising to observe two distinct proton beams separated by an amount of the order of each sheath diameter. 
The darker area seen at the intersection of the two proton beams in Fig.~\ref{fig:RCF_proton}(c1) simply originates from the addition of their respective dose depositions in the RCF; it is observed to disappear at higher proton energies [see Fig.~\ref{fig:RCF_proton}(c2)], as the corresponding protons have their angular opening reduced \cite{snavely2000intense, bolton2014instrumentation}. Moreover, the stripe structures visible in the raw RCF data of Figs.~\ref{fig:RCF_proton}(c) and (d) are due to modulations imprinted on the target backside \cite{cowan2004ultralow}, thus demonstrating that these protons are indeed accelerated from this side of the target.

The cutoff energy of TNSA protons is known to increase with the hot-electron temperature and density \cite{fuchs2006laser, mora2003plasma, mora2005thin}. The observed increase in proton cutoff energy therefore points to a more efficient conversion of the laser energy into hot electrons. It is also known that, due to their extremely low emittance, the proton beams detected on RCFs are a magnified projection of the surface of the accelerating sheath from which they originate \cite{cowan2004ultralow}. Hence, the reduced area of the fastest protons that we report here would be consistent with an accelerating sheath narrower than that generated under standard conditions, which results from the typical $30-40\degree$ divergence of the hot electrons driven by a single laser pulse \cite{adam2006dispersion, green2008effect}. Our RCF data thus suggest that the protons have been accelerated by a beam of higher-energy, lower-divergence electrons, which is consistent with the IP measurements of the hot-electron source [see Fig.~\ref{fig:IPs} (b)]. This is also supported by our analysis of the hot-electron transport within the target, described below.  



\section{Numerical simulations}
\label{PIC}

To pinpoint the mechanisms enhancing the generation and collimation of the hot-electron beam (HEB), we have carried out a series of particle-in-cell (PIC) simulations with the fully relativistic kinetic code \textsc{smilei} \cite{derouillat2018smilei}. Due to difficulties in simulating the multidimensional dynamics of the laser-driven solid Au target on a picosecond timescale while taking into account both dynamic ionization and collisions, those simulations were performed into two stages.

\begin{description}
    \item[Stage 1] addressed the enhanced HEB generation during the laser beam propagation in the preplasma. Since the main potential mechanism is the MR induced in the dilute fully ionized preplasma, a 3D geometry without neither ionization nor collisions was used.
    \item[Stage 2] focused on the enhanced HEB collimation. Since the main potential mechanism involves the resistive magnetic field induced during the HEB transport through the target bulk, both collisions and ionization processes were considered in 2D geometry.
\end{description}

\subsection{Stage 1: MR-enhanced HEB generation in the preplasma}
\label{stage1}

The computational cost of the 3D simulations in Stage~1 forced us to run them under down-scaled conditions, i.e., with a similar reduction factor for both the laser separation distance and the laser spot size. In addition, we considered a target made of helium ions instead of gold. Despite such limitations, these simulations could capture the essence of the main physical mechanisms at play.

\begin{figure}[ht]
    \centering
    \includegraphics[width=0.45\textwidth]{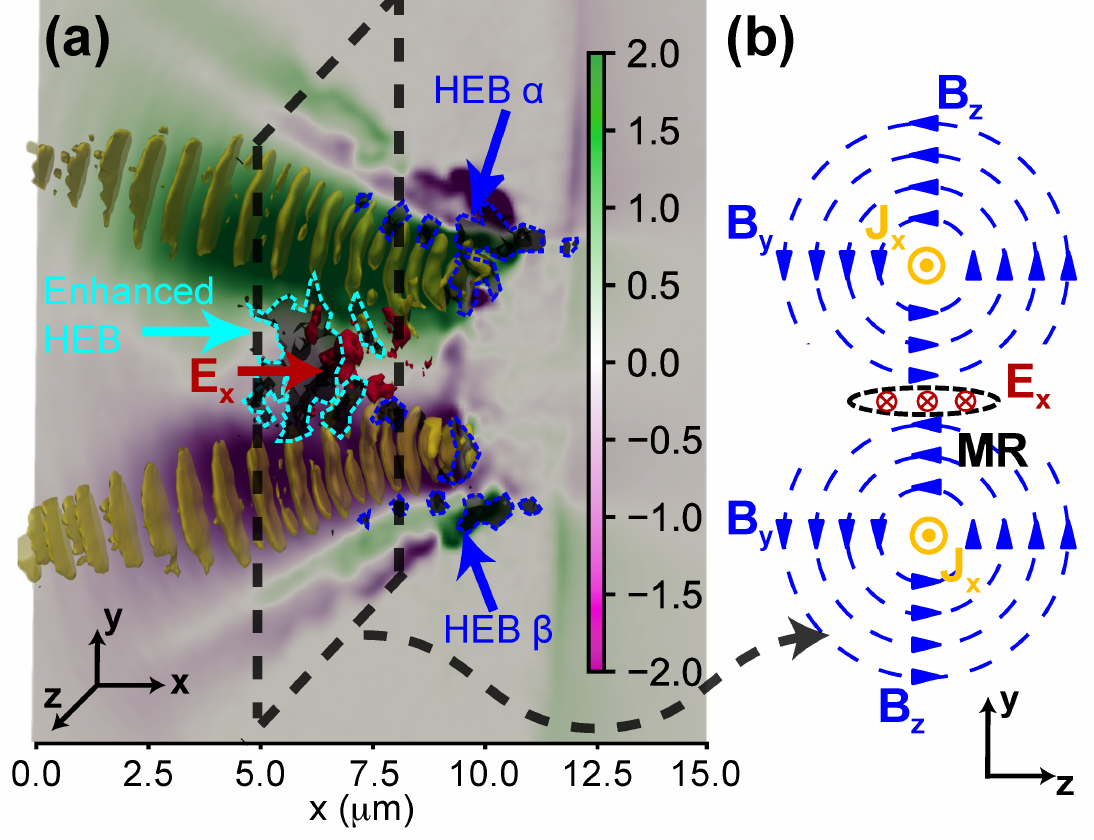}
    \caption{\textbf{Electron acceleration, magnetic field generation and magnetic reconnection in the laser-driven preplasma corresponding to Case 2.} 
    (a) 3D rendering (yellow) of \textcolor{black}{($\vert E_y \vert$)}, depicting the two laser pulses coming from the left and focused onto the target surface ($x=12\,\rm \mu m$). Overlaid is the 2D $yz$ projection of the self-generated, \textcolor{black}{ laser-cycle-averaged $\langle B_z \rangle$} magnetic field (colormap on the right, in $m_e\omega_{pe}/e$ units). The HEBs (grey volume rendering surrounded by blue dashed lines) generated (with energies above 5~MeV) by the $\alpha$ and $\beta$ laser beams are indicated by blue arrows.
    The enhanced HEB (grey volume rendering surrounded by cyan dashed lines) around the MR region (between the laser beams) is marked by the cyan arrow (around $x = 6.5\,\rm \mu m$), while the MR-induced $E_x$ field (red) is highlighted by the red arrow. (b) Sketch of MR in a $yz$ plasma slice ahead of the target surface, as indicated by the dashed black box in (a). $J_x$ represents the electron current density along the longitudinal $x$ direction, $B_y$ and $B_z$ are the in-plane magnetic fields, and $E_x$ is the MR-driven, out-of-plane electric field. The MR region is delineated by black dashed lines.}
    \label{fig:simu_directional}
\end{figure}

The 3D simulations employed a box size of $L_x \times L_y \times L_z = 20 \times 30 \times 30\,\rm \mu m^3$. \textcolor{black}{The target was modeled as a fully ionized He plasma of electron density profile
\begin{align}
     n_e(x) &= n_{e,\rm max} \nonumber \\
     & \times
\begin{cases}
      \exp \left[\ln\left (\frac{n_{e,\rm max}}{n_{e,\rm min}}\right)\frac{x-l_0}{l_0}\right] & x\le 12\,\rm \mu m \\
      1 & 12 \le x \le 15\,\rm \mu m \\
      0 & x > 15\,\rm \mu m 
    \end{cases}    
    \label{eq:prep}
\end{align}}
with $n_{e,\rm min} = 0.2 n_c$, $n_{e,\rm max} = 7.5 n_c$, and $l_0= 12\,\rm \mu m$. Here $n_c \equiv \epsilon_0 m_e(2 \pi c)^2 /(e \lambda)^2 = 1.1 \times 10^{21}\,\rm cm^{-3}$ denotes the critical density ($m_e$ is the electron mass, $e$ the elementary charge, and $\epsilon_0$ the vacuum permittivity). To save computational time, only the He$^{2+}$ ions from the dilute preplasma region ($x\le 12\,\rm \mu m$) were allowed to move. The exponential density profile of the latter is characterized by a scale-length $L_n \simeq 4\,\rm \mu m$; this value is much shorter than in the experiment due to our limited computational resources, but we checked that the simulation results remained qualitatively similar when doubling it. The electron critical surface was located at $x = 6.5\,\rm \mu m$ [see Fig.~\ref{fig:simu_directional} (a)]. The density profile was taken to be uniform in the transverse ($yz$) plane. The plasma was initialized with a temperature of $0.5\,\rm keV$.

The two laser beams, injected from the left boundary ($x=0$) and $p$-polarized (i.e., with their electric field lying in the $xy$ propagation plane), were focused at oblique incidence ($\pm 15^\degree$ relative to the target normal) on the dense plasma surface ($x=12\,\rm \mu m$). They both had a wavelength of $\lambda = 1 \,\rm \mu m$ and a Gaussian intensity profile of waist $\sigma_L = 2\,\rm \mu m$. Their peak intensity (in vacuum) was $I = 3.5 \times 10^{19}\,\rm W\,cm^{-2}$, corresponding to a dimensionless field strength of $a_0 = [I\lambda^2\mu_0 e^2/(2 \pi^2 m_e^2 c^3)]^{1/2} \simeq 5.0$. The preplasma was large enough to prevent the formation of a coherent beating wave pattern between the laser beams \cite{ferri2019enhanced, ferri2020effects}. Their angular separation was smaller than in the experiment, a consequence of the limited simulation domain; however, it corresponded to that in compact beam stacking geometry, i.e., one where two $\sim f/2$ laser beams would irradiate the target side by side. To simplify the analysis, the temporal laser intensity profile consisted of a one-cycle ($t_0 = 3.3\,\rm fs$) long ramp followed by a plateau of $\tau_L = 300t_0 = 1.0\,\rm ps$ duration. A simulation run with a Gaussian temporal profile of $\sim 500\,\rm fs$ FWHM yielded quite similar results. The transverse separation distance between the two focal spots was set to $20\,\rm \mu m$ in Case 1 and $5.5\,\rm \mu m$ in Case 2.

\begin{figure*}[ht]
    \centering
    \includegraphics[width=0.7\textwidth]{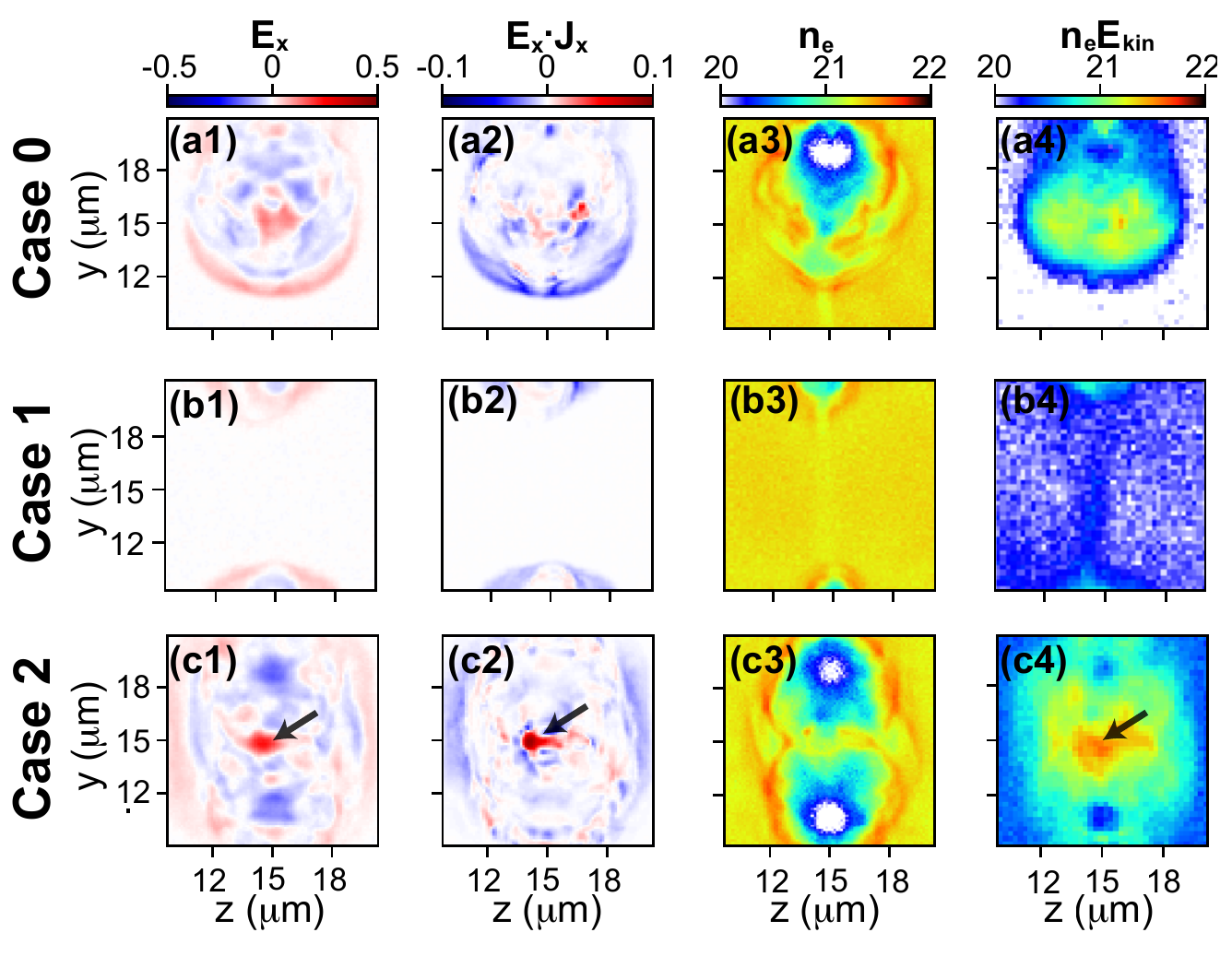}
    \caption{\textbf{Features of MR and resulting enhanced HEB generation} (a1)-(c1) Out-of-plane electric field $E_x$ ($m_ec\omega_{pe}/e$ units). (a2)-(c2) Time-averaged power density due to out-of-plane electric field, $\langle J_x\, E_x \rangle$ ($m_e c^2 \omega_{pe} n_c$ units). (a3)-(c3) Electron number density $n_e$ ($\rm cm^{-3}$ units) in log$_{10}$ scale. 
    (a4)-(c4) Kinetic energy density $n_e E_{\rm kin}$ ($\rm MeV\,cm^{-3}$ units) of electrons above 1~MeV in log$_{10}$ scale. Top row: one laser beam (Case 0). Middle row: two nonoverlapping laser beams (Case 1). Bottom row: two overlapping laser beams (Case 2). All panels show $yz$ slices taken at $t=440\,\rm fs$ and averaged longitudinally over $6< x < 10\,\rm \mu m$, i.e., in front of the dense plasma region ($12 < x < 15\,\rm \mu m$).
    }
    \label{fig:simu_enhancement}
\end{figure*}

The mesh size was set to $\Delta x = \Delta y = \Delta z = d_e$, where $d_e \equiv c /\omega_{pe} = c\sqrt{m_e \epsilon_0 / n_{e,\rm max} e^2}$ is the electron inertial length (corresponding to 16 cells per laser wavelength). The temporal resolution was of $\Delta t = 0.5 \Delta x/c$. The plasma electrons were initially represented by 8 particles per cell, with a fourth-order shape function. Boundary conditions for both particles and fields were absorbing along $x$ and periodic in the other directions. Binary collisions were not included, a reasonable approximation given the $>100\,\rm keV$ electron temperatures reached in the interaction region where MR arises.

Figure~\ref{fig:simu_directional}(a) displays the regions of HEB generation in the preplasma traversed by the two laser beams in Case 2. In line with the experimental data (Fig.~\ref{fig:IPs}), each laser beam generates its own HEB, mainly directed along the laser propagation direction. The magnetic fields self-induced around the laser beams are strong enough \textcolor{black}{($\sim m_e \omega_{pe}/e \sim 2.7\times 10^4\,\rm T$}, 
consistent with previous measurements on the same laser system \cite{sarri2012dynamics}) to confine the MeV-range electrons around the laser paths \cite{perez2013deflection}. In addition to the HEBs directly originating from the laser beams, electron energization also takes place in the region between the laser paths, a phenomenon which we attribute to MR.  

Figure~\ref{fig:simu_directional}(b) shows schematically how MR can arise in a transverse ($yz$) plane [represented by the dashed black box in panel (a)] if the laser beams are close to each other ahead of the target bulk. MR will induce an out-of-plane ($E_x$) electric field pointing to the target bulk. This polarity differs from that reported in Ref.~\cite{raymond2018relativistic}, because the reconnecting magnetic fields were then produced by electron currents outgoing from the target surface.

Features supporting the occurrence of MR and enhanced HEB generation, are provided in Fig.~\ref{fig:simu_enhancement}.
We start by comparing Cases 0 and 2. The first column of Fig.~\ref{fig:simu_enhancement} depicts the out-of-plane ($E_x$) electric field in the transverse $yz$-plane. In Case 2 (c1), it reaches $0.3\,m_ec\omega_{pe}/e$ around the midpoint (indicated by the black arrow), while in Case 0 (a1), it remains below $0.1\,m_ec\omega_{pe}/e$. This twofold increase in $E_x$ suggests that an additional source of $E_x$, other than the sheath field at the target front, is operative in Case 2. The second column presents the laser-cycle-averaged power density transferred to the electrons by the $E_x$ field, i.e., $\langle J_x E_x \rangle$ (with $J_x$ as the electron current density)
In Case 2 (c2), a significant positive signal is visible around the midpoint (indicated by the black arrow), indicating net energy transfer from the electromagnetic fields to the electrons, hence accounting for the local HEB generation observed in Fig.~\ref{fig:simu_directional}(a). Conversely, no such signal is observed for Case 0 [panel (a2)]. \textcolor{black}{The features revealed in panels (c1) and (c2) are consistent with findings from \citet{xu2016characterization}, supporting the conclusion that MR is responsible for the peak in $E_x$ around the midpoint region.}


\begin{figure*}[ht]
   \centering
   \includegraphics[width=0.6\textwidth]{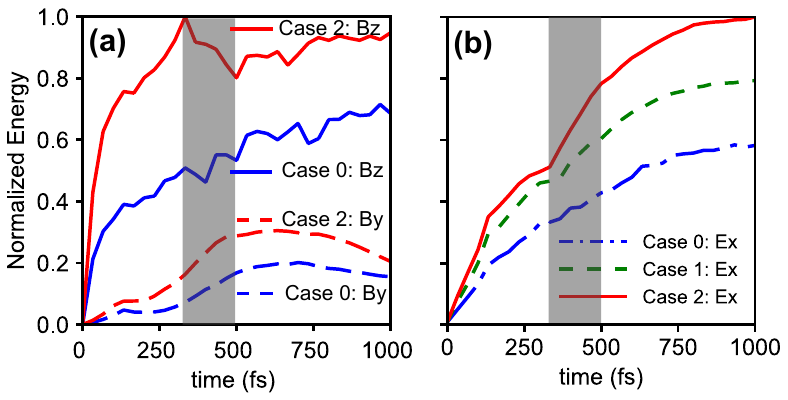}
   \caption{\textcolor{black}{\textbf{Time evolution of the spatially integrated electromagnetic field energies.}
   (a) Energies associated with $B_y$ and $B_z$ in Cases 0 and 2. All curves are normalized to the maximum of the $B_z$ energy in Case 2.
   (b) Energies associated with $E_x$ in Cases 1-3. All curves are normalized to the maximum of the $E_x$ energy in Case 2.
   In each panel, the dark vertical band indicates the period ($330\lesssim t \lesssim 500\,\rm fs$) during which MR is effective in Case 2, i.e., when a portion of the $B_z$ energy [red solid line in (a)] is transferred to both $B_y$ [red dashed line in (a)] and $E_x$ [red solid line in (b)]. The history of the $E_x$ energy in Case~1 is plotted to quantify the contribution from the additional laser beam.
   }}
   \label{fig:simu_energy_partition}
\end{figure*}

The occurrence of MR coincides with an enhancement of HEB generation. Upon comparing Cases 0 and 2, as illustrated in Figs.~\ref{fig:simu_enhancement}(a3) and \ref{fig:simu_enhancement}(c3), we observe that around the two laser spots, the preplasma electrons have been pushed away, leading to the formation of electron density holes (with a number density lower than $0.1n_c$ in the third column). However, the electron density at the center of the $yz$-plane ($y = z= 15\,\rm \mu m$) remains close to $\sim n_c$ in both cases. Regarding the energy density (fourth column), at the same midpoint of the $yz$-plane (indicated by the black arrow), the energy density of the HEB in Case 2 (c4) exhibits a substantial enhancement compared to that in Case 0 (a4). In short, the clear positive spatiotemporal correlation between the occurrence of MR and the enhancement of the HEB suggests that the former could be responsible for the latter. In turn, the higher energy density of the fast electrons injected within the target is likely to boost proton acceleration from the target backside.

We now proceed to compare the characteristics of MR and HEB generation between Cases 1 and 2. In Case 1, where the two focal spots are relatively distant from each other, both $E_x$ and its associated work are weak, making them barely discernible in Figs.~\ref{fig:simu_enhancement}(b1) and (b2). Furthermore, an almost uniform electron number density distribution between $10 \leq (y,z) \leq 20\,\rm \mu m$ is observed in (b3), with no HEB generation evident in (b4). The absence of electron-depleted regions is due to the laser spots lying outside the plotted ranges. At a later time ($t\simeq 900\,\rm fs$), when the self-generated magnetic structures have extended close to each other, their strength has dropped with their expansion radius ($r$) as $1/r$. Consequently, no MR features and HEB enhancement are observed.

\textcolor{black}{In addition to the previous analysis, another significant aspect of MR manifests in the temporal evolution of the simulated energy distribution. As is shown in Fig.~\ref{fig:simu_energy_partition}(a), in Case 2 where MR occurs, the energy associated with $B_z$ (red solid line) exhibits a decline from $t=330\,\rm fs$ to $t=500\,\rm fs$ (indicated by the dark vertical band), while the energy of $B_y$ (red dashed line) demonstrates a concurrent increase. This trend aligns with the MR process illustrated in Fig.~\ref{fig:simu_directional}(b), involving the dissipation of magnetic field energy along the $z$-axis and the reconnection of magnetic field lines along the $y$-axis. Conversely, such a behavior is absent in Case 0, where a single laser beam is employed. Here, both the $B_z$ and $B_y$ energies are increasing due to the laser's energy input into the simulation box. The higher $B_z$ energy results from the contribution of the laser's own $B_z$ field. The final decrease in $B_y$ at late times arises from the imposed boundary conditions of the finite simulation box.}

\begin{figure}[ht]
    \centering
    \includegraphics[width=0.4\textwidth]{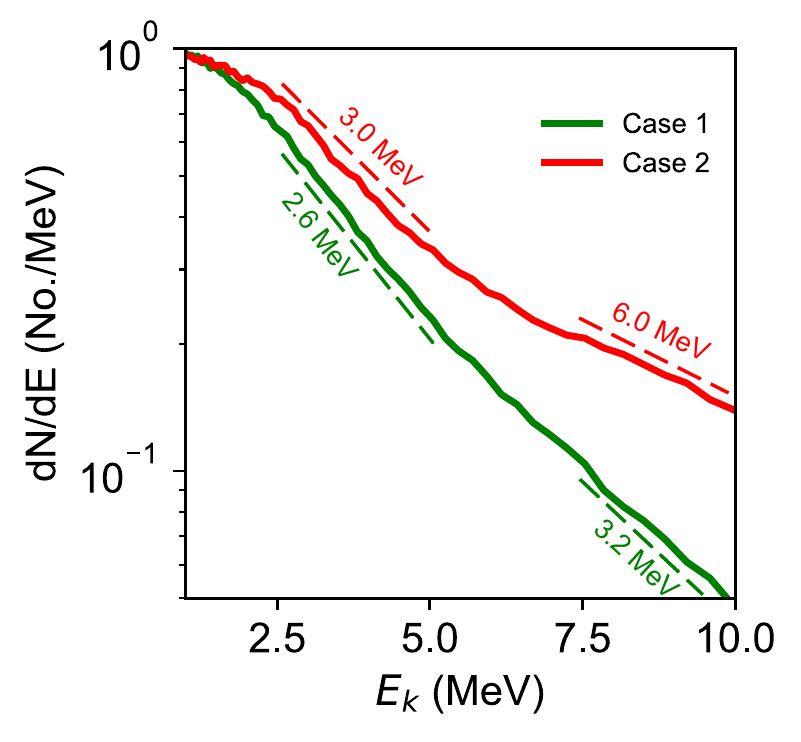}
    \caption{\textbf{Comparison of the outgoing hot-electron spectra in simulation Cases 1 and 2}. The energy spectra are recorded behind the target, across the $x=18\, \rm \mu m$ plane.
    Each curve is normalized to its maximum value for better comparison, and fitted by two temperatures over energy ranges indicated by the thin dashed lines.}
    \label{fig:simu_spectrum}
\end{figure}

\begin{figure*}
    \centering
    \includegraphics[width=0.6\textwidth]{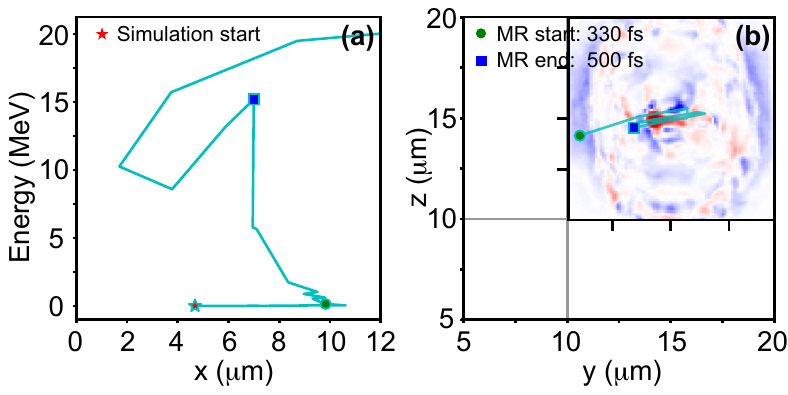}
    \caption{\textbf{Dynamics of a representative MR-accelerated electron.} Electron trajectory as recorded in the (a) $x$-energy and (b) $y$-$z$ spaces. In (b) the trajectory is superimposed on the $\langle J_x E_x \rangle$ map of Fig.~\ref{fig:simu_enhancement}(c2).
    In both panels, the red star marks the starting point of the trajectory indicates the approximate onset (respectively, end) of MR. A large energy gain is observed within the region and time period of MR activity. 
    }
    \label{fig:trace}
\end{figure*}

\textcolor{black}{Simultaneously, as depicted in Fig.~\ref{fig:simu_energy_partition}(b), the energy associated with $E_x$ undergoes a rapid increase in Case 2 (red solid line) during the MR period. This contrasts with the steady increase observed in Case 0 (blue dash-dotted line), which is attributed to the continuous laser input. Additionally, the evolution of the $E_x$ energy in Case 1 (green dashed line) demonstrates that the enhancement observed in Case 2 cannot solely be attributed to the additional laser beam. The higher energy of $E_x$ in Case 2 compared to Case 1 suggests the operation of an additional generation mechanism for $E_x$, namely MR, when the two laser beams overlap.}

Figure~\ref{fig:simu_spectrum} represents the simulated outgoing electron spectra as recorded across a plane located at $x=18\,\rm \mu m$. They exhibit a trend similar to the experimental results along the target normal direction [see Fig.~\ref{fig:IPs}(b)]. Specifically, in Case 2, there is a noticeable increase in the hot electron temperature compared to Case 1. Note, however, that the simulated temperatures can only be qualitatively compared to the experimental data due to several factors: (i) the use of down-scaled simulation parameters constrained by computational limits, (ii) a laser incidence angle of 15$\degree$ in the simulation, rather than 36$\degree$ in the experiment due to spatial limitations of the down-scaled model, and (iii) the use of helium ions, instead of gold ions, in the simulated preplasma. Thus, the simulated temperatures tend to exceed the experimental ones, as anticipated.

To illustrate the role of MR in the electron energization, we plot in Fig.~\ref{fig:trace} the trajectory of a representative accelerated electron in both the $x$-energy and $y$-$z$ spaces.
One can see that prior to the onset of MR (from the red star to the green circle), the electron kinetic energy remains quite weak (below 1~MeV), but rapidly rises to $\sim 15\,\rm MeV$ during MR (from the green circle to the blue square). 
Throughout this phase, the electron moves in the preplasma towards $x<0$ (away from the target surface), consistent with the $E_x>0$ field induced by MR.
\textcolor{black}{The observed energy gain is consistent with the mean power density transferred to the plasma electrons in the MR region [Fig.~\ref{fig:simu_enhancement}(c2)]. Taking $n_e \sim n_c$ gives an energy increase rate of $0.1 m_ec^2\omega_{pe}$ for a relativistic ($v\sim c$) electron. Over the $\sim 100\,\rm fs$ period spent by the electron in the MR region, one therefore expects an energy gain of $\sim 10\,\rm MeV$, close to what is observed.}

Later on, the tracked electron ends up moving towards $x>0$ (and hence can contribute to the spectrum of Fig.~\ref{fig:simu_spectrum}), possibly due to a combination of the laser field and self-generated magnetic field. \textcolor{black}{The latter indeed attains a strength of $\sim 10^4\,\rm T$ [Fig.~\ref{fig:simu_directional}(a)], resulting in a $\sim 1\,\rm \mu m$ Larmor radius for a typical 10~MeV electron. This value is consistent with the late stage of the trajectory plotted in Fig.~\ref{fig:trace}(a).} 

Although other electron energization mechanisms like direct laser acceleration (DLA) may operate under interaction conditions similar to those in our experiment, involving ps-scale pulse durations and large-scale preplasmas \cite{kemp2020direct}, that the electron energy jump seen in Fig.~\ref{fig:trace} is only reproduced in Case~2 suggests that MR -- instead of DLA -- plays a key role in the energy boost observed with overlapping beams. This conclusion is further supported by the consistent results obtained using a larger simulation box ($L_x = 30\,\rm \mu m$), a doubled preplasma scale-length ($L_n \simeq 8\,\rm \mu m$) or a Gaussian temporal laser profile.

\begin{figure}
    \centering
    \includegraphics[width=0.45\textwidth]{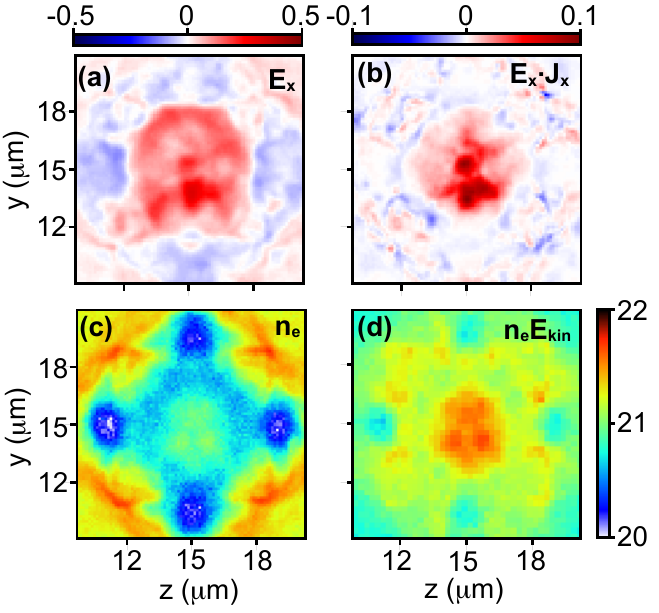}
    \caption{\textbf{Four-beam case.} (a) Out-of-plane electric field $E_x$ ($m_e c\omega_{pe}/e$ units). (b) Time-averaged power density $\langle J_x E_x \rangle$ ($m_e c^2 \omega_{pe} n_c$ units). (c) Electron number density $n_e$ ($\rm cm^{-3}$ units) in log$_{10}$ scale. (d) Kinetic energy density $n_e E_{kin}$ ($\rm MeV\,cm^{-3}$ units) of electrons above 1~MeV in log$_{10}$ scale. All panels show $yz$ slices recorded at $t = 440\,\rm fs$ and averaged over $6 < x < 10 \,\rm \mu m$ in front of the solid target . Panels (c) and (d) share the same colormap.}
    \label{fig:simu_fourbeam}
\end{figure}

To evaluate the impact of additional beams on the HEB generation, we have simulated a four-beam scenario. Figure~\ref{fig:simu_fourbeam}(a) shows that the accelerating electric ($E_x >0$) reconnection field then extends over a much larger area than in the corresponding two-beam case [Fig.~\ref{fig:simu_enhancement}(c1)], and so does the associated power density distributions [compare Fig.~\ref{fig:simu_enhancement}(c2)] and Fig.~\ref{fig:simu_fourbeam}(b)]. 
More quantitatively, we have calculated the electron energy enhancement factor, defined as the ratio of the integrated kinetic energy of the preplasma electrons between different configurations. When changing from single-beam to double-beam irradiation [compare Fig.~\ref{fig:simu_enhancement}(a4) and Fig.~\ref{fig:simu_enhancement}(c4)] and from double-beam to four-beam [compare Fig.~\ref{fig:simu_enhancement}(c4) and Fig.~\ref{fig:simu_fourbeam}(d)], that factor reaches $\sim 3$ during the laser interaction and then stabilizes to $\sim ~2$ (single-beam vs. double-beam) or $\sim 1.5$ (double-beam vs. four-beam) at later times. In detail, increasing the number of beams both enhances the mean energy and number of the hot electrons accelerated within a larger reconnection region. 

In summary, our down-scaled 3D PIC simulations provide numerical evidence of boosted HEB generation when two (or four) laser beams overlap in a mirror-like geometry. While these findings align with the experimental data, they do not demonstrate any improved collimation of the HEB through the dense target region. To tackle this issue, we now turn to larger-scale 2D simulations that treat both collisional and ionization effects.

\subsection{Stage 2: Magnetically collimated HEB transport in the resistive target bulk}
\label{stage2}

The propagation of the HEB in the collisional solid target in Stage 2 was simulated in a 2D domain of size $L_x\times L_y =10 \times 40\,\rm \mu m^2$. The target density profile obeyed Eq.~\eqref{eq:prep}. The preplasma profile, characterized by  $l_0 = 4\,\rm \mu m$, increased from $n_{i,\rm min} = 0.01 n_c$ at $x = 1\,\rm \mu m$ to $n_{i,\rm max} = 50 n_c$ at $x = 5\,\rm \mu m$, and was followed by a $4\,\rm \mu m$-long plateau at $n_{i,\rm max} = 50 n_c$ over $5 \le x \le 9\,\rm \mu m$. 
To reduce the computational cost, an aluminium target was used instead of gold. The initial charge state and temperature of the Al ions (of mass $m_i = 49572\,m_e$) were set to $Z^\star = 5$ and $T_i = 160\,\rm eV$ in the preplasma, and $Z^\star = 3$ and $T_i = 30\,\rm eV$ in the dense region.
Each plasma species was modeled by 32 particles per cell, with fourth-order interpolation functions. 
Coulomb collisions between all particle species and electron impact ionization \cite{perez2012improved, higginson2020corrected} were described together with field-induced ionization \cite{nuter2011field}. The spatial resolution was $\Delta x = \Delta y = d_e$, where $d_e$ represents the electron inertial length in the fully ionized dense region ($Z^\star = 13$, $n_{e,\rm max} = 650n_c$). This corresponds to about 160 cells per laser wavelength. 
The temporal resolution was set to $\Delta t = 0.5 \Delta x/c$, in order to safely use the Friedman temporal filter \cite{greenwood2004elimination}. A multi-pass binomial filter was also applied to particle current densities to further mitigate numerical heating \cite{vay2011numerical}. 

We considered the laser parameters of Case~0 (single beam) and Case~2 (two beams with $\delta_{\rm front}=0$), but with a larger waist ($\sigma_L = 6\,\rm \mu m$) and a longer up-ramp (33~fs). Particles were absorbed across the $\pm x$ boundaries and thermally reinjected across the $\pm y$ boundaries. Absorbing boundary conditions were used for the fields in all directions.


\begin{figure}
    \centering
    \includegraphics[width=0.45\textwidth]{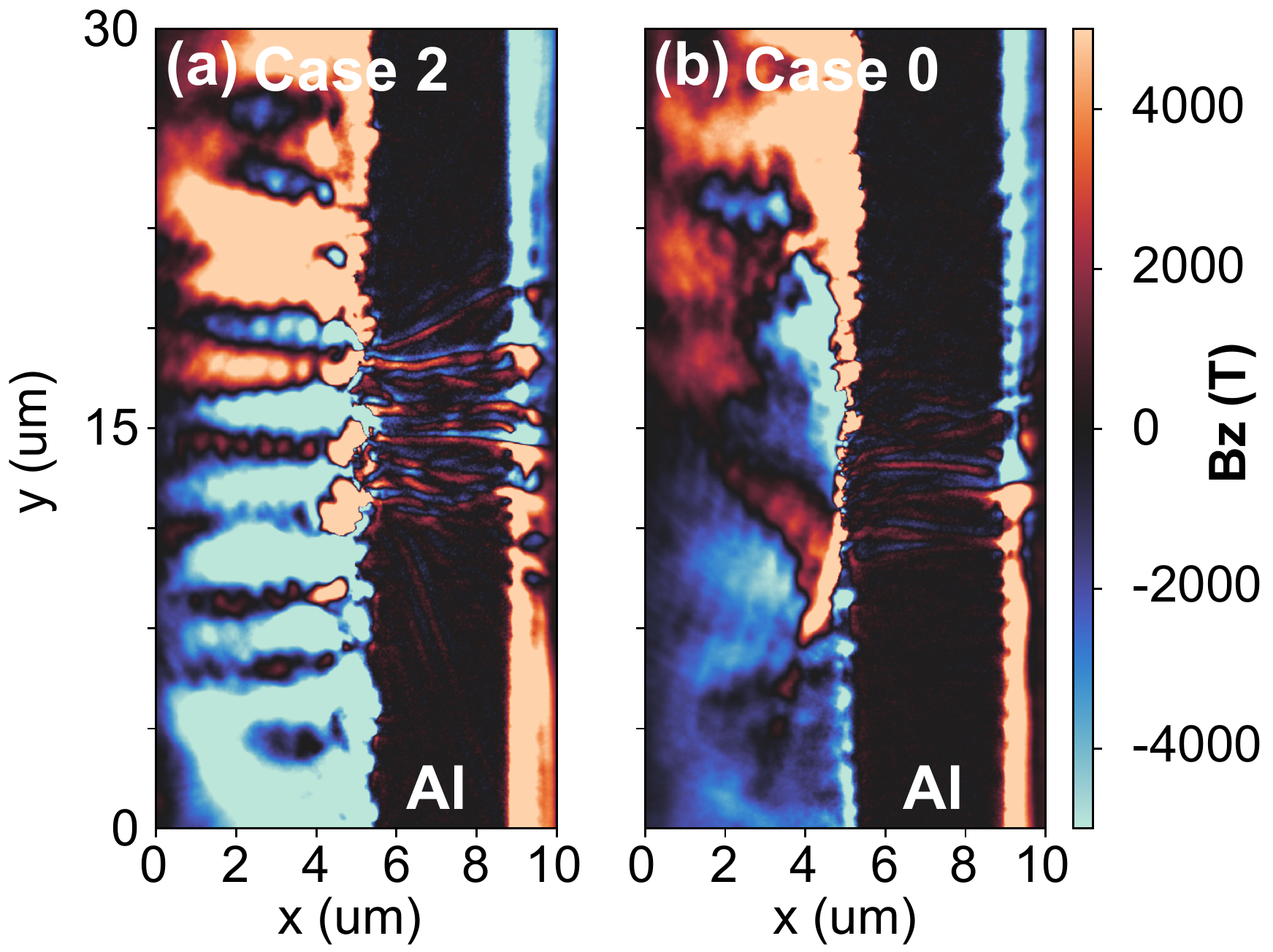}
    \caption{\textbf{Resistive magnetic field generation in the collisional dense target.} (a) Case 2: two laser beams with $\delta_{\rm front} = 0$. (b) Case 0: single beam. Both snapshots are taken 1~ps after the start of the simulation. \textcolor{black}{The magnetic field is averaged over the laser cycle.}}
    \label{fig:resistive}
\end{figure}

It is well known that the finite collisionality of the target electrons, which controls the Ohmic electric field associated with the return electron current, and thus the generation of the resistive magnetic field, can help recollimate the otherwise divergent HEB \cite{Davies_1997, bell2003resistive, Evans_2006, debayle2010divergence, Robinson_2014}. Given the simplified Ohm's law, $\mathbf{E} \simeq \eta \mathbf{J_p} \simeq -\eta \mathbf{J_h}$ ($\eta$ is the electrical resistivity, $\mathbf{J_h}$ is the HEB current density and $\mathbf{J}_p \simeq - \mathbf{J_h}$ the return current density due to thermal electrons), the resistive magnetic field should evolve as \cite{Davies_1997, sentoku2011dynamic}:
\begin{equation}
    \frac{\partial \mathbf{B}}{\partial t} \simeq \eta \boldsymbol{\nabla} \times \mathbf{J_h} + \boldsymbol{\nabla} \eta \times \mathbf{J_h} \,.
\end{equation}
In this equation, the electrical resistivity dynamically changes as a function of the space- and time-varying properties (i.e. temperature and density) of the bulk plasma particles. Note, however, that the use (due to computational constraints) of a target thinner than in the experiment ($4\,\rm \mu m$ vs. $30\,\rm \mu m$) and the substitution of Al ions for Au ions tend to overestimate the target heating. This, in turn, should underestimate the electrical resistivity (in the Spitzer regime \cite{perez2012improved}) and therefore weaken the resistive magnetic field.

Figure~\ref{fig:resistive} displays snapshots of the out-of-plane magnetic field ($B_z$) distribution at $t=1\,\rm ps$ in the two simulation cases. Both configurations lead to the formation of filamentary structures (with \textcolor{black}{$\sim 1\,\rm \mu m$}-scale wavelength) within the bulk target \cite{Gremillet_2002, sentoku2011dynamic}. The resistive magnetic field reaches a peak strength of $B_z \simeq 3000\,\rm T$ in the double-beam case (Case 2), which is about $3\times$ higher than in the single-beam case (Case 0). This difference originates from the stronger HEB current generated in Case 2.

\begin{figure}
    \centering
    \includegraphics[width=0.45\textwidth]{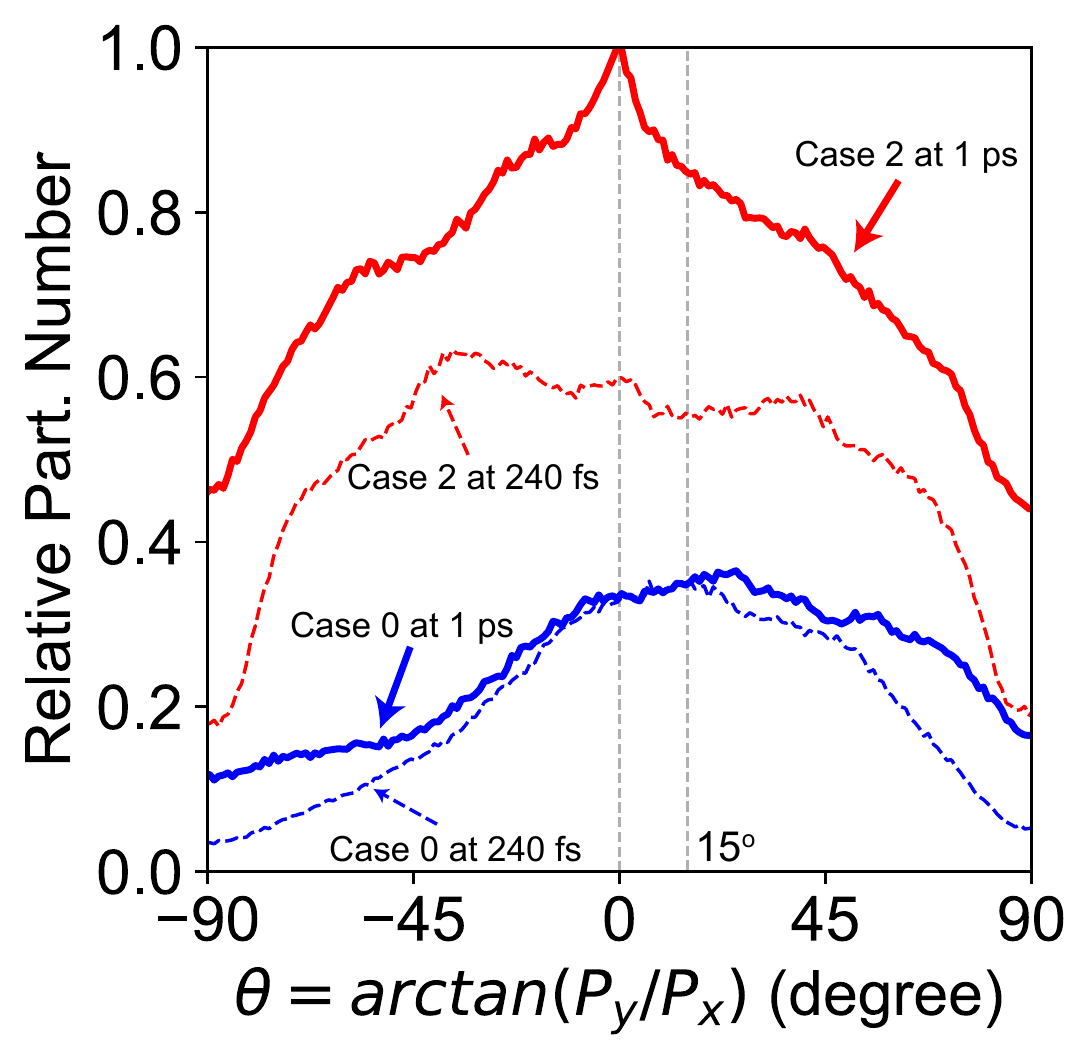}
    \caption{\textbf{Angular distributions of the HEB within the dense target.} The distributions of propagation angle are computed at times $t=240\,\rm fs$ (dashed curves) and $t=1\,\rm ps$ (solid curves) from the higher-energy ($>1.5\,\rm MeV$) electrons contained in the area $5 < x < 9 \, \rm \mu m$ and $15 < y <25 \, \rm \mu m$ (to suppress numerical boundary effects). Blue curves: single laser beam (Case 0). Red curves: two overlapping beams (Case 2). All curves are normalized to the maximum of the Case 2 curve at $t=1\,\rm ps$. The grey dashed lines indicate the target normal ($0\degree$) and laser incidence ($15\degree$) directions.
    }
    \label{fig:collimation}
\end{figure}

Although the magnetic collimation of the HEB is likely underestimated by the limited spatiotemporal scales of our simulations, it can still be captured, as illustrated in Fig.~\ref{fig:collimation}. This figure plots the angular distribution of the higher-energy ($>1.5\,\rm MeV$) electrons as measured at two successive times ($t=240\,\rm fs$ and $t=1\,\rm ps$) within the target region $5<x< 9\,\rm \mu m$ and $15 < y < 25\,\rm \mu m$. For a single laser beam, this distribution hardly changes over time, with a broad maximum at $\theta_{\rm max} \simeq +15\degree$, matching the laser incidence angle, and \textcolor{black}{about 10\% of the electrons having $\vert \theta - \theta_{\rm max} \vert \lesssim 30\degree$.} When overlapping two beams, the number of hot electrons more than doubles. At $t=240\,\rm fs$, their distribution exhibits a broad plateau ($\vert \theta \vert \lesssim 45\degree$), which significantly narrows by $t=1\,\rm ps$, with a single peak around $\theta_{\rm max} \simeq 0^\degree$ and approximately a quarter of the electrons concentrated within $\vert \theta \vert < 30\degree$.


The above 2D and 3D simulation results, which reveal enhanced production and collimation of the HEB (through various mechanisms investigated separately due to computational limitations) with overlapping laser beams are qualitatively consistent with our measurements. The predicted emission of the HEB along the laser propagation direction in Case 0 also aligns with the experimental observation. In Case 2, we can even expect that the boosted HEB generation due to MR will strengthen the magnetic collimation effect across the bulk target. 


\section{Conclusions}
\label{Conc}

Our experimental findings and supporting numerical modeling are of particular interest for the next generation of multi-PW-class laser platforms, which will be composed of multiple beamlets. Our 3D PIC simulations demonstrate that, when the beams are properly distributed on the front side of a solid target, magnetic reconnection can arise and boost the electron energization, and hence also the subsequent ion acceleration. Separate 2D PIC simulations including ionization and collisional effects show another benefit of overlapping beams: the generation of a stronger resistive magnetic field in the bulk target, which acts to reduce the HEB divergence. Our comprehensive investigation into the physics of multi-beam-solid interactions paves the way for optimizing particle and radiation sources at PW-class picosecond laser facilities by adjusting the multi-beam irradiation setup.

\bigskip
\textbf{Acknowledgements}

The authors would like to thank the teams of the Vulcan laser facility for their expert support. We also thank the \textsc{smilei} dev-team for technical support.  W.Y. would like to thank X. F. Shen (HHU, Germany) for useful discussions.
This work was supported by the European Research Council (ERC) under the European Union's Horizon 2020 research and innovation program (Grant Agreement No. 787539). We also acknowledge funding from EPRSC (grants EP/E035728, EP/C003586 and EP/P010059/1). The computational resources of this work were supported by the National Sciences and Engineering Research Council of Canada (NSERC) and Compute Canada (Job: pve-323-ac, PA). The research leading to these results is supported by Extreme Light Infrastructure Nuclear Physics (ELI-NP) Phase II, a project co-financed by the Romanian Government and European Union through the European Regional Development Fund.

\bigskip
\textbf{Author Contributions}

J.F. conceived the project. M.N., S.B., P.A., M.B., R.H., M.Q., L.R., G.S., H.P.S., T.T., O.T., L.V., O.W., and J.F. performed the experiments. W.Y., S.B., S.N.C., M.N., and J.F. analyzed the data. W.Y. performed and analyzed the \textsc{smilei} simulations with discussions with R.R., Y.S., V.H., E.d.H., L.G., and J.F. W.Y., S.N.C., L.G. and J.F. wrote the bulk of the paper, with major contributions from E.d.H. All authors commented and revised the paper.

\bigskip
\textbf{Conflict of Interest}

The authors declare no competing interests.

\bigskip
\textbf{{Data availability}}

All data needed to evaluate the conclusions in the paper are present in the paper.
Experimental data and simulations are archived on servers at LULI laboratory and are available from the corresponding author upon reasonable request.

\bigskip
\textbf{Code availability}

The code used to generate the data shown in Figures \ref{fig:simu_directional}, \ref{fig:simu_enhancement}, \ref{fig:simu_spectrum}, \ref{fig:trace}, \ref{fig:simu_fourbeam}, \ref{fig:resistive}, and \ref{fig:collimation} is \textsc{smilei}. The code used to generate the dashed line data shown in Figs.~\ref{fig:IPs} is \textsc{fluka}.

\bibliography{refs}

\begin{thebibliography}{83}%
\makeatletter
\providecommand \@ifxundefined [1]{%
 \@ifx{#1\undefined}
}%
\providecommand \@ifnum [1]{%
 \ifnum #1\expandafter \@firstoftwo
 \else \expandafter \@secondoftwo
 \fi
}%
\providecommand \@ifx [1]{%
 \ifx #1\expandafter \@firstoftwo
 \else \expandafter \@secondoftwo
 \fi
}%
\providecommand \natexlab [1]{#1}%
\providecommand \enquote  [1]{``#1''}%
\providecommand \bibnamefont  [1]{#1}%
\providecommand \bibfnamefont [1]{#1}%
\providecommand \citenamefont [1]{#1}%
\providecommand \href@noop [0]{\@secondoftwo}%
\providecommand \href [0]{\begingroup \@sanitize@url \@href}%
\providecommand \@href[1]{\@@startlink{#1}\@@href}%
\providecommand \@@href[1]{\endgroup#1\@@endlink}%
\providecommand \@sanitize@url [0]{\catcode `\\12\catcode `\$12\catcode
  `\&12\catcode `\#12\catcode `\^12\catcode `\_12\catcode `\%12\relax}%
\providecommand \@@startlink[1]{}%
\providecommand \@@endlink[0]{}%
\providecommand \url  [0]{\begingroup\@sanitize@url \@url }%
\providecommand \@url [1]{\endgroup\@href {#1}{\urlprefix }}%
\providecommand \urlprefix  [0]{URL }%
\providecommand \Eprint [0]{\href }%
\providecommand \doibase [0]{https://doi.org/}%
\providecommand \selectlanguage [0]{\@gobble}%
\providecommand \bibinfo  [0]{\@secondoftwo}%
\providecommand \bibfield  [0]{\@secondoftwo}%
\providecommand \translation [1]{[#1]}%
\providecommand \BibitemOpen [0]{}%
\providecommand \bibitemStop [0]{}%
\providecommand \bibitemNoStop [0]{.\EOS\space}%
\providecommand \EOS [0]{\spacefactor3000\relax}%
\providecommand \BibitemShut  [1]{\csname bibitem#1\endcsname}%
\let\auto@bib@innerbib\@empty
\bibitem [{\citenamefont {{Danson}}\ \emph {et~al.}(2019)\citenamefont
  {{Danson}}, \citenamefont {{Haefner}}, \citenamefont {{Bromage}},
  \citenamefont {{Butcher}}, \citenamefont {{Chanteloup}}, \citenamefont
  {{Chowdhury}}, \citenamefont {{Galvanauskas}}, \citenamefont {{Gizzi}},
  \citenamefont {{Hein}}, \citenamefont {{Hillier}}, \citenamefont {{Hopps}},
  \citenamefont {{Kato}}, \citenamefont {{Khazanov}}, \citenamefont {{Kodama}},
  \citenamefont {{Korn}}, \citenamefont {{Li}}, \citenamefont {{Li}},
  \citenamefont {{Limpert}}, \citenamefont {{Ma}}, \citenamefont {{Nam}},
  \citenamefont {{Neely}}, \citenamefont {{Papadopoulos}}, \citenamefont
  {{Penman}}, \citenamefont {{Qian}}, \citenamefont {{Rocca}}, \citenamefont
  {{Shaykin}}, \citenamefont {{Siders}}, \citenamefont {{Spindloe}},
  \citenamefont {{Szatm{\'a}ri}}, \citenamefont {{Trines}}, \citenamefont
  {{Zhu}}, \citenamefont {{Zhu}},\ and\ \citenamefont
  {{Zuegel}}}]{danson2019petawatt}%
  \BibitemOpen
  \bibfield  {author} {\bibinfo {author} {\bibfnamefont {C.~N.}\ \bibnamefont
  {{Danson}}}, \bibinfo {author} {\bibfnamefont {C.}~\bibnamefont {{Haefner}}},
  \bibinfo {author} {\bibfnamefont {J.}~\bibnamefont {{Bromage}}}, \bibinfo
  {author} {\bibfnamefont {T.}~\bibnamefont {{Butcher}}}, \bibinfo {author}
  {\bibfnamefont {J.-C.~F.}\ \bibnamefont {{Chanteloup}}}, \bibinfo {author}
  {\bibfnamefont {E.~A.}\ \bibnamefont {{Chowdhury}}}, \bibinfo {author}
  {\bibfnamefont {A.}~\bibnamefont {{Galvanauskas}}}, \bibinfo {author}
  {\bibfnamefont {L.~A.}\ \bibnamefont {{Gizzi}}}, \bibinfo {author}
  {\bibfnamefont {J.}~\bibnamefont {{Hein}}}, \bibinfo {author} {\bibfnamefont
  {D.~I.}\ \bibnamefont {{Hillier}}}, \bibinfo {author} {\bibfnamefont {N.~W.}\
  \bibnamefont {{Hopps}}}, \bibinfo {author} {\bibfnamefont {Y.}~\bibnamefont
  {{Kato}}}, \bibinfo {author} {\bibfnamefont {E.~A.}\ \bibnamefont
  {{Khazanov}}}, \bibinfo {author} {\bibfnamefont {R.}~\bibnamefont
  {{Kodama}}}, \bibinfo {author} {\bibfnamefont {G.}~\bibnamefont {{Korn}}},
  \bibinfo {author} {\bibfnamefont {R.}~\bibnamefont {{Li}}}, \bibinfo {author}
  {\bibfnamefont {Y.}~\bibnamefont {{Li}}}, \bibinfo {author} {\bibfnamefont
  {J.}~\bibnamefont {{Limpert}}}, \bibinfo {author} {\bibfnamefont
  {J.}~\bibnamefont {{Ma}}}, \bibinfo {author} {\bibfnamefont {C.~H.}\
  \bibnamefont {{Nam}}}, \bibinfo {author} {\bibfnamefont {D.}~\bibnamefont
  {{Neely}}}, \bibinfo {author} {\bibfnamefont {D.}~\bibnamefont
  {{Papadopoulos}}}, \bibinfo {author} {\bibfnamefont {R.~R.}\ \bibnamefont
  {{Penman}}}, \bibinfo {author} {\bibfnamefont {L.}~\bibnamefont {{Qian}}},
  \bibinfo {author} {\bibfnamefont {J.~J.}\ \bibnamefont {{Rocca}}}, \bibinfo
  {author} {\bibfnamefont {A.~A.}\ \bibnamefont {{Shaykin}}}, \bibinfo {author}
  {\bibfnamefont {C.~W.}\ \bibnamefont {{Siders}}}, \bibinfo {author}
  {\bibfnamefont {C.}~\bibnamefont {{Spindloe}}}, \bibinfo {author}
  {\bibfnamefont {S.}~\bibnamefont {{Szatm{\'a}ri}}}, \bibinfo {author}
  {\bibfnamefont {R.~M.~G.~M.}\ \bibnamefont {{Trines}}}, \bibinfo {author}
  {\bibfnamefont {J.}~\bibnamefont {{Zhu}}}, \bibinfo {author} {\bibfnamefont
  {P.}~\bibnamefont {{Zhu}}},\ and\ \bibinfo {author} {\bibfnamefont {J.~D.}\
  \bibnamefont {{Zuegel}}},\ }\bibfield  {title} {\bibinfo {title} {{Petawatt
  and exawatt class lasers worldwide}},\ }\href
  {https://doi.org/10.1017/hpl.2019.36} {\bibfield  {journal} {\bibinfo
  {journal} {High Power Laser Science and Engineering}\ }\textbf {\bibinfo
  {volume} {7}},\ \bibinfo {eid} {e54} (\bibinfo {year} {2019})}\BibitemShut
  {NoStop}%
\bibitem [{\citenamefont {{Fuchs}}\ \emph {et~al.}(2006)\citenamefont
  {{Fuchs}}, \citenamefont {{Antici}}, \citenamefont {{D'Humi{\`e}res}},
  \citenamefont {{Lefebvre}}, \citenamefont {{Borghesi}}, \citenamefont
  {{Brambrink}}, \citenamefont {{Cecchetti}}, \citenamefont {{Kaluza}},
  \citenamefont {{Malka}}, \citenamefont {{Manclossi}}, \citenamefont
  {{Meyroneinc}}, \citenamefont {{Mora}}, \citenamefont {{Schreiber}},
  \citenamefont {{Toncian}}, \citenamefont {{P{\'e}pin}},\ and\ \citenamefont
  {{Audebert}}}]{fuchs2006laser}%
  \BibitemOpen
  \bibfield  {author} {\bibinfo {author} {\bibfnamefont {J.}~\bibnamefont
  {{Fuchs}}}, \bibinfo {author} {\bibfnamefont {P.}~\bibnamefont {{Antici}}},
  \bibinfo {author} {\bibfnamefont {E.}~\bibnamefont {{D'Humi{\`e}res}}},
  \bibinfo {author} {\bibfnamefont {E.}~\bibnamefont {{Lefebvre}}}, \bibinfo
  {author} {\bibfnamefont {M.}~\bibnamefont {{Borghesi}}}, \bibinfo {author}
  {\bibfnamefont {E.}~\bibnamefont {{Brambrink}}}, \bibinfo {author}
  {\bibfnamefont {C.~A.}\ \bibnamefont {{Cecchetti}}}, \bibinfo {author}
  {\bibfnamefont {M.}~\bibnamefont {{Kaluza}}}, \bibinfo {author}
  {\bibfnamefont {V.}~\bibnamefont {{Malka}}}, \bibinfo {author} {\bibfnamefont
  {M.}~\bibnamefont {{Manclossi}}}, \bibinfo {author} {\bibfnamefont
  {S.}~\bibnamefont {{Meyroneinc}}}, \bibinfo {author} {\bibfnamefont
  {P.}~\bibnamefont {{Mora}}}, \bibinfo {author} {\bibfnamefont
  {J.}~\bibnamefont {{Schreiber}}}, \bibinfo {author} {\bibfnamefont
  {T.}~\bibnamefont {{Toncian}}}, \bibinfo {author} {\bibfnamefont
  {H.}~\bibnamefont {{P{\'e}pin}}},\ and\ \bibinfo {author} {\bibfnamefont
  {P.}~\bibnamefont {{Audebert}}},\ }\bibfield  {title} {\bibinfo {title}
  {{Laser-driven proton scaling laws and new paths towards energy increase}},\
  }\href {https://doi.org/10.1038/nphys199} {\bibfield  {journal} {\bibinfo
  {journal} {Nature Physics}\ }\textbf {\bibinfo {volume} {2}},\ \bibinfo
  {pages} {48} (\bibinfo {year} {2006})}\BibitemShut {NoStop}%
\bibitem [{\citenamefont {{Roth}}\ \emph {et~al.}(2013)\citenamefont {{Roth}},
  \citenamefont {{Jung}}, \citenamefont {{Falk}}, \citenamefont {{Guler}},
  \citenamefont {{Deppert}}, \citenamefont {{Devlin}}, \citenamefont
  {{Favalli}}, \citenamefont {{Fernandez}}, \citenamefont {{Gautier}},
  \citenamefont {{Geissel}}, \citenamefont {{Haight}}, \citenamefont
  {{Hamilton}}, \citenamefont {{Hegelich}}, \citenamefont {{Johnson}},
  \citenamefont {{Merrill}}, \citenamefont {{Schaumann}}, \citenamefont
  {{Schoenberg}}, \citenamefont {{Schollmeier}}, \citenamefont {{Shimada}},
  \citenamefont {{Taddeucci}}, \citenamefont {{Tybo}}, \citenamefont
  {{Wagner}}, \citenamefont {{Wender}}, \citenamefont {{Wilde}},\ and\
  \citenamefont {{Wurden}}}]{roth2013bright}%
  \BibitemOpen
  \bibfield  {author} {\bibinfo {author} {\bibfnamefont {M.}~\bibnamefont
  {{Roth}}}, \bibinfo {author} {\bibfnamefont {D.}~\bibnamefont {{Jung}}},
  \bibinfo {author} {\bibfnamefont {K.}~\bibnamefont {{Falk}}}, \bibinfo
  {author} {\bibfnamefont {N.}~\bibnamefont {{Guler}}}, \bibinfo {author}
  {\bibfnamefont {O.}~\bibnamefont {{Deppert}}}, \bibinfo {author}
  {\bibfnamefont {M.}~\bibnamefont {{Devlin}}}, \bibinfo {author}
  {\bibfnamefont {A.}~\bibnamefont {{Favalli}}}, \bibinfo {author}
  {\bibfnamefont {J.}~\bibnamefont {{Fernandez}}}, \bibinfo {author}
  {\bibfnamefont {D.}~\bibnamefont {{Gautier}}}, \bibinfo {author}
  {\bibfnamefont {M.}~\bibnamefont {{Geissel}}}, \bibinfo {author}
  {\bibfnamefont {R.}~\bibnamefont {{Haight}}}, \bibinfo {author}
  {\bibfnamefont {C.~E.}\ \bibnamefont {{Hamilton}}}, \bibinfo {author}
  {\bibfnamefont {B.~M.}\ \bibnamefont {{Hegelich}}}, \bibinfo {author}
  {\bibfnamefont {R.~P.}\ \bibnamefont {{Johnson}}}, \bibinfo {author}
  {\bibfnamefont {F.}~\bibnamefont {{Merrill}}}, \bibinfo {author}
  {\bibfnamefont {G.}~\bibnamefont {{Schaumann}}}, \bibinfo {author}
  {\bibfnamefont {K.}~\bibnamefont {{Schoenberg}}}, \bibinfo {author}
  {\bibfnamefont {M.}~\bibnamefont {{Schollmeier}}}, \bibinfo {author}
  {\bibfnamefont {T.}~\bibnamefont {{Shimada}}}, \bibinfo {author}
  {\bibfnamefont {T.}~\bibnamefont {{Taddeucci}}}, \bibinfo {author}
  {\bibfnamefont {J.~L.}\ \bibnamefont {{Tybo}}}, \bibinfo {author}
  {\bibfnamefont {F.}~\bibnamefont {{Wagner}}}, \bibinfo {author}
  {\bibfnamefont {S.~A.}\ \bibnamefont {{Wender}}}, \bibinfo {author}
  {\bibfnamefont {C.~H.}\ \bibnamefont {{Wilde}}},\ and\ \bibinfo {author}
  {\bibfnamefont {G.~A.}\ \bibnamefont {{Wurden}}},\ }\bibfield  {title}
  {\bibinfo {title} {{Bright Laser-Driven Neutron Source Based on the
  Relativistic Transparency of Solids}},\ }\href
  {https://doi.org/10.1103/PhysRevLett.110.044802} {\bibfield  {journal}
  {\bibinfo  {journal} {Physical Review Letters}\ }\textbf {\bibinfo {volume}
  {110}},\ \bibinfo {eid} {044802} (\bibinfo {year} {2013})}\BibitemShut
  {NoStop}%
\bibitem [{\citenamefont {{Higginson}}\ \emph {et~al.}(2018)\citenamefont
  {{Higginson}}, \citenamefont {{Gray}}, \citenamefont {{King}}, \citenamefont
  {{Dance}}, \citenamefont {{Williamson}}, \citenamefont {{Butler}},
  \citenamefont {{Wilson}}, \citenamefont {{Capdessus}}, \citenamefont
  {{Armstrong}}, \citenamefont {{Green}}, \citenamefont {{Hawkes}},
  \citenamefont {{Martin}}, \citenamefont {{Wei}}, \citenamefont {{Mirfayzi}},
  \citenamefont {{Yuan}}, \citenamefont {{Kar}}, \citenamefont {{Borghesi}},
  \citenamefont {{Clarke}}, \citenamefont {{Neely}},\ and\ \citenamefont
  {{McKenna}}}]{higginson2018near}%
  \BibitemOpen
  \bibfield  {author} {\bibinfo {author} {\bibfnamefont {A.}~\bibnamefont
  {{Higginson}}}, \bibinfo {author} {\bibfnamefont {R.~J.}\ \bibnamefont
  {{Gray}}}, \bibinfo {author} {\bibfnamefont {M.}~\bibnamefont {{King}}},
  \bibinfo {author} {\bibfnamefont {R.~J.}\ \bibnamefont {{Dance}}}, \bibinfo
  {author} {\bibfnamefont {S.~D.~R.}\ \bibnamefont {{Williamson}}}, \bibinfo
  {author} {\bibfnamefont {N.~M.~H.}\ \bibnamefont {{Butler}}}, \bibinfo
  {author} {\bibfnamefont {R.}~\bibnamefont {{Wilson}}}, \bibinfo {author}
  {\bibfnamefont {R.}~\bibnamefont {{Capdessus}}}, \bibinfo {author}
  {\bibfnamefont {C.}~\bibnamefont {{Armstrong}}}, \bibinfo {author}
  {\bibfnamefont {J.~S.}\ \bibnamefont {{Green}}}, \bibinfo {author}
  {\bibfnamefont {S.~J.}\ \bibnamefont {{Hawkes}}}, \bibinfo {author}
  {\bibfnamefont {P.}~\bibnamefont {{Martin}}}, \bibinfo {author}
  {\bibfnamefont {W.~Q.}\ \bibnamefont {{Wei}}}, \bibinfo {author}
  {\bibfnamefont {S.~R.}\ \bibnamefont {{Mirfayzi}}}, \bibinfo {author}
  {\bibfnamefont {X.~H.}\ \bibnamefont {{Yuan}}}, \bibinfo {author}
  {\bibfnamefont {S.}~\bibnamefont {{Kar}}}, \bibinfo {author} {\bibfnamefont
  {M.}~\bibnamefont {{Borghesi}}}, \bibinfo {author} {\bibfnamefont {R.~J.}\
  \bibnamefont {{Clarke}}}, \bibinfo {author} {\bibfnamefont {D.}~\bibnamefont
  {{Neely}}},\ and\ \bibinfo {author} {\bibfnamefont {P.}~\bibnamefont
  {{McKenna}}},\ }\bibfield  {title} {\bibinfo {title} {{Near-100 MeV protons
  via a laser-driven transparency-enhanced hybrid acceleration scheme}},\
  }\href {https://doi.org/10.1038/s41467-018-03063-9} {\bibfield  {journal}
  {\bibinfo  {journal} {Nature Communications}\ }\textbf {\bibinfo {volume}
  {9}},\ \bibinfo {eid} {724} (\bibinfo {year} {2018})}\BibitemShut {NoStop}%
\bibitem [{\citenamefont {{Barberio}}\ \emph {et~al.}(2018)\citenamefont
  {{Barberio}}, \citenamefont {{Scisci{\`o}}}, \citenamefont {{Valli{\`e}res}},
  \citenamefont {{Cardelli}}, \citenamefont {{Chen}}, \citenamefont
  {{Famulari}}, \citenamefont {{Gangolf}}, \citenamefont {{Revet}},
  \citenamefont {{Schiavi}}, \citenamefont {{Senzacqua}},\ and\ \citenamefont
  {{Antici}}}]{barberio2018laser}%
  \BibitemOpen
  \bibfield  {author} {\bibinfo {author} {\bibfnamefont {M.}~\bibnamefont
  {{Barberio}}}, \bibinfo {author} {\bibfnamefont {M.}~\bibnamefont
  {{Scisci{\`o}}}}, \bibinfo {author} {\bibfnamefont {S.}~\bibnamefont
  {{Valli{\`e}res}}}, \bibinfo {author} {\bibfnamefont {F.}~\bibnamefont
  {{Cardelli}}}, \bibinfo {author} {\bibfnamefont {S.~N.}\ \bibnamefont
  {{Chen}}}, \bibinfo {author} {\bibfnamefont {G.}~\bibnamefont {{Famulari}}},
  \bibinfo {author} {\bibfnamefont {T.}~\bibnamefont {{Gangolf}}}, \bibinfo
  {author} {\bibfnamefont {G.}~\bibnamefont {{Revet}}}, \bibinfo {author}
  {\bibfnamefont {A.}~\bibnamefont {{Schiavi}}}, \bibinfo {author}
  {\bibfnamefont {M.}~\bibnamefont {{Senzacqua}}},\ and\ \bibinfo {author}
  {\bibfnamefont {P.}~\bibnamefont {{Antici}}},\ }\bibfield  {title} {\bibinfo
  {title} {{Laser-accelerated particle beams for stress testing of
  materials}},\ }\href {https://doi.org/10.1038/s41467-017-02675-x} {\bibfield
  {journal} {\bibinfo  {journal} {Nature Communications}\ }\textbf {\bibinfo
  {volume} {9}},\ \bibinfo {eid} {372} (\bibinfo {year} {2018})}\BibitemShut
  {NoStop}%
\bibitem [{\citenamefont {{Glinec}}\ \emph {et~al.}(2005)\citenamefont
  {{Glinec}}, \citenamefont {{Faure}}, \citenamefont {{Dain}}, \citenamefont
  {{Darbon}}, \citenamefont {{Hosokai}}, \citenamefont {{Santos}},
  \citenamefont {{Lefebvre}}, \citenamefont {{Rousseau}}, \citenamefont
  {{Burgy}}, \citenamefont {{Mercier}},\ and\ \citenamefont
  {{Malka}}}]{glinec2005high}%
  \BibitemOpen
  \bibfield  {author} {\bibinfo {author} {\bibfnamefont {Y.}~\bibnamefont
  {{Glinec}}}, \bibinfo {author} {\bibfnamefont {J.}~\bibnamefont {{Faure}}},
  \bibinfo {author} {\bibfnamefont {L.~L.}\ \bibnamefont {{Dain}}}, \bibinfo
  {author} {\bibfnamefont {S.}~\bibnamefont {{Darbon}}}, \bibinfo {author}
  {\bibfnamefont {T.}~\bibnamefont {{Hosokai}}}, \bibinfo {author}
  {\bibfnamefont {J.~J.}\ \bibnamefont {{Santos}}}, \bibinfo {author}
  {\bibfnamefont {E.}~\bibnamefont {{Lefebvre}}}, \bibinfo {author}
  {\bibfnamefont {J.~P.}\ \bibnamefont {{Rousseau}}}, \bibinfo {author}
  {\bibfnamefont {F.}~\bibnamefont {{Burgy}}}, \bibinfo {author} {\bibfnamefont
  {B.}~\bibnamefont {{Mercier}}},\ and\ \bibinfo {author} {\bibfnamefont
  {V.}~\bibnamefont {{Malka}}},\ }\bibfield  {title} {\bibinfo {title}
  {{High-Resolution {\ensuremath{\gamma}}-Ray Radiography Produced by a
  Laser-Plasma Driven Electron Source}},\ }\href
  {https://doi.org/10.1103/PhysRevLett.94.025003} {\bibfield  {journal}
  {\bibinfo  {journal} {Physical Review Letters}\ }\textbf {\bibinfo {volume}
  {94}},\ \bibinfo {eid} {025003} (\bibinfo {year} {2005})}\BibitemShut
  {NoStop}%
\bibitem [{\citenamefont {{Man{\v{c}}i{\'c}}}\ \emph
  {et~al.}(2010)\citenamefont {{Man{\v{c}}i{\'c}}}, \citenamefont {{L{\'e}vy}},
  \citenamefont {{Harmand}}, \citenamefont {{Nakatsutsumi}}, \citenamefont
  {{Antici}}, \citenamefont {{Audebert}}, \citenamefont {{Combis}},
  \citenamefont {{Fourmaux}}, \citenamefont {{Mazevet}}, \citenamefont
  {{Peyrusse}}, \citenamefont {{Recoules}}, \citenamefont {{Renaudin}},
  \citenamefont {{Robiche}}, \citenamefont {{Dorchies}},\ and\ \citenamefont
  {{Fuchs}}}]{manvcic2010picosecond}%
  \BibitemOpen
  \bibfield  {author} {\bibinfo {author} {\bibfnamefont {A.}~\bibnamefont
  {{Man{\v{c}}i{\'c}}}}, \bibinfo {author} {\bibfnamefont {A.}~\bibnamefont
  {{L{\'e}vy}}}, \bibinfo {author} {\bibfnamefont {M.}~\bibnamefont
  {{Harmand}}}, \bibinfo {author} {\bibfnamefont {M.}~\bibnamefont
  {{Nakatsutsumi}}}, \bibinfo {author} {\bibfnamefont {P.}~\bibnamefont
  {{Antici}}}, \bibinfo {author} {\bibfnamefont {P.}~\bibnamefont
  {{Audebert}}}, \bibinfo {author} {\bibfnamefont {P.}~\bibnamefont
  {{Combis}}}, \bibinfo {author} {\bibfnamefont {S.}~\bibnamefont
  {{Fourmaux}}}, \bibinfo {author} {\bibfnamefont {S.}~\bibnamefont
  {{Mazevet}}}, \bibinfo {author} {\bibfnamefont {O.}~\bibnamefont
  {{Peyrusse}}}, \bibinfo {author} {\bibfnamefont {V.}~\bibnamefont
  {{Recoules}}}, \bibinfo {author} {\bibfnamefont {P.}~\bibnamefont
  {{Renaudin}}}, \bibinfo {author} {\bibfnamefont {J.}~\bibnamefont
  {{Robiche}}}, \bibinfo {author} {\bibfnamefont {F.}~\bibnamefont
  {{Dorchies}}},\ and\ \bibinfo {author} {\bibfnamefont {J.}~\bibnamefont
  {{Fuchs}}},\ }\bibfield  {title} {\bibinfo {title} {{Picosecond Short-Range
  Disordering in Isochorically Heated Aluminum at Solid Density}},\ }\href
  {https://doi.org/10.1103/PhysRevLett.104.035002} {\bibfield  {journal}
  {\bibinfo  {journal} {Physical Review Letters}\ }\textbf {\bibinfo {volume}
  {104}},\ \bibinfo {eid} {035002} (\bibinfo {year} {2010})}\BibitemShut
  {NoStop}%
\bibitem [{\citenamefont {{Mahieu}}\ \emph {et~al.}(2018)\citenamefont
  {{Mahieu}}, \citenamefont {{Jourdain}}, \citenamefont {{Ta Phuoc}},
  \citenamefont {{Dorchies}}, \citenamefont {{Goddet}}, \citenamefont
  {{Lifschitz}}, \citenamefont {{Renaudin}},\ and\ \citenamefont
  {{Lecherbourg}}}]{mahieu2018probing}%
  \BibitemOpen
  \bibfield  {author} {\bibinfo {author} {\bibfnamefont {B.}~\bibnamefont
  {{Mahieu}}}, \bibinfo {author} {\bibfnamefont {N.}~\bibnamefont
  {{Jourdain}}}, \bibinfo {author} {\bibfnamefont {K.}~\bibnamefont {{Ta
  Phuoc}}}, \bibinfo {author} {\bibfnamefont {F.}~\bibnamefont {{Dorchies}}},
  \bibinfo {author} {\bibfnamefont {J.~P.}\ \bibnamefont {{Goddet}}}, \bibinfo
  {author} {\bibfnamefont {A.}~\bibnamefont {{Lifschitz}}}, \bibinfo {author}
  {\bibfnamefont {P.}~\bibnamefont {{Renaudin}}},\ and\ \bibinfo {author}
  {\bibfnamefont {L.}~\bibnamefont {{Lecherbourg}}},\ }\bibfield  {title}
  {\bibinfo {title} {{Probing warm dense matter using femtosecond X-ray
  absorption spectroscopy with a laser-produced betatron source}},\ }\href
  {https://doi.org/10.1038/s41467-018-05791-4} {\bibfield  {journal} {\bibinfo
  {journal} {Nature Communications}\ }\textbf {\bibinfo {volume} {9}},\
  \bibinfo {eid} {3276} (\bibinfo {year} {2018})}\BibitemShut {NoStop}%
\bibitem [{\citenamefont {{Chen}}\ \emph {et~al.}(2015)\citenamefont {{Chen}},
  \citenamefont {{Fiuza}}, \citenamefont {{Link}}, \citenamefont {{Hazi}},
  \citenamefont {{Hill}}, \citenamefont {{Hoarty}}, \citenamefont {{James}},
  \citenamefont {{Kerr}}, \citenamefont {{Meyerhofer}}, \citenamefont
  {{Myatt}}, \citenamefont {{Park}}, \citenamefont {{Sentoku}},\ and\
  \citenamefont {{Williams}}}]{chen2015scaling}%
  \BibitemOpen
  \bibfield  {author} {\bibinfo {author} {\bibfnamefont {H.}~\bibnamefont
  {{Chen}}}, \bibinfo {author} {\bibfnamefont {F.}~\bibnamefont {{Fiuza}}},
  \bibinfo {author} {\bibfnamefont {A.}~\bibnamefont {{Link}}}, \bibinfo
  {author} {\bibfnamefont {A.}~\bibnamefont {{Hazi}}}, \bibinfo {author}
  {\bibfnamefont {M.}~\bibnamefont {{Hill}}}, \bibinfo {author} {\bibfnamefont
  {D.}~\bibnamefont {{Hoarty}}}, \bibinfo {author} {\bibfnamefont
  {S.}~\bibnamefont {{James}}}, \bibinfo {author} {\bibfnamefont
  {S.}~\bibnamefont {{Kerr}}}, \bibinfo {author} {\bibfnamefont {D.~D.}\
  \bibnamefont {{Meyerhofer}}}, \bibinfo {author} {\bibfnamefont
  {J.}~\bibnamefont {{Myatt}}}, \bibinfo {author} {\bibfnamefont
  {J.}~\bibnamefont {{Park}}}, \bibinfo {author} {\bibfnamefont
  {Y.}~\bibnamefont {{Sentoku}}},\ and\ \bibinfo {author} {\bibfnamefont
  {G.~J.}\ \bibnamefont {{Williams}}},\ }\bibfield  {title} {\bibinfo {title}
  {{Scaling the Yield of Laser-Driven Electron-Positron Jets to Laboratory
  Astrophysical Applications}},\ }\href
  {https://doi.org/10.1103/PhysRevLett.114.215001} {\bibfield  {journal}
  {\bibinfo  {journal} {Physical Review Letters}\ }\textbf {\bibinfo {volume}
  {114}},\ \bibinfo {eid} {215001} (\bibinfo {year} {2015})}\BibitemShut
  {NoStop}%
\bibitem [{\citenamefont {Higginson}\ \emph {et~al.}(2019)\citenamefont
  {Higginson}, \citenamefont {Korneev}, \citenamefont {Ruyer}, \citenamefont
  {Riquier}, \citenamefont {Moreno}, \citenamefont {B{\'e}ard}, \citenamefont
  {Chen}, \citenamefont {Grassi}, \citenamefont {Grech}, \citenamefont
  {Gremillet} \emph {et~al.}}]{higginson2019laboratory}%
  \BibitemOpen
  \bibfield  {author} {\bibinfo {author} {\bibfnamefont {D.}~\bibnamefont
  {Higginson}}, \bibinfo {author} {\bibfnamefont {P.}~\bibnamefont {Korneev}},
  \bibinfo {author} {\bibfnamefont {C.}~\bibnamefont {Ruyer}}, \bibinfo
  {author} {\bibfnamefont {R.}~\bibnamefont {Riquier}}, \bibinfo {author}
  {\bibfnamefont {Q.}~\bibnamefont {Moreno}}, \bibinfo {author} {\bibfnamefont
  {J.}~\bibnamefont {B{\'e}ard}}, \bibinfo {author} {\bibfnamefont
  {S.}~\bibnamefont {Chen}}, \bibinfo {author} {\bibfnamefont {A.}~\bibnamefont
  {Grassi}}, \bibinfo {author} {\bibfnamefont {M.}~\bibnamefont {Grech}},
  \bibinfo {author} {\bibfnamefont {L.}~\bibnamefont {Gremillet}}, \emph
  {et~al.},\ }\bibfield  {title} {\bibinfo {title} {Laboratory investigation of
  particle acceleration and magnetic field compression in collisionless
  colliding fast plasma flows},\ }\href@noop {} {\bibfield  {journal} {\bibinfo
   {journal} {Communications Physics}\ }\textbf {\bibinfo {volume} {2}},\
  \bibinfo {pages} {1} (\bibinfo {year} {2019})}\BibitemShut {NoStop}%
\bibitem [{\citenamefont {Prasselsperger}\ \emph {et~al.}(2021)\citenamefont
  {Prasselsperger}, \citenamefont {Coughlan}, \citenamefont {Breslin},
  \citenamefont {Yeung}, \citenamefont {Arthur}, \citenamefont {Donnelly},
  \citenamefont {White}, \citenamefont {Afshari}, \citenamefont {Speicher},
  \citenamefont {Yang} \emph {et~al.}}]{prasselsperger2021real}%
  \BibitemOpen
  \bibfield  {author} {\bibinfo {author} {\bibfnamefont {A.}~\bibnamefont
  {Prasselsperger}}, \bibinfo {author} {\bibfnamefont {M.}~\bibnamefont
  {Coughlan}}, \bibinfo {author} {\bibfnamefont {N.}~\bibnamefont {Breslin}},
  \bibinfo {author} {\bibfnamefont {M.}~\bibnamefont {Yeung}}, \bibinfo
  {author} {\bibfnamefont {C.}~\bibnamefont {Arthur}}, \bibinfo {author}
  {\bibfnamefont {H.}~\bibnamefont {Donnelly}}, \bibinfo {author}
  {\bibfnamefont {S.}~\bibnamefont {White}}, \bibinfo {author} {\bibfnamefont
  {M.}~\bibnamefont {Afshari}}, \bibinfo {author} {\bibfnamefont
  {M.}~\bibnamefont {Speicher}}, \bibinfo {author} {\bibfnamefont
  {R.}~\bibnamefont {Yang}}, \emph {et~al.},\ }\bibfield  {title} {\bibinfo
  {title} {Real-time electron solvation induced by bursts of laser-accelerated
  protons in liquid water},\ }\href@noop {} {\bibfield  {journal} {\bibinfo
  {journal} {Physical Review Letters}\ }\textbf {\bibinfo {volume} {127}},\
  \bibinfo {pages} {186001} (\bibinfo {year} {2021})}\BibitemShut {NoStop}%
\bibitem [{\citenamefont {Nguyen}\ \emph {et~al.}(2006)\citenamefont {Nguyen},
  \citenamefont {Britten}, \citenamefont {Carlson}, \citenamefont {Nissen},
  \citenamefont {Summers}, \citenamefont {Hoaglan}, \citenamefont {Aasen},
  \citenamefont {Peterson},\ and\ \citenamefont
  {Jovanovic}}]{nguyen2006gratings}%
  \BibitemOpen
  \bibfield  {author} {\bibinfo {author} {\bibfnamefont {H.}~\bibnamefont
  {Nguyen}}, \bibinfo {author} {\bibfnamefont {J.}~\bibnamefont {Britten}},
  \bibinfo {author} {\bibfnamefont {T.}~\bibnamefont {Carlson}}, \bibinfo
  {author} {\bibfnamefont {J.}~\bibnamefont {Nissen}}, \bibinfo {author}
  {\bibfnamefont {L.}~\bibnamefont {Summers}}, \bibinfo {author} {\bibfnamefont
  {C.}~\bibnamefont {Hoaglan}}, \bibinfo {author} {\bibfnamefont
  {M.}~\bibnamefont {Aasen}}, \bibinfo {author} {\bibfnamefont
  {J.}~\bibnamefont {Peterson}},\ and\ \bibinfo {author} {\bibfnamefont
  {I.}~\bibnamefont {Jovanovic}},\ }\bibfield  {title} {\bibinfo {title}
  {Gratings for high-energy petawatt lasers},\ }\href@noop {} {\bibfield
  {journal} {\bibinfo  {journal} {Proceedings Volume 5991, Laser-Induced Damage
  in Optical Materials: 2005}\ ,\ \bibinfo {pages} {59911M}} (\bibinfo {year}
  {2006})}\BibitemShut {NoStop}%
\bibitem [{\citenamefont {Chorel}\ \emph {et~al.}(2018)\citenamefont {Chorel},
  \citenamefont {Lanternier}, \citenamefont {Lavastre}, \citenamefont {Bonod},
  \citenamefont {Bousquet},\ and\ \citenamefont
  {N{\'e}auport}}]{chorel2018robust}%
  \BibitemOpen
  \bibfield  {author} {\bibinfo {author} {\bibfnamefont {M.}~\bibnamefont
  {Chorel}}, \bibinfo {author} {\bibfnamefont {T.}~\bibnamefont {Lanternier}},
  \bibinfo {author} {\bibfnamefont {E.}~\bibnamefont {Lavastre}}, \bibinfo
  {author} {\bibfnamefont {N.}~\bibnamefont {Bonod}}, \bibinfo {author}
  {\bibfnamefont {B.}~\bibnamefont {Bousquet}},\ and\ \bibinfo {author}
  {\bibfnamefont {J.}~\bibnamefont {N{\'e}auport}},\ }\bibfield  {title}
  {\bibinfo {title} {Robust optimization of the laser induced damage threshold
  of dielectric mirrors for high power lasers},\ }\href@noop {} {\bibfield
  {journal} {\bibinfo  {journal} {Optics Express}\ }\textbf {\bibinfo {volume}
  {26}},\ \bibinfo {pages} {11764} (\bibinfo {year} {2018})}\BibitemShut
  {NoStop}%
\bibitem [{\citenamefont {Batani}\ \emph {et~al.}(2014)\citenamefont {Batani},
  \citenamefont {Koenig}, \citenamefont {Miquel}, \citenamefont {Ducret},
  \citenamefont {d'Humieres}, \citenamefont {Hulin}, \citenamefont {Caron},
  \citenamefont {Feugeas}, \citenamefont {Nicolai}, \citenamefont {Tikhonchuk}
  \emph {et~al.}}]{batani2014development}%
  \BibitemOpen
  \bibfield  {author} {\bibinfo {author} {\bibfnamefont {D.}~\bibnamefont
  {Batani}}, \bibinfo {author} {\bibfnamefont {M.}~\bibnamefont {Koenig}},
  \bibinfo {author} {\bibfnamefont {J.}~\bibnamefont {Miquel}}, \bibinfo
  {author} {\bibfnamefont {J.}~\bibnamefont {Ducret}}, \bibinfo {author}
  {\bibfnamefont {E.}~\bibnamefont {d'Humieres}}, \bibinfo {author}
  {\bibfnamefont {S.}~\bibnamefont {Hulin}}, \bibinfo {author} {\bibfnamefont
  {J.}~\bibnamefont {Caron}}, \bibinfo {author} {\bibfnamefont
  {J.}~\bibnamefont {Feugeas}}, \bibinfo {author} {\bibfnamefont
  {P.}~\bibnamefont {Nicolai}}, \bibinfo {author} {\bibfnamefont
  {V.}~\bibnamefont {Tikhonchuk}}, \emph {et~al.},\ }\bibfield  {title}
  {\bibinfo {title} {Development of the petawatt aquitaine laser system and new
  perspectives in physics},\ }\href@noop {} {\bibfield  {journal} {\bibinfo
  {journal} {Physica Scripta}\ }\textbf {\bibinfo {volume} {2014}},\ \bibinfo
  {pages} {014016} (\bibinfo {year} {2014})}\BibitemShut {NoStop}%
\bibitem [{\citenamefont {Di~Nicola}\ \emph {et~al.}(2015)\citenamefont
  {Di~Nicola}, \citenamefont {Yang}, \citenamefont {Boley}, \citenamefont
  {Crane}, \citenamefont {Heebner}, \citenamefont {Spinka}, \citenamefont
  {Arnold}, \citenamefont {Barty}, \citenamefont {Bowers}, \citenamefont
  {Budge} \emph {et~al.}}]{di2015commissioning}%
  \BibitemOpen
  \bibfield  {author} {\bibinfo {author} {\bibfnamefont {J.}~\bibnamefont
  {Di~Nicola}}, \bibinfo {author} {\bibfnamefont {S.}~\bibnamefont {Yang}},
  \bibinfo {author} {\bibfnamefont {C.}~\bibnamefont {Boley}}, \bibinfo
  {author} {\bibfnamefont {J.~K.}\ \bibnamefont {Crane}}, \bibinfo {author}
  {\bibfnamefont {J.}~\bibnamefont {Heebner}}, \bibinfo {author} {\bibfnamefont
  {T.~M.}\ \bibnamefont {Spinka}}, \bibinfo {author} {\bibfnamefont
  {P.}~\bibnamefont {Arnold}}, \bibinfo {author} {\bibfnamefont
  {C.}~\bibnamefont {Barty}}, \bibinfo {author} {\bibfnamefont
  {M.}~\bibnamefont {Bowers}}, \bibinfo {author} {\bibfnamefont
  {T.}~\bibnamefont {Budge}}, \emph {et~al.},\ }\bibfield  {title} {\bibinfo
  {title} {The commissioning of the advanced radiographic capability laser
  system: experimental and modeling results at the main laser output},\
  }\href@noop {} {\bibfield  {journal} {\bibinfo  {journal} {High Power Lasers
  for Fusion Research III}\ }\textbf {\bibinfo {volume} {9345}},\ \bibinfo
  {pages} {93450I} (\bibinfo {year} {2015})}\BibitemShut {NoStop}%
\bibitem [{\citenamefont {Arikawa}\ \emph {et~al.}(2016)\citenamefont
  {Arikawa}, \citenamefont {Kojima}, \citenamefont {Morace}, \citenamefont
  {Sakata}, \citenamefont {Gawa}, \citenamefont {Taguchi}, \citenamefont {Abe},
  \citenamefont {Zhang}, \citenamefont {Vaisseau}, \citenamefont {Lee} \emph
  {et~al.}}]{arikawa2016ultrahigh}%
  \BibitemOpen
  \bibfield  {author} {\bibinfo {author} {\bibfnamefont {Y.}~\bibnamefont
  {Arikawa}}, \bibinfo {author} {\bibfnamefont {S.}~\bibnamefont {Kojima}},
  \bibinfo {author} {\bibfnamefont {A.}~\bibnamefont {Morace}}, \bibinfo
  {author} {\bibfnamefont {S.}~\bibnamefont {Sakata}}, \bibinfo {author}
  {\bibfnamefont {T.}~\bibnamefont {Gawa}}, \bibinfo {author} {\bibfnamefont
  {Y.}~\bibnamefont {Taguchi}}, \bibinfo {author} {\bibfnamefont
  {Y.}~\bibnamefont {Abe}}, \bibinfo {author} {\bibfnamefont {Z.}~\bibnamefont
  {Zhang}}, \bibinfo {author} {\bibfnamefont {X.}~\bibnamefont {Vaisseau}},
  \bibinfo {author} {\bibfnamefont {S.~H.}\ \bibnamefont {Lee}}, \emph
  {et~al.},\ }\bibfield  {title} {\bibinfo {title} {Ultrahigh-contrast
  kilojoule-class petawatt lfex laser using a plasma mirror},\ }\href@noop {}
  {\bibfield  {journal} {\bibinfo  {journal} {Applied Optics}\ }\textbf
  {\bibinfo {volume} {55}},\ \bibinfo {pages} {6850} (\bibinfo {year}
  {2016})}\BibitemShut {NoStop}%
\bibitem [{\citenamefont {Liang}\ \emph {et~al.}(2020)\citenamefont {Liang},
  \citenamefont {Leng}, \citenamefont {Li},\ and\ \citenamefont
  {Xu}}]{liang2020recent}%
  \BibitemOpen
  \bibfield  {author} {\bibinfo {author} {\bibfnamefont {X.}~\bibnamefont
  {Liang}}, \bibinfo {author} {\bibfnamefont {Y.}~\bibnamefont {Leng}},
  \bibinfo {author} {\bibfnamefont {R.}~\bibnamefont {Li}},\ and\ \bibinfo
  {author} {\bibfnamefont {Z.}~\bibnamefont {Xu}},\ }\bibfield  {title}
  {\bibinfo {title} {Recent progress on the shanghai superintense ultrafast
  laser facility (sulf) at siom},\ }\href@noop {} {\bibfield  {journal}
  {\bibinfo  {journal} {OSA High-brightness Sources and Light-driven
  Interactions Congress 2020 (EUVXRAY, HILAS, MICS)}\ ,\ \bibinfo {pages}
  {HTh2B.2}} (\bibinfo {year} {2020})}\BibitemShut {NoStop}%
\bibitem [{\citenamefont {Steinke}\ \emph {et~al.}(2016)\citenamefont
  {Steinke}, \citenamefont {Van~Tilborg}, \citenamefont {Benedetti},
  \citenamefont {Geddes}, \citenamefont {Schroeder}, \citenamefont {Daniels},
  \citenamefont {Swanson}, \citenamefont {Gonsalves}, \citenamefont {Nakamura},
  \citenamefont {Matlis} \emph {et~al.}}]{steinke2016multistage}%
  \BibitemOpen
  \bibfield  {author} {\bibinfo {author} {\bibfnamefont {S.}~\bibnamefont
  {Steinke}}, \bibinfo {author} {\bibfnamefont {J.}~\bibnamefont
  {Van~Tilborg}}, \bibinfo {author} {\bibfnamefont {C.}~\bibnamefont
  {Benedetti}}, \bibinfo {author} {\bibfnamefont {C.}~\bibnamefont {Geddes}},
  \bibinfo {author} {\bibfnamefont {C.}~\bibnamefont {Schroeder}}, \bibinfo
  {author} {\bibfnamefont {J.}~\bibnamefont {Daniels}}, \bibinfo {author}
  {\bibfnamefont {K.}~\bibnamefont {Swanson}}, \bibinfo {author} {\bibfnamefont
  {A.}~\bibnamefont {Gonsalves}}, \bibinfo {author} {\bibfnamefont
  {K.}~\bibnamefont {Nakamura}}, \bibinfo {author} {\bibfnamefont
  {N.}~\bibnamefont {Matlis}}, \emph {et~al.},\ }\bibfield  {title} {\bibinfo
  {title} {Multistage coupling of independent laser-plasma accelerators},\
  }\href@noop {} {\bibfield  {journal} {\bibinfo  {journal} {Nature}\ }\textbf
  {\bibinfo {volume} {530}},\ \bibinfo {pages} {190} (\bibinfo {year}
  {2016})}\BibitemShut {NoStop}%
\bibitem [{\citenamefont {Debus}\ \emph {et~al.}(2019)\citenamefont {Debus},
  \citenamefont {Pausch}, \citenamefont {Huebl}, \citenamefont {Steiniger},
  \citenamefont {Widera}, \citenamefont {Cowan}, \citenamefont {Schramm},\ and\
  \citenamefont {Bussmann}}]{debus2019circumventing}%
  \BibitemOpen
  \bibfield  {author} {\bibinfo {author} {\bibfnamefont {A.}~\bibnamefont
  {Debus}}, \bibinfo {author} {\bibfnamefont {R.}~\bibnamefont {Pausch}},
  \bibinfo {author} {\bibfnamefont {A.}~\bibnamefont {Huebl}}, \bibinfo
  {author} {\bibfnamefont {K.}~\bibnamefont {Steiniger}}, \bibinfo {author}
  {\bibfnamefont {R.}~\bibnamefont {Widera}}, \bibinfo {author} {\bibfnamefont
  {T.~E.}\ \bibnamefont {Cowan}}, \bibinfo {author} {\bibfnamefont
  {U.}~\bibnamefont {Schramm}},\ and\ \bibinfo {author} {\bibfnamefont
  {M.}~\bibnamefont {Bussmann}},\ }\bibfield  {title} {\bibinfo {title}
  {Circumventing the dephasing and depletion limits of laser-wakefield
  acceleration},\ }\href@noop {} {\bibfield  {journal} {\bibinfo  {journal}
  {Physical Review X}\ }\textbf {\bibinfo {volume} {9}},\ \bibinfo {pages}
  {031044} (\bibinfo {year} {2019})}\BibitemShut {NoStop}%
\bibitem [{\citenamefont {Santala}\ \emph {et~al.}(2000)\citenamefont
  {Santala}, \citenamefont {Zepf}, \citenamefont {Watts}, \citenamefont {Beg},
  \citenamefont {Clark}, \citenamefont {Tatarakis}, \citenamefont
  {Krushelnick}, \citenamefont {Dangor}, \citenamefont {McCanny}, \citenamefont
  {Spencer} \emph {et~al.}}]{santala2000effect}%
  \BibitemOpen
  \bibfield  {author} {\bibinfo {author} {\bibfnamefont {M.}~\bibnamefont
  {Santala}}, \bibinfo {author} {\bibfnamefont {M.}~\bibnamefont {Zepf}},
  \bibinfo {author} {\bibfnamefont {I.}~\bibnamefont {Watts}}, \bibinfo
  {author} {\bibfnamefont {F.}~\bibnamefont {Beg}}, \bibinfo {author}
  {\bibfnamefont {E.}~\bibnamefont {Clark}}, \bibinfo {author} {\bibfnamefont
  {M.}~\bibnamefont {Tatarakis}}, \bibinfo {author} {\bibfnamefont
  {K.}~\bibnamefont {Krushelnick}}, \bibinfo {author} {\bibfnamefont
  {A.}~\bibnamefont {Dangor}}, \bibinfo {author} {\bibfnamefont
  {T.}~\bibnamefont {McCanny}}, \bibinfo {author} {\bibfnamefont
  {I.}~\bibnamefont {Spencer}}, \emph {et~al.},\ }\bibfield  {title} {\bibinfo
  {title} {Effect of the plasma density scale length on the direction of fast
  electrons in relativistic laser-solid interactions},\ }\href@noop {}
  {\bibfield  {journal} {\bibinfo  {journal} {Physical Review Letters}\
  }\textbf {\bibinfo {volume} {84}},\ \bibinfo {pages} {1459} (\bibinfo {year}
  {2000})}\BibitemShut {NoStop}%
\bibitem [{\citenamefont {Mackinnon}\ \emph {et~al.}(2002)\citenamefont
  {Mackinnon}, \citenamefont {Sentoku}, \citenamefont {Patel}, \citenamefont
  {Price}, \citenamefont {Hatchett}, \citenamefont {Key}, \citenamefont
  {Andersen}, \citenamefont {Snavely},\ and\ \citenamefont
  {Freeman}}]{mackinnon2002enhancement}%
  \BibitemOpen
  \bibfield  {author} {\bibinfo {author} {\bibfnamefont {A.}~\bibnamefont
  {Mackinnon}}, \bibinfo {author} {\bibfnamefont {Y.}~\bibnamefont {Sentoku}},
  \bibinfo {author} {\bibfnamefont {P.}~\bibnamefont {Patel}}, \bibinfo
  {author} {\bibfnamefont {D.}~\bibnamefont {Price}}, \bibinfo {author}
  {\bibfnamefont {S.}~\bibnamefont {Hatchett}}, \bibinfo {author}
  {\bibfnamefont {M.}~\bibnamefont {Key}}, \bibinfo {author} {\bibfnamefont
  {C.}~\bibnamefont {Andersen}}, \bibinfo {author} {\bibfnamefont
  {R.}~\bibnamefont {Snavely}},\ and\ \bibinfo {author} {\bibfnamefont
  {R.}~\bibnamefont {Freeman}},\ }\bibfield  {title} {\bibinfo {title}
  {Enhancement of proton acceleration by hot-electron recirculation in thin
  foils irradiated by ultraintense laser pulses},\ }\href@noop {} {\bibfield
  {journal} {\bibinfo  {journal} {Physical review letters}\ }\textbf {\bibinfo
  {volume} {88}},\ \bibinfo {pages} {215006} (\bibinfo {year}
  {2002})}\BibitemShut {NoStop}%
\bibitem [{\citenamefont {Green}\ \emph {et~al.}(2008)\citenamefont {Green},
  \citenamefont {Ovchinnikov}, \citenamefont {Evans}, \citenamefont {Akli},
  \citenamefont {Azechi}, \citenamefont {Beg}, \citenamefont {Bellei},
  \citenamefont {Freeman}, \citenamefont {Habara}, \citenamefont {Heathcote}
  \emph {et~al.}}]{green2008effect}%
  \BibitemOpen
  \bibfield  {author} {\bibinfo {author} {\bibfnamefont {J.}~\bibnamefont
  {Green}}, \bibinfo {author} {\bibfnamefont {V.}~\bibnamefont {Ovchinnikov}},
  \bibinfo {author} {\bibfnamefont {R.}~\bibnamefont {Evans}}, \bibinfo
  {author} {\bibfnamefont {K.}~\bibnamefont {Akli}}, \bibinfo {author}
  {\bibfnamefont {H.}~\bibnamefont {Azechi}}, \bibinfo {author} {\bibfnamefont
  {F.}~\bibnamefont {Beg}}, \bibinfo {author} {\bibfnamefont {C.}~\bibnamefont
  {Bellei}}, \bibinfo {author} {\bibfnamefont {R.}~\bibnamefont {Freeman}},
  \bibinfo {author} {\bibfnamefont {H.}~\bibnamefont {Habara}}, \bibinfo
  {author} {\bibfnamefont {R.}~\bibnamefont {Heathcote}}, \emph {et~al.},\
  }\bibfield  {title} {\bibinfo {title} {Effect of laser intensity on
  fast-electron-beam divergence in solid-density plasmas},\ }\href@noop {}
  {\bibfield  {journal} {\bibinfo  {journal} {Physical Review Letters}\
  }\textbf {\bibinfo {volume} {100}},\ \bibinfo {pages} {015003} (\bibinfo
  {year} {2008})}\BibitemShut {NoStop}%
\bibitem [{\citenamefont {Chawla}\ \emph {et~al.}(2013)\citenamefont {Chawla},
  \citenamefont {Wei}, \citenamefont {Mishra}, \citenamefont {Akli},
  \citenamefont {Chen}, \citenamefont {McLean}, \citenamefont {Morace},
  \citenamefont {Patel}, \citenamefont {Sawada}, \citenamefont {Sentoku} \emph
  {et~al.}}]{chawla2013effect}%
  \BibitemOpen
  \bibfield  {author} {\bibinfo {author} {\bibfnamefont {S.}~\bibnamefont
  {Chawla}}, \bibinfo {author} {\bibfnamefont {M.}~\bibnamefont {Wei}},
  \bibinfo {author} {\bibfnamefont {R.}~\bibnamefont {Mishra}}, \bibinfo
  {author} {\bibfnamefont {K.}~\bibnamefont {Akli}}, \bibinfo {author}
  {\bibfnamefont {C.}~\bibnamefont {Chen}}, \bibinfo {author} {\bibfnamefont
  {H.}~\bibnamefont {McLean}}, \bibinfo {author} {\bibfnamefont
  {A.}~\bibnamefont {Morace}}, \bibinfo {author} {\bibfnamefont
  {P.}~\bibnamefont {Patel}}, \bibinfo {author} {\bibfnamefont
  {H.}~\bibnamefont {Sawada}}, \bibinfo {author} {\bibfnamefont
  {Y.}~\bibnamefont {Sentoku}}, \emph {et~al.},\ }\bibfield  {title} {\bibinfo
  {title} {Effect of target material on fast-electron transport and resistive
  collimation},\ }\href@noop {} {\bibfield  {journal} {\bibinfo  {journal}
  {Physical Review Letters}\ }\textbf {\bibinfo {volume} {110}},\ \bibinfo
  {pages} {025001} (\bibinfo {year} {2013})}\BibitemShut {NoStop}%
\bibitem [{\citenamefont {Chen}\ \emph {et~al.}(2013)\citenamefont {Chen},
  \citenamefont {Kemp}, \citenamefont {Perez}, \citenamefont {Link},
  \citenamefont {Beg}, \citenamefont {Chawla}, \citenamefont {Key},
  \citenamefont {McLean}, \citenamefont {Morace}, \citenamefont {Ping} \emph
  {et~al.}}]{chen2013comparisons}%
  \BibitemOpen
  \bibfield  {author} {\bibinfo {author} {\bibfnamefont {C.}~\bibnamefont
  {Chen}}, \bibinfo {author} {\bibfnamefont {A.}~\bibnamefont {Kemp}}, \bibinfo
  {author} {\bibfnamefont {F.}~\bibnamefont {Perez}}, \bibinfo {author}
  {\bibfnamefont {A.}~\bibnamefont {Link}}, \bibinfo {author} {\bibfnamefont
  {F.}~\bibnamefont {Beg}}, \bibinfo {author} {\bibfnamefont {S.}~\bibnamefont
  {Chawla}}, \bibinfo {author} {\bibfnamefont {M.}~\bibnamefont {Key}},
  \bibinfo {author} {\bibfnamefont {H.}~\bibnamefont {McLean}}, \bibinfo
  {author} {\bibfnamefont {A.}~\bibnamefont {Morace}}, \bibinfo {author}
  {\bibfnamefont {Y.}~\bibnamefont {Ping}}, \emph {et~al.},\ }\bibfield
  {title} {\bibinfo {title} {Comparisons of angularly and spectrally resolved
  bremsstrahlung measurements to two-dimensional multi-stage simulations of
  short-pulse laser-plasma interactions},\ }\href@noop {} {\bibfield  {journal}
  {\bibinfo  {journal} {Physics of Plasmas}\ }\textbf {\bibinfo {volume}
  {20}},\ \bibinfo {pages} {052703} (\bibinfo {year} {2013})}\BibitemShut
  {NoStop}%
\bibitem [{\citenamefont {Fujioka}\ \emph {et~al.}(2015)\citenamefont
  {Fujioka}, \citenamefont {Johzaki}, \citenamefont {Arikawa}, \citenamefont
  {Zhang}, \citenamefont {Morace}, \citenamefont {Ikenouchi}, \citenamefont
  {Ozaki}, \citenamefont {Nagai}, \citenamefont {Abe}, \citenamefont {Kojima}
  \emph {et~al.}}]{fujioka2015heating}%
  \BibitemOpen
  \bibfield  {author} {\bibinfo {author} {\bibfnamefont {S.}~\bibnamefont
  {Fujioka}}, \bibinfo {author} {\bibfnamefont {T.}~\bibnamefont {Johzaki}},
  \bibinfo {author} {\bibfnamefont {Y.}~\bibnamefont {Arikawa}}, \bibinfo
  {author} {\bibfnamefont {Z.}~\bibnamefont {Zhang}}, \bibinfo {author}
  {\bibfnamefont {A.}~\bibnamefont {Morace}}, \bibinfo {author} {\bibfnamefont
  {T.}~\bibnamefont {Ikenouchi}}, \bibinfo {author} {\bibfnamefont
  {T.}~\bibnamefont {Ozaki}}, \bibinfo {author} {\bibfnamefont
  {T.}~\bibnamefont {Nagai}}, \bibinfo {author} {\bibfnamefont
  {Y.}~\bibnamefont {Abe}}, \bibinfo {author} {\bibfnamefont {S.}~\bibnamefont
  {Kojima}}, \emph {et~al.},\ }\bibfield  {title} {\bibinfo {title} {Heating
  efficiency evaluation with mimicking plasma conditions of integrated
  fast-ignition experiment},\ }\href@noop {} {\bibfield  {journal} {\bibinfo
  {journal} {Physical Review E}\ }\textbf {\bibinfo {volume} {91}},\ \bibinfo
  {pages} {063102} (\bibinfo {year} {2015})}\BibitemShut {NoStop}%
\bibitem [{\citenamefont {Ziegler}\ \emph {et~al.}(2021)\citenamefont
  {Ziegler}, \citenamefont {Albach}, \citenamefont {Bernert}, \citenamefont
  {Bock}, \citenamefont {Brack}, \citenamefont {Cowan}, \citenamefont {Dover},
  \citenamefont {Garten}, \citenamefont {Gaus}, \citenamefont {Gebhardt} \emph
  {et~al.}}]{ziegler2021proton}%
  \BibitemOpen
  \bibfield  {author} {\bibinfo {author} {\bibfnamefont {T.}~\bibnamefont
  {Ziegler}}, \bibinfo {author} {\bibfnamefont {D.}~\bibnamefont {Albach}},
  \bibinfo {author} {\bibfnamefont {C.}~\bibnamefont {Bernert}}, \bibinfo
  {author} {\bibfnamefont {S.}~\bibnamefont {Bock}}, \bibinfo {author}
  {\bibfnamefont {F.-E.}\ \bibnamefont {Brack}}, \bibinfo {author}
  {\bibfnamefont {T.}~\bibnamefont {Cowan}}, \bibinfo {author} {\bibfnamefont
  {N.}~\bibnamefont {Dover}}, \bibinfo {author} {\bibfnamefont
  {M.}~\bibnamefont {Garten}}, \bibinfo {author} {\bibfnamefont
  {L.}~\bibnamefont {Gaus}}, \bibinfo {author} {\bibfnamefont {R.}~\bibnamefont
  {Gebhardt}}, \emph {et~al.},\ }\bibfield  {title} {\bibinfo {title} {Proton
  beam quality enhancement by spectral phase control of a pw-class laser
  system},\ }\href@noop {} {\bibfield  {journal} {\bibinfo  {journal}
  {Scientific Reports}\ }\textbf {\bibinfo {volume} {11}},\ \bibinfo {pages}
  {1} (\bibinfo {year} {2021})}\BibitemShut {NoStop}%
\bibitem [{\citenamefont {Scott}\ \emph
  {et~al.}(2012{\natexlab{a}})\citenamefont {Scott}, \citenamefont {Beaucourt},
  \citenamefont {Schlenvoigt}, \citenamefont {Markey}, \citenamefont
  {Lancaster}, \citenamefont {Ridgers}, \citenamefont {Brenner}, \citenamefont
  {Pasley}, \citenamefont {Gray}, \citenamefont {Musgrave} \emph
  {et~al.}}]{scott2012controlling}%
  \BibitemOpen
  \bibfield  {author} {\bibinfo {author} {\bibfnamefont {R.}~\bibnamefont
  {Scott}}, \bibinfo {author} {\bibfnamefont {C.}~\bibnamefont {Beaucourt}},
  \bibinfo {author} {\bibfnamefont {H.-P.}\ \bibnamefont {Schlenvoigt}},
  \bibinfo {author} {\bibfnamefont {K.}~\bibnamefont {Markey}}, \bibinfo
  {author} {\bibfnamefont {K.}~\bibnamefont {Lancaster}}, \bibinfo {author}
  {\bibfnamefont {C.}~\bibnamefont {Ridgers}}, \bibinfo {author} {\bibfnamefont
  {C.}~\bibnamefont {Brenner}}, \bibinfo {author} {\bibfnamefont
  {J.}~\bibnamefont {Pasley}}, \bibinfo {author} {\bibfnamefont
  {R.}~\bibnamefont {Gray}}, \bibinfo {author} {\bibfnamefont {I.}~\bibnamefont
  {Musgrave}}, \emph {et~al.},\ }\bibfield  {title} {\bibinfo {title}
  {Controlling fast-electron-beam divergence using two laser pulses},\
  }\href@noop {} {\bibfield  {journal} {\bibinfo  {journal} {Physical Review
  Letters}\ }\textbf {\bibinfo {volume} {109}},\ \bibinfo {pages} {015001}
  (\bibinfo {year} {2012}{\natexlab{a}})}\BibitemShut {NoStop}%
\bibitem [{\citenamefont {{Malko}}\ \emph {et~al.}(2019)\citenamefont
  {{Malko}}, \citenamefont {{Vaisseau}}, \citenamefont {{Perez}}, \citenamefont
  {{Batani}}, \citenamefont {{Curcio}}, \citenamefont {{Ehret}}, \citenamefont
  {{Honrubia}}, \citenamefont {{Jakubowska}}, \citenamefont {{Morace}},
  \citenamefont {{Santos}},\ and\ \citenamefont {{Volpe}}}]{malko2019enhanced}%
  \BibitemOpen
  \bibfield  {author} {\bibinfo {author} {\bibfnamefont {S.}~\bibnamefont
  {{Malko}}}, \bibinfo {author} {\bibfnamefont {X.}~\bibnamefont {{Vaisseau}}},
  \bibinfo {author} {\bibfnamefont {F.}~\bibnamefont {{Perez}}}, \bibinfo
  {author} {\bibfnamefont {D.}~\bibnamefont {{Batani}}}, \bibinfo {author}
  {\bibfnamefont {A.}~\bibnamefont {{Curcio}}}, \bibinfo {author}
  {\bibfnamefont {M.}~\bibnamefont {{Ehret}}}, \bibinfo {author} {\bibfnamefont
  {J.}~\bibnamefont {{Honrubia}}}, \bibinfo {author} {\bibfnamefont
  {K.}~\bibnamefont {{Jakubowska}}}, \bibinfo {author} {\bibfnamefont
  {A.}~\bibnamefont {{Morace}}}, \bibinfo {author} {\bibfnamefont {J.~J.}\
  \bibnamefont {{Santos}}},\ and\ \bibinfo {author} {\bibfnamefont
  {L.}~\bibnamefont {{Volpe}}},\ }\bibfield  {title} {\bibinfo {title}
  {{Enhanced relativistic-electron beam collimation using two consecutive laser
  pulses}},\ }\href {https://doi.org/10.1038/s41598-019-50401-y} {\bibfield
  {journal} {\bibinfo  {journal} {Scientific Reports}\ }\textbf {\bibinfo
  {volume} {9}},\ \bibinfo {eid} {14061} (\bibinfo {year} {2019})}\BibitemShut
  {NoStop}%
\bibitem [{\citenamefont {{Wilks}}\ \emph {et~al.}(2001)\citenamefont
  {{Wilks}}, \citenamefont {{Langdon}}, \citenamefont {{Cowan}}, \citenamefont
  {{Roth}}, \citenamefont {{Singh}}, \citenamefont {{Hatchett}}, \citenamefont
  {{Key}}, \citenamefont {{Pennington}}, \citenamefont {{MacKinnon}},\ and\
  \citenamefont {{Snavely}}}]{wilks2001energetic}%
  \BibitemOpen
  \bibfield  {author} {\bibinfo {author} {\bibfnamefont {S.~C.}\ \bibnamefont
  {{Wilks}}}, \bibinfo {author} {\bibfnamefont {A.~B.}\ \bibnamefont
  {{Langdon}}}, \bibinfo {author} {\bibfnamefont {T.~E.}\ \bibnamefont
  {{Cowan}}}, \bibinfo {author} {\bibfnamefont {M.}~\bibnamefont {{Roth}}},
  \bibinfo {author} {\bibfnamefont {M.}~\bibnamefont {{Singh}}}, \bibinfo
  {author} {\bibfnamefont {S.}~\bibnamefont {{Hatchett}}}, \bibinfo {author}
  {\bibfnamefont {M.~H.}\ \bibnamefont {{Key}}}, \bibinfo {author}
  {\bibfnamefont {D.}~\bibnamefont {{Pennington}}}, \bibinfo {author}
  {\bibfnamefont {A.}~\bibnamefont {{MacKinnon}}},\ and\ \bibinfo {author}
  {\bibfnamefont {R.~A.}\ \bibnamefont {{Snavely}}},\ }\bibfield  {title}
  {\bibinfo {title} {{Energetic proton generation in ultra-intense laser-solid
  interactions}},\ }\href {https://doi.org/10.1063/1.1333697} {\bibfield
  {journal} {\bibinfo  {journal} {Physics of Plasmas}\ }\textbf {\bibinfo
  {volume} {8}},\ \bibinfo {pages} {542} (\bibinfo {year} {2001})}\BibitemShut
  {NoStop}%
\bibitem [{\citenamefont {Mora}(2003)}]{mora2003plasma}%
  \BibitemOpen
  \bibfield  {author} {\bibinfo {author} {\bibfnamefont {P.}~\bibnamefont
  {Mora}},\ }\bibfield  {title} {\bibinfo {title} {Plasma expansion into a
  vacuum},\ }\href@noop {} {\bibfield  {journal} {\bibinfo  {journal} {Physical
  Review Letters}\ }\textbf {\bibinfo {volume} {90}},\ \bibinfo {pages}
  {185002} (\bibinfo {year} {2003})}\BibitemShut {NoStop}%
\bibitem [{\citenamefont {Markey}\ \emph {et~al.}(2010)\citenamefont {Markey},
  \citenamefont {McKenna}, \citenamefont {Brenner}, \citenamefont {Carroll},
  \citenamefont {G{\"u}nther}, \citenamefont {Harres}, \citenamefont {Kar},
  \citenamefont {Lancaster}, \citenamefont {N{\"u}rnberg}, \citenamefont
  {Quinn} \emph {et~al.}}]{markey2010spectral}%
  \BibitemOpen
  \bibfield  {author} {\bibinfo {author} {\bibfnamefont {K.}~\bibnamefont
  {Markey}}, \bibinfo {author} {\bibfnamefont {P.}~\bibnamefont {McKenna}},
  \bibinfo {author} {\bibfnamefont {C.}~\bibnamefont {Brenner}}, \bibinfo
  {author} {\bibfnamefont {D.}~\bibnamefont {Carroll}}, \bibinfo {author}
  {\bibfnamefont {M.}~\bibnamefont {G{\"u}nther}}, \bibinfo {author}
  {\bibfnamefont {K.}~\bibnamefont {Harres}}, \bibinfo {author} {\bibfnamefont
  {S.}~\bibnamefont {Kar}}, \bibinfo {author} {\bibfnamefont {K.}~\bibnamefont
  {Lancaster}}, \bibinfo {author} {\bibfnamefont {F.}~\bibnamefont
  {N{\"u}rnberg}}, \bibinfo {author} {\bibfnamefont {M.}~\bibnamefont {Quinn}},
  \emph {et~al.},\ }\bibfield  {title} {\bibinfo {title} {Spectral enhancement
  in the double pulse regime of laser proton acceleration},\ }\href@noop {}
  {\bibfield  {journal} {\bibinfo  {journal} {Physical Review Letters}\
  }\textbf {\bibinfo {volume} {105}},\ \bibinfo {pages} {195008} (\bibinfo
  {year} {2010})}\BibitemShut {NoStop}%
\bibitem [{\citenamefont {Scott}\ \emph
  {et~al.}(2012{\natexlab{b}})\citenamefont {Scott}, \citenamefont {Green},
  \citenamefont {Bagnoud}, \citenamefont {Brabetz}, \citenamefont {Brenner},
  \citenamefont {Carroll}, \citenamefont {MacLellan}, \citenamefont {Robinson},
  \citenamefont {Roth}, \citenamefont {Spindloe} \emph
  {et~al.}}]{scott2012multi}%
  \BibitemOpen
  \bibfield  {author} {\bibinfo {author} {\bibfnamefont {G.}~\bibnamefont
  {Scott}}, \bibinfo {author} {\bibfnamefont {J.}~\bibnamefont {Green}},
  \bibinfo {author} {\bibfnamefont {V.}~\bibnamefont {Bagnoud}}, \bibinfo
  {author} {\bibfnamefont {C.}~\bibnamefont {Brabetz}}, \bibinfo {author}
  {\bibfnamefont {C.}~\bibnamefont {Brenner}}, \bibinfo {author} {\bibfnamefont
  {D.}~\bibnamefont {Carroll}}, \bibinfo {author} {\bibfnamefont
  {D.}~\bibnamefont {MacLellan}}, \bibinfo {author} {\bibfnamefont
  {A.}~\bibnamefont {Robinson}}, \bibinfo {author} {\bibfnamefont
  {M.}~\bibnamefont {Roth}}, \bibinfo {author} {\bibfnamefont {C.}~\bibnamefont
  {Spindloe}}, \emph {et~al.},\ }\bibfield  {title} {\bibinfo {title}
  {Multi-pulse enhanced laser ion acceleration using plasma half cavity
  targets},\ }\href@noop {} {\bibfield  {journal} {\bibinfo  {journal} {Applied
  Physics Letters}\ }\textbf {\bibinfo {volume} {101}},\ \bibinfo {pages}
  {024101} (\bibinfo {year} {2012}{\natexlab{b}})}\BibitemShut {NoStop}%
\bibitem [{\citenamefont {Morace}\ \emph {et~al.}(2019)\citenamefont {Morace},
  \citenamefont {Iwata}, \citenamefont {Sentoku}, \citenamefont {Mima},
  \citenamefont {Arikawa}, \citenamefont {Yogo}, \citenamefont {Andreev},
  \citenamefont {Tosaki}, \citenamefont {Vaisseau}, \citenamefont {Abe} \emph
  {et~al.}}]{morace2019enhancing}%
  \BibitemOpen
  \bibfield  {author} {\bibinfo {author} {\bibfnamefont {A.}~\bibnamefont
  {Morace}}, \bibinfo {author} {\bibfnamefont {N.}~\bibnamefont {Iwata}},
  \bibinfo {author} {\bibfnamefont {Y.}~\bibnamefont {Sentoku}}, \bibinfo
  {author} {\bibfnamefont {K.}~\bibnamefont {Mima}}, \bibinfo {author}
  {\bibfnamefont {Y.}~\bibnamefont {Arikawa}}, \bibinfo {author} {\bibfnamefont
  {A.}~\bibnamefont {Yogo}}, \bibinfo {author} {\bibfnamefont {A.}~\bibnamefont
  {Andreev}}, \bibinfo {author} {\bibfnamefont {S.}~\bibnamefont {Tosaki}},
  \bibinfo {author} {\bibfnamefont {X.}~\bibnamefont {Vaisseau}}, \bibinfo
  {author} {\bibfnamefont {Y.}~\bibnamefont {Abe}}, \emph {et~al.},\ }\bibfield
   {title} {\bibinfo {title} {Enhancing laser beam performance by interfering
  intense laser beamlets},\ }\href@noop {} {\bibfield  {journal} {\bibinfo
  {journal} {Nature Communications}\ }\textbf {\bibinfo {volume} {10}},\
  \bibinfo {pages} {1} (\bibinfo {year} {2019})}\BibitemShut {NoStop}%
\bibitem [{\citenamefont {{Yogo}}\ \emph {et~al.}(2017)\citenamefont {{Yogo}},
  \citenamefont {{Mima}}, \citenamefont {{Iwata}}, \citenamefont {{Tosaki}},
  \citenamefont {{Morace}}, \citenamefont {{Arikawa}}, \citenamefont
  {{Fujioka}}, \citenamefont {{Johzaki}}, \citenamefont {{Sentoku}},
  \citenamefont {{Nishimura}}, \citenamefont {{Sagisaka}}, \citenamefont
  {{Matsuo}}, \citenamefont {{Kamitsukasa}}, \citenamefont {{Kojima}},
  \citenamefont {{Nagatomo}}, \citenamefont {{Nakai}}, \citenamefont
  {{Shiraga}}, \citenamefont {{Murakami}}, \citenamefont {{Tokita}},
  \citenamefont {{Kawanaka}}, \citenamefont {{Miyanaga}}, \citenamefont
  {{Yamanoi}}, \citenamefont {{Norimatsu}}, \citenamefont {{Sakagami}},
  \citenamefont {{Bulanov}}, \citenamefont {{Kondo}},\ and\ \citenamefont
  {{Azechi}}}]{yogo2017boosting}%
  \BibitemOpen
  \bibfield  {author} {\bibinfo {author} {\bibfnamefont {A.}~\bibnamefont
  {{Yogo}}}, \bibinfo {author} {\bibfnamefont {K.}~\bibnamefont {{Mima}}},
  \bibinfo {author} {\bibfnamefont {N.}~\bibnamefont {{Iwata}}}, \bibinfo
  {author} {\bibfnamefont {S.}~\bibnamefont {{Tosaki}}}, \bibinfo {author}
  {\bibfnamefont {A.}~\bibnamefont {{Morace}}}, \bibinfo {author}
  {\bibfnamefont {Y.}~\bibnamefont {{Arikawa}}}, \bibinfo {author}
  {\bibfnamefont {S.}~\bibnamefont {{Fujioka}}}, \bibinfo {author}
  {\bibfnamefont {T.}~\bibnamefont {{Johzaki}}}, \bibinfo {author}
  {\bibfnamefont {Y.}~\bibnamefont {{Sentoku}}}, \bibinfo {author}
  {\bibfnamefont {H.}~\bibnamefont {{Nishimura}}}, \bibinfo {author}
  {\bibfnamefont {A.}~\bibnamefont {{Sagisaka}}}, \bibinfo {author}
  {\bibfnamefont {K.}~\bibnamefont {{Matsuo}}}, \bibinfo {author}
  {\bibfnamefont {N.}~\bibnamefont {{Kamitsukasa}}}, \bibinfo {author}
  {\bibfnamefont {S.}~\bibnamefont {{Kojima}}}, \bibinfo {author}
  {\bibfnamefont {H.}~\bibnamefont {{Nagatomo}}}, \bibinfo {author}
  {\bibfnamefont {M.}~\bibnamefont {{Nakai}}}, \bibinfo {author} {\bibfnamefont
  {H.}~\bibnamefont {{Shiraga}}}, \bibinfo {author} {\bibfnamefont
  {M.}~\bibnamefont {{Murakami}}}, \bibinfo {author} {\bibfnamefont
  {S.}~\bibnamefont {{Tokita}}}, \bibinfo {author} {\bibfnamefont
  {J.}~\bibnamefont {{Kawanaka}}}, \bibinfo {author} {\bibfnamefont
  {N.}~\bibnamefont {{Miyanaga}}}, \bibinfo {author} {\bibfnamefont
  {K.}~\bibnamefont {{Yamanoi}}}, \bibinfo {author} {\bibfnamefont
  {T.}~\bibnamefont {{Norimatsu}}}, \bibinfo {author} {\bibfnamefont
  {H.}~\bibnamefont {{Sakagami}}}, \bibinfo {author} {\bibfnamefont {S.~V.}\
  \bibnamefont {{Bulanov}}}, \bibinfo {author} {\bibfnamefont {K.}~\bibnamefont
  {{Kondo}}},\ and\ \bibinfo {author} {\bibfnamefont {H.}~\bibnamefont
  {{Azechi}}},\ }\bibfield  {title} {\bibinfo {title} {{Boosting laser-ion
  acceleration with multi-picosecond pulses}},\ }\href
  {https://doi.org/10.1038/srep42451} {\bibfield  {journal} {\bibinfo
  {journal} {Scientific Reports}\ }\textbf {\bibinfo {volume} {7}},\ \bibinfo
  {eid} {42451} (\bibinfo {year} {2017})}\BibitemShut {NoStop}%
\bibitem [{\citenamefont {Raymond}\ \emph {et~al.}(2018)\citenamefont
  {Raymond}, \citenamefont {Dong}, \citenamefont {McKelvey}, \citenamefont
  {Zulick}, \citenamefont {Alexander}, \citenamefont {Bhattacharjee},
  \citenamefont {Campbell}, \citenamefont {Chen}, \citenamefont {Chvykov},
  \citenamefont {Del~Rio} \emph {et~al.}}]{raymond2018relativistic}%
  \BibitemOpen
  \bibfield  {author} {\bibinfo {author} {\bibfnamefont {A.}~\bibnamefont
  {Raymond}}, \bibinfo {author} {\bibfnamefont {C.}~\bibnamefont {Dong}},
  \bibinfo {author} {\bibfnamefont {A.}~\bibnamefont {McKelvey}}, \bibinfo
  {author} {\bibfnamefont {C.}~\bibnamefont {Zulick}}, \bibinfo {author}
  {\bibfnamefont {N.}~\bibnamefont {Alexander}}, \bibinfo {author}
  {\bibfnamefont {A.}~\bibnamefont {Bhattacharjee}}, \bibinfo {author}
  {\bibfnamefont {P.}~\bibnamefont {Campbell}}, \bibinfo {author}
  {\bibfnamefont {H.}~\bibnamefont {Chen}}, \bibinfo {author} {\bibfnamefont
  {V.}~\bibnamefont {Chvykov}}, \bibinfo {author} {\bibfnamefont
  {E.}~\bibnamefont {Del~Rio}}, \emph {et~al.},\ }\bibfield  {title} {\bibinfo
  {title} {Relativistic-electron-driven magnetic reconnection in the
  laboratory},\ }\href@noop {} {\bibfield  {journal} {\bibinfo  {journal}
  {Physical Review E}\ }\textbf {\bibinfo {volume} {98}},\ \bibinfo {pages}
  {043207} (\bibinfo {year} {2018})}\BibitemShut {NoStop}%
\bibitem [{\citenamefont {Palmer}\ \emph {et~al.}(2019)\citenamefont {Palmer},
  \citenamefont {Campbell}, \citenamefont {Ma}, \citenamefont {Antonelli},
  \citenamefont {Bott}, \citenamefont {Gregori}, \citenamefont {Halliday},
  \citenamefont {Katzir}, \citenamefont {Kordell}, \citenamefont {Krushelnick}
  \emph {et~al.}}]{palmer2019field}%
  \BibitemOpen
  \bibfield  {author} {\bibinfo {author} {\bibfnamefont {C.}~\bibnamefont
  {Palmer}}, \bibinfo {author} {\bibfnamefont {P.}~\bibnamefont {Campbell}},
  \bibinfo {author} {\bibfnamefont {Y.}~\bibnamefont {Ma}}, \bibinfo {author}
  {\bibfnamefont {L.}~\bibnamefont {Antonelli}}, \bibinfo {author}
  {\bibfnamefont {A.}~\bibnamefont {Bott}}, \bibinfo {author} {\bibfnamefont
  {G.}~\bibnamefont {Gregori}}, \bibinfo {author} {\bibfnamefont
  {J.}~\bibnamefont {Halliday}}, \bibinfo {author} {\bibfnamefont
  {Y.}~\bibnamefont {Katzir}}, \bibinfo {author} {\bibfnamefont
  {P.}~\bibnamefont {Kordell}}, \bibinfo {author} {\bibfnamefont
  {K.}~\bibnamefont {Krushelnick}}, \emph {et~al.},\ }\bibfield  {title}
  {\bibinfo {title} {Field reconstruction from proton radiography of intense
  laser driven magnetic reconnection},\ }\href@noop {} {\bibfield  {journal}
  {\bibinfo  {journal} {Physics of Plasmas}\ }\textbf {\bibinfo {volume}
  {26}},\ \bibinfo {pages} {083109} (\bibinfo {year} {2019})}\BibitemShut
  {NoStop}%
\bibitem [{\citenamefont {Sarri}\ \emph {et~al.}(2012)\citenamefont {Sarri},
  \citenamefont {Macchi}, \citenamefont {Cecchetti}, \citenamefont {Kar},
  \citenamefont {Liseykina}, \citenamefont {Yang}, \citenamefont {Dieckmann},
  \citenamefont {Fuchs}, \citenamefont {Galimberti}, \citenamefont {Gizzi}
  \emph {et~al.}}]{sarri2012dynamics}%
  \BibitemOpen
  \bibfield  {author} {\bibinfo {author} {\bibfnamefont {G.}~\bibnamefont
  {Sarri}}, \bibinfo {author} {\bibfnamefont {A.}~\bibnamefont {Macchi}},
  \bibinfo {author} {\bibfnamefont {C.}~\bibnamefont {Cecchetti}}, \bibinfo
  {author} {\bibfnamefont {S.}~\bibnamefont {Kar}}, \bibinfo {author}
  {\bibfnamefont {T.}~\bibnamefont {Liseykina}}, \bibinfo {author}
  {\bibfnamefont {X.}~\bibnamefont {Yang}}, \bibinfo {author} {\bibfnamefont
  {M.~E.}\ \bibnamefont {Dieckmann}}, \bibinfo {author} {\bibfnamefont
  {J.}~\bibnamefont {Fuchs}}, \bibinfo {author} {\bibfnamefont
  {M.}~\bibnamefont {Galimberti}}, \bibinfo {author} {\bibfnamefont
  {L.}~\bibnamefont {Gizzi}}, \emph {et~al.},\ }\bibfield  {title} {\bibinfo
  {title} {Dynamics of self-generated, large amplitude magnetic fields
  following high-intensity laser matter interaction},\ }\href@noop {}
  {\bibfield  {journal} {\bibinfo  {journal} {Physical Review Letters}\
  }\textbf {\bibinfo {volume} {109}},\ \bibinfo {pages} {205002} (\bibinfo
  {year} {2012})}\BibitemShut {NoStop}%
\bibitem [{\citenamefont {Schumaker}\ \emph {et~al.}(2013)\citenamefont
  {Schumaker}, \citenamefont {Nakanii}, \citenamefont {McGuffey}, \citenamefont
  {Zulick}, \citenamefont {Chyvkov}, \citenamefont {Dollar}, \citenamefont
  {Habara}, \citenamefont {Kalintchenko}, \citenamefont {Maksimchuk},
  \citenamefont {Tanaka} \emph {et~al.}}]{schumaker2013ultrafast}%
  \BibitemOpen
  \bibfield  {author} {\bibinfo {author} {\bibfnamefont {W.}~\bibnamefont
  {Schumaker}}, \bibinfo {author} {\bibfnamefont {N.}~\bibnamefont {Nakanii}},
  \bibinfo {author} {\bibfnamefont {C.}~\bibnamefont {McGuffey}}, \bibinfo
  {author} {\bibfnamefont {C.}~\bibnamefont {Zulick}}, \bibinfo {author}
  {\bibfnamefont {V.}~\bibnamefont {Chyvkov}}, \bibinfo {author} {\bibfnamefont
  {F.}~\bibnamefont {Dollar}}, \bibinfo {author} {\bibfnamefont
  {H.}~\bibnamefont {Habara}}, \bibinfo {author} {\bibfnamefont
  {G.}~\bibnamefont {Kalintchenko}}, \bibinfo {author} {\bibfnamefont
  {A.}~\bibnamefont {Maksimchuk}}, \bibinfo {author} {\bibfnamefont
  {K.}~\bibnamefont {Tanaka}}, \emph {et~al.},\ }\bibfield  {title} {\bibinfo
  {title} {Ultrafast electron radiography of magnetic fields in high-intensity
  laser-solid interactions},\ }\href@noop {} {\bibfield  {journal} {\bibinfo
  {journal} {Physical Review Letters}\ }\textbf {\bibinfo {volume} {110}},\
  \bibinfo {pages} {015003} (\bibinfo {year} {2013})}\BibitemShut {NoStop}%
\bibitem [{\citenamefont {Nilson}\ \emph {et~al.}(2006)\citenamefont {Nilson},
  \citenamefont {Willingale}, \citenamefont {Kaluza}, \citenamefont
  {Kamperidis}, \citenamefont {Minardi}, \citenamefont {Wei}, \citenamefont
  {Fernandes}, \citenamefont {Notley}, \citenamefont {Bandyopadhyay},
  \citenamefont {Sherlock} \emph {et~al.}}]{nilson2006magnetic}%
  \BibitemOpen
  \bibfield  {author} {\bibinfo {author} {\bibfnamefont {P.}~\bibnamefont
  {Nilson}}, \bibinfo {author} {\bibfnamefont {L.}~\bibnamefont {Willingale}},
  \bibinfo {author} {\bibfnamefont {M.}~\bibnamefont {Kaluza}}, \bibinfo
  {author} {\bibfnamefont {C.}~\bibnamefont {Kamperidis}}, \bibinfo {author}
  {\bibfnamefont {S.}~\bibnamefont {Minardi}}, \bibinfo {author} {\bibfnamefont
  {M.}~\bibnamefont {Wei}}, \bibinfo {author} {\bibfnamefont {P.}~\bibnamefont
  {Fernandes}}, \bibinfo {author} {\bibfnamefont {M.}~\bibnamefont {Notley}},
  \bibinfo {author} {\bibfnamefont {S.}~\bibnamefont {Bandyopadhyay}}, \bibinfo
  {author} {\bibfnamefont {M.}~\bibnamefont {Sherlock}}, \emph {et~al.},\
  }\bibfield  {title} {\bibinfo {title} {Magnetic reconnection and plasma
  dynamics in two-beam laser-solid interactions},\ }\href@noop {} {\bibfield
  {journal} {\bibinfo  {journal} {Physical Review Letters}\ }\textbf {\bibinfo
  {volume} {97}},\ \bibinfo {pages} {255001} (\bibinfo {year}
  {2006})}\BibitemShut {NoStop}%
\bibitem [{\citenamefont {Rosenberg}\ \emph {et~al.}(2015)\citenamefont
  {Rosenberg}, \citenamefont {Li}, \citenamefont {Fox}, \citenamefont
  {Zylstra}, \citenamefont {Stoeckl}, \citenamefont {S{\'e}guin}, \citenamefont
  {Frenje},\ and\ \citenamefont {Petrasso}}]{rosenberg2015slowing}%
  \BibitemOpen
  \bibfield  {author} {\bibinfo {author} {\bibfnamefont {M.}~\bibnamefont
  {Rosenberg}}, \bibinfo {author} {\bibfnamefont {C.}~\bibnamefont {Li}},
  \bibinfo {author} {\bibfnamefont {W.}~\bibnamefont {Fox}}, \bibinfo {author}
  {\bibfnamefont {A.}~\bibnamefont {Zylstra}}, \bibinfo {author} {\bibfnamefont
  {C.}~\bibnamefont {Stoeckl}}, \bibinfo {author} {\bibfnamefont
  {F.}~\bibnamefont {S{\'e}guin}}, \bibinfo {author} {\bibfnamefont
  {J.}~\bibnamefont {Frenje}},\ and\ \bibinfo {author} {\bibfnamefont
  {R.}~\bibnamefont {Petrasso}},\ }\bibfield  {title} {\bibinfo {title}
  {Slowing of magnetic reconnection concurrent with weakening plasma inflows
  and increasing collisionality in strongly driven laser-plasma experiments},\
  }\href@noop {} {\bibfield  {journal} {\bibinfo  {journal} {Physical Review
  Letters}\ }\textbf {\bibinfo {volume} {114}},\ \bibinfo {pages} {205004}
  (\bibinfo {year} {2015})}\BibitemShut {NoStop}%
\bibitem [{\citenamefont {Golovin}\ \emph {et~al.}(2020)\citenamefont
  {Golovin}, \citenamefont {Mirfayzi}, \citenamefont {Gu}, \citenamefont {Abe},
  \citenamefont {Honoki}, \citenamefont {Mori}, \citenamefont {Nagatomo},
  \citenamefont {Okamoto}, \citenamefont {Shokita}, \citenamefont {Yamanoi}
  \emph {et~al.}}]{golovin2020enhancement}%
  \BibitemOpen
  \bibfield  {author} {\bibinfo {author} {\bibfnamefont {D.~O.}\ \bibnamefont
  {Golovin}}, \bibinfo {author} {\bibfnamefont {S.~R.}\ \bibnamefont
  {Mirfayzi}}, \bibinfo {author} {\bibfnamefont {Y.~J.}\ \bibnamefont {Gu}},
  \bibinfo {author} {\bibfnamefont {Y.}~\bibnamefont {Abe}}, \bibinfo {author}
  {\bibfnamefont {Y.}~\bibnamefont {Honoki}}, \bibinfo {author} {\bibfnamefont
  {T.}~\bibnamefont {Mori}}, \bibinfo {author} {\bibfnamefont {H.}~\bibnamefont
  {Nagatomo}}, \bibinfo {author} {\bibfnamefont {K.}~\bibnamefont {Okamoto}},
  \bibinfo {author} {\bibfnamefont {S.}~\bibnamefont {Shokita}}, \bibinfo
  {author} {\bibfnamefont {K.}~\bibnamefont {Yamanoi}}, \emph {et~al.},\
  }\bibfield  {title} {\bibinfo {title} {Enhancement of ion energy and flux by
  the influence of magnetic reconnection in foam targets},\ }\href@noop {}
  {\bibfield  {journal} {\bibinfo  {journal} {High Energy Density Physics}\
  }\textbf {\bibinfo {volume} {36}},\ \bibinfo {pages} {100840} (\bibinfo
  {year} {2020})}\BibitemShut {NoStop}%
\bibitem [{\citenamefont {{Kim}}\ \emph {et~al.}(2022)\citenamefont {{Kim}},
  \citenamefont {{Wilks}}, \citenamefont {{Kemp}}, \citenamefont {{Sherlock}},
  \citenamefont {{Ma}}, \citenamefont {{Beg}},\ and\ \citenamefont
  {{Mariscal}}}]{kim2022efficient}%
  \BibitemOpen
  \bibfield  {author} {\bibinfo {author} {\bibfnamefont {J.}~\bibnamefont
  {{Kim}}}, \bibinfo {author} {\bibfnamefont {S.}~\bibnamefont {{Wilks}}},
  \bibinfo {author} {\bibfnamefont {A.}~\bibnamefont {{Kemp}}}, \bibinfo
  {author} {\bibfnamefont {M.}~\bibnamefont {{Sherlock}}}, \bibinfo {author}
  {\bibfnamefont {T.}~\bibnamefont {{Ma}}}, \bibinfo {author} {\bibfnamefont
  {F.}~\bibnamefont {{Beg}}},\ and\ \bibinfo {author} {\bibfnamefont
  {D.}~\bibnamefont {{Mariscal}}},\ }\bibfield  {title} {\bibinfo {title}
  {{Efficient ion acceleration by multistaged intense short laser pulses}},\
  }\href {https://doi.org/10.1103/PhysRevResearch.4.L032003} {\bibfield
  {journal} {\bibinfo  {journal} {Physical Review Research}\ }\textbf {\bibinfo
  {volume} {4}},\ \bibinfo {eid} {L032003} (\bibinfo {year}
  {2022})}\BibitemShut {NoStop}%
\bibitem [{\citenamefont {Gu}\ \emph {et~al.}(2019)\citenamefont {Gu},
  \citenamefont {Pegoraro}, \citenamefont {Sasorov}, \citenamefont {Golovin},
  \citenamefont {Yogo}, \citenamefont {Korn},\ and\ \citenamefont
  {Bulanov}}]{gu2019electromagnetic}%
  \BibitemOpen
  \bibfield  {author} {\bibinfo {author} {\bibfnamefont {Y.}~\bibnamefont
  {Gu}}, \bibinfo {author} {\bibfnamefont {F.}~\bibnamefont {Pegoraro}},
  \bibinfo {author} {\bibfnamefont {P.}~\bibnamefont {Sasorov}}, \bibinfo
  {author} {\bibfnamefont {D.}~\bibnamefont {Golovin}}, \bibinfo {author}
  {\bibfnamefont {A.}~\bibnamefont {Yogo}}, \bibinfo {author} {\bibfnamefont
  {G.}~\bibnamefont {Korn}},\ and\ \bibinfo {author} {\bibfnamefont
  {S.}~\bibnamefont {Bulanov}},\ }\bibfield  {title} {\bibinfo {title}
  {Electromagnetic burst generation during annihilation of magnetic field in
  relativistic laser-plasma interaction},\ }\href@noop {} {\bibfield  {journal}
  {\bibinfo  {journal} {Scientific Reports}\ }\textbf {\bibinfo {volume} {9}},\
  \bibinfo {pages} {1} (\bibinfo {year} {2019})}\BibitemShut {NoStop}%
\bibitem [{\citenamefont {Ferri}\ \emph {et~al.}(2019)\citenamefont {Ferri},
  \citenamefont {Siminos},\ and\ \citenamefont
  {F{\"u}l{\"o}p}}]{ferri2019enhanced}%
  \BibitemOpen
  \bibfield  {author} {\bibinfo {author} {\bibfnamefont {J.}~\bibnamefont
  {Ferri}}, \bibinfo {author} {\bibfnamefont {E.}~\bibnamefont {Siminos}},\
  and\ \bibinfo {author} {\bibfnamefont {T.}~\bibnamefont {F{\"u}l{\"o}p}},\
  }\bibfield  {title} {\bibinfo {title} {Enhanced target normal sheath
  acceleration using colliding laser pulses},\ }\href@noop {} {\bibfield
  {journal} {\bibinfo  {journal} {Communications Physics}\ }\textbf {\bibinfo
  {volume} {2}},\ \bibinfo {pages} {1} (\bibinfo {year} {2019})}\BibitemShut
  {NoStop}%
\bibitem [{\citenamefont {Ferri}\ \emph {et~al.}(2020)\citenamefont {Ferri},
  \citenamefont {Siminos}, \citenamefont {Gremillet},\ and\ \citenamefont
  {F{\"u}l{\"o}p}}]{ferri2020effects}%
  \BibitemOpen
  \bibfield  {author} {\bibinfo {author} {\bibfnamefont {J.}~\bibnamefont
  {Ferri}}, \bibinfo {author} {\bibfnamefont {E.}~\bibnamefont {Siminos}},
  \bibinfo {author} {\bibfnamefont {L.}~\bibnamefont {Gremillet}},\ and\
  \bibinfo {author} {\bibfnamefont {T.}~\bibnamefont {F{\"u}l{\"o}p}},\
  }\bibfield  {title} {\bibinfo {title} {Effects of oblique incidence and
  colliding pulses on laser-driven proton acceleration from relativistically
  transparent ultrathin targets},\ }\href@noop {} {\bibfield  {journal}
  {\bibinfo  {journal} {Journal of Plasma Physics}\ }\textbf {\bibinfo {volume}
  {86}} (\bibinfo {year} {2020})}\BibitemShut {NoStop}%
\bibitem [{\citenamefont {Rahman}\ \emph {et~al.}(2021)\citenamefont {Rahman},
  \citenamefont {Smith}, \citenamefont {Ngirmang},\ and\ \citenamefont
  {Orban}}]{rahman2021particle}%
  \BibitemOpen
  \bibfield  {author} {\bibinfo {author} {\bibfnamefont {N.}~\bibnamefont
  {Rahman}}, \bibinfo {author} {\bibfnamefont {J.~R.}\ \bibnamefont {Smith}},
  \bibinfo {author} {\bibfnamefont {G.~K.}\ \bibnamefont {Ngirmang}},\ and\
  \bibinfo {author} {\bibfnamefont {C.}~\bibnamefont {Orban}},\ }\bibfield
  {title} {\bibinfo {title} {Particle-in-cell modeling of a potential
  demonstration experiment for double pulse enhanced target normal sheath
  acceleration},\ }\href@noop {} {\bibfield  {journal} {\bibinfo  {journal}
  {Physics of Plasmas}\ }\textbf {\bibinfo {volume} {28}},\ \bibinfo {pages}
  {073103} (\bibinfo {year} {2021})}\BibitemShut {NoStop}%
\bibitem [{\citenamefont {Burdonov}\ \emph {et~al.}(2021)\citenamefont
  {Burdonov}, \citenamefont {Fazzini}, \citenamefont {Lelasseux}, \citenamefont
  {Albrecht}, \citenamefont {Antici}, \citenamefont {Ayoul}, \citenamefont
  {Beluze}, \citenamefont {Cavanna}, \citenamefont {Ceccotti}, \citenamefont
  {Chabanis} \emph {et~al.}}]{burdonov2021characterization}%
  \BibitemOpen
  \bibfield  {author} {\bibinfo {author} {\bibfnamefont {K.}~\bibnamefont
  {Burdonov}}, \bibinfo {author} {\bibfnamefont {A.}~\bibnamefont {Fazzini}},
  \bibinfo {author} {\bibfnamefont {V.}~\bibnamefont {Lelasseux}}, \bibinfo
  {author} {\bibfnamefont {J.}~\bibnamefont {Albrecht}}, \bibinfo {author}
  {\bibfnamefont {P.}~\bibnamefont {Antici}}, \bibinfo {author} {\bibfnamefont
  {Y.}~\bibnamefont {Ayoul}}, \bibinfo {author} {\bibfnamefont
  {A.}~\bibnamefont {Beluze}}, \bibinfo {author} {\bibfnamefont
  {D.}~\bibnamefont {Cavanna}}, \bibinfo {author} {\bibfnamefont
  {T.}~\bibnamefont {Ceccotti}}, \bibinfo {author} {\bibfnamefont
  {M.}~\bibnamefont {Chabanis}}, \emph {et~al.},\ }\bibfield  {title} {\bibinfo
  {title} {Characterization and performance of the apollon short-focal-area
  facility following its commissioning at 1 pw level},\ }\href@noop {}
  {\bibfield  {journal} {\bibinfo  {journal} {Matter and Radiation at
  Extremes}\ }\textbf {\bibinfo {volume} {6}},\ \bibinfo {pages} {064402}
  (\bibinfo {year} {2021})}\BibitemShut {NoStop}%
\bibitem [{\citenamefont {Raffestin}\ \emph {et~al.}(2021)\citenamefont
  {Raffestin}, \citenamefont {Lecherbourg}, \citenamefont {Lantu{\'e}joul},
  \citenamefont {Vauzour}, \citenamefont {Masson-Laborde}, \citenamefont
  {Davoine}, \citenamefont {Blanchot}, \citenamefont {Dubois}, \citenamefont
  {Vaisseau}, \citenamefont {d’Humi{\`e}res} \emph
  {et~al.}}]{raffestin2021enhanced}%
  \BibitemOpen
  \bibfield  {author} {\bibinfo {author} {\bibfnamefont {D.}~\bibnamefont
  {Raffestin}}, \bibinfo {author} {\bibfnamefont {L.}~\bibnamefont
  {Lecherbourg}}, \bibinfo {author} {\bibfnamefont {I.}~\bibnamefont
  {Lantu{\'e}joul}}, \bibinfo {author} {\bibfnamefont {B.}~\bibnamefont
  {Vauzour}}, \bibinfo {author} {\bibfnamefont {P.}~\bibnamefont
  {Masson-Laborde}}, \bibinfo {author} {\bibfnamefont {X.}~\bibnamefont
  {Davoine}}, \bibinfo {author} {\bibfnamefont {N.}~\bibnamefont {Blanchot}},
  \bibinfo {author} {\bibfnamefont {J.}~\bibnamefont {Dubois}}, \bibinfo
  {author} {\bibfnamefont {X.}~\bibnamefont {Vaisseau}}, \bibinfo {author}
  {\bibfnamefont {E.}~\bibnamefont {d’Humi{\`e}res}}, \emph {et~al.},\
  }\bibfield  {title} {\bibinfo {title} {Enhanced ion acceleration using the
  high-energy petawatt petal laser},\ }\href@noop {} {\bibfield  {journal}
  {\bibinfo  {journal} {Matter and Radiation at Extremes}\ }\textbf {\bibinfo
  {volume} {6}},\ \bibinfo {pages} {056901} (\bibinfo {year}
  {2021})}\BibitemShut {NoStop}%
\bibitem [{\citenamefont {Bolton}\ \emph {et~al.}(2014)\citenamefont {Bolton},
  \citenamefont {Borghesi}, \citenamefont {Brenner}, \citenamefont {Carroll},
  \citenamefont {De~Martinis}, \citenamefont {Fiorini}, \citenamefont {Flacco},
  \citenamefont {Floquet}, \citenamefont {Fuchs}, \citenamefont {Gallegos}
  \emph {et~al.}}]{bolton2014instrumentation}%
  \BibitemOpen
  \bibfield  {author} {\bibinfo {author} {\bibfnamefont {P.}~\bibnamefont
  {Bolton}}, \bibinfo {author} {\bibfnamefont {M.}~\bibnamefont {Borghesi}},
  \bibinfo {author} {\bibfnamefont {C.}~\bibnamefont {Brenner}}, \bibinfo
  {author} {\bibfnamefont {D.}~\bibnamefont {Carroll}}, \bibinfo {author}
  {\bibfnamefont {C.}~\bibnamefont {De~Martinis}}, \bibinfo {author}
  {\bibfnamefont {F.}~\bibnamefont {Fiorini}}, \bibinfo {author} {\bibfnamefont
  {A.}~\bibnamefont {Flacco}}, \bibinfo {author} {\bibfnamefont
  {V.}~\bibnamefont {Floquet}}, \bibinfo {author} {\bibfnamefont
  {J.}~\bibnamefont {Fuchs}}, \bibinfo {author} {\bibfnamefont
  {P.}~\bibnamefont {Gallegos}}, \emph {et~al.},\ }\bibfield  {title} {\bibinfo
  {title} {Instrumentation for diagnostics and control of laser-accelerated
  proton (ion) beams},\ }\href@noop {} {\bibfield  {journal} {\bibinfo
  {journal} {Physica Medica}\ }\textbf {\bibinfo {volume} {30}},\ \bibinfo
  {pages} {255} (\bibinfo {year} {2014})}\BibitemShut {NoStop}%
\bibitem [{\citenamefont {{Link}}\ \emph {et~al.}(2011)\citenamefont {{Link}},
  \citenamefont {{Freeman}}, \citenamefont {{Schumacher}},\ and\ \citenamefont
  {{Van Woerkom}}}]{link2011effects}%
  \BibitemOpen
  \bibfield  {author} {\bibinfo {author} {\bibfnamefont {A.}~\bibnamefont
  {{Link}}}, \bibinfo {author} {\bibfnamefont {R.~R.}\ \bibnamefont
  {{Freeman}}}, \bibinfo {author} {\bibfnamefont {D.~W.}\ \bibnamefont
  {{Schumacher}}},\ and\ \bibinfo {author} {\bibfnamefont {L.~D.}\ \bibnamefont
  {{Van Woerkom}}},\ }\bibfield  {title} {\bibinfo {title} {{Effects of target
  charging and ion emission on the energy spectrum of emitted electrons}},\
  }\href {https://doi.org/10.1063/1.3587123} {\bibfield  {journal} {\bibinfo
  {journal} {Physics of Plasmas}\ }\textbf {\bibinfo {volume} {18}},\ \bibinfo
  {eid} {053107} (\bibinfo {year} {2011})}\BibitemShut {NoStop}%
\bibitem [{\citenamefont {Rusby}\ \emph {et~al.}(2015)\citenamefont {Rusby},
  \citenamefont {Wilson}, \citenamefont {Gray}, \citenamefont {Dance},
  \citenamefont {Butler}, \citenamefont {MacLellan}, \citenamefont {Scott},
  \citenamefont {Bagnoud}, \citenamefont {Zielbauer}, \citenamefont {McKenna}
  \emph {et~al.}}]{rusby2015measurement}%
  \BibitemOpen
  \bibfield  {author} {\bibinfo {author} {\bibfnamefont {D.}~\bibnamefont
  {Rusby}}, \bibinfo {author} {\bibfnamefont {L.}~\bibnamefont {Wilson}},
  \bibinfo {author} {\bibfnamefont {R.}~\bibnamefont {Gray}}, \bibinfo {author}
  {\bibfnamefont {R.}~\bibnamefont {Dance}}, \bibinfo {author} {\bibfnamefont
  {N.}~\bibnamefont {Butler}}, \bibinfo {author} {\bibfnamefont
  {D.}~\bibnamefont {MacLellan}}, \bibinfo {author} {\bibfnamefont
  {G.}~\bibnamefont {Scott}}, \bibinfo {author} {\bibfnamefont
  {V.}~\bibnamefont {Bagnoud}}, \bibinfo {author} {\bibfnamefont
  {B.}~\bibnamefont {Zielbauer}}, \bibinfo {author} {\bibfnamefont
  {P.}~\bibnamefont {McKenna}}, \emph {et~al.},\ }\bibfield  {title} {\bibinfo
  {title} {Measurement of the angle, temperature and flux of fast electrons
  emitted from intense laser--solid interactions},\ }\href@noop {} {\bibfield
  {journal} {\bibinfo  {journal} {Journal of Plasma Physics}\ }\textbf
  {\bibinfo {volume} {81}},\ \bibinfo {pages} {475810505} (\bibinfo {year}
  {2015})}\BibitemShut {NoStop}%
\bibitem [{\citenamefont {Wagner}\ \emph {et~al.}(2014)\citenamefont {Wagner},
  \citenamefont {Bedacht}, \citenamefont {Ortner}, \citenamefont {Roth},
  \citenamefont {Tauschwitz}, \citenamefont {Zielbauer},\ and\ \citenamefont
  {Bagnoud}}]{wagner2014pre}%
  \BibitemOpen
  \bibfield  {author} {\bibinfo {author} {\bibfnamefont {F.}~\bibnamefont
  {Wagner}}, \bibinfo {author} {\bibfnamefont {S.}~\bibnamefont {Bedacht}},
  \bibinfo {author} {\bibfnamefont {A.}~\bibnamefont {Ortner}}, \bibinfo
  {author} {\bibfnamefont {M.}~\bibnamefont {Roth}}, \bibinfo {author}
  {\bibfnamefont {A.}~\bibnamefont {Tauschwitz}}, \bibinfo {author}
  {\bibfnamefont {B.}~\bibnamefont {Zielbauer}},\ and\ \bibinfo {author}
  {\bibfnamefont {V.}~\bibnamefont {Bagnoud}},\ }\bibfield  {title} {\bibinfo
  {title} {Pre-plasma formation in experiments using petawatt lasers},\
  }\href@noop {} {\bibfield  {journal} {\bibinfo  {journal} {Optics Express}\
  }\textbf {\bibinfo {volume} {22}},\ \bibinfo {pages} {29505} (\bibinfo {year}
  {2014})}\BibitemShut {NoStop}%
\bibitem [{\citenamefont {Chen}\ \emph {et~al.}(2007)\citenamefont {Chen},
  \citenamefont {Gregori}, \citenamefont {Patel}, \citenamefont {Chung},
  \citenamefont {Evans}, \citenamefont {Freeman}, \citenamefont {Garcia~Saiz},
  \citenamefont {Glenzer}, \citenamefont {Hansen}, \citenamefont {Khattak}
  \emph {et~al.}}]{chen2007creation}%
  \BibitemOpen
  \bibfield  {author} {\bibinfo {author} {\bibfnamefont {S.}~\bibnamefont
  {Chen}}, \bibinfo {author} {\bibfnamefont {G.}~\bibnamefont {Gregori}},
  \bibinfo {author} {\bibfnamefont {P.}~\bibnamefont {Patel}}, \bibinfo
  {author} {\bibfnamefont {H.-K.}\ \bibnamefont {Chung}}, \bibinfo {author}
  {\bibfnamefont {R.}~\bibnamefont {Evans}}, \bibinfo {author} {\bibfnamefont
  {R.}~\bibnamefont {Freeman}}, \bibinfo {author} {\bibfnamefont
  {E.}~\bibnamefont {Garcia~Saiz}}, \bibinfo {author} {\bibfnamefont
  {S.}~\bibnamefont {Glenzer}}, \bibinfo {author} {\bibfnamefont
  {S.}~\bibnamefont {Hansen}}, \bibinfo {author} {\bibfnamefont
  {F.}~\bibnamefont {Khattak}}, \emph {et~al.},\ }\bibfield  {title} {\bibinfo
  {title} {Creation of hot dense matter in short-pulse laser-plasma interaction
  with tamped titanium foils},\ }\href@noop {} {\bibfield  {journal} {\bibinfo
  {journal} {Physics of Plasmas}\ }\textbf {\bibinfo {volume} {14}},\ \bibinfo
  {pages} {102701} (\bibinfo {year} {2007})}\BibitemShut {NoStop}%
\bibitem [{\citenamefont {{Ramis}}\ \emph {et~al.}(1988)\citenamefont
  {{Ramis}}, \citenamefont {{Schmalz}},\ and\ \citenamefont
  {{Meyer-Ter-Vehn}}}]{Ramis_1988}%
  \BibitemOpen
  \bibfield  {author} {\bibinfo {author} {\bibfnamefont {R.}~\bibnamefont
  {{Ramis}}}, \bibinfo {author} {\bibfnamefont {R.}~\bibnamefont {{Schmalz}}},\
  and\ \bibinfo {author} {\bibfnamefont {J.}~\bibnamefont {{Meyer-Ter-Vehn}}},\
  }\bibfield  {title} {\bibinfo {title} {{MULTI {\textemdash} A computer code
  for one-dimensional multigroup radiation hydrodynamics}},\ }\href
  {https://doi.org/10.1016/0010-4655(88)90008-2} {\bibfield  {journal}
  {\bibinfo  {journal} {Comp. Phys. Commun.}\ }\textbf {\bibinfo {volume}
  {49}},\ \bibinfo {pages} {475} (\bibinfo {year} {1988})}\BibitemShut
  {NoStop}%
\bibitem [{\citenamefont {Michaelis}\ and\ \citenamefont
  {Willi}(1981)}]{michaelis1981refractive}%
  \BibitemOpen
  \bibfield  {author} {\bibinfo {author} {\bibfnamefont {M.}~\bibnamefont
  {Michaelis}}\ and\ \bibinfo {author} {\bibfnamefont {O.}~\bibnamefont
  {Willi}},\ }\bibfield  {title} {\bibinfo {title} {Refractive fringe
  diagnostics of laser produced plasmas},\ }\href@noop {} {\bibfield  {journal}
  {\bibinfo  {journal} {Optics Communications}\ }\textbf {\bibinfo {volume}
  {36}},\ \bibinfo {pages} {153} (\bibinfo {year} {1981})}\BibitemShut
  {NoStop}%
\bibitem [{\citenamefont {Wilks}\ \emph {et~al.}(1992)\citenamefont {Wilks},
  \citenamefont {Kruer}, \citenamefont {Tabak},\ and\ \citenamefont
  {Langdon}}]{wilks1992absorption}%
  \BibitemOpen
  \bibfield  {author} {\bibinfo {author} {\bibfnamefont {S.}~\bibnamefont
  {Wilks}}, \bibinfo {author} {\bibfnamefont {W.}~\bibnamefont {Kruer}},
  \bibinfo {author} {\bibfnamefont {M.}~\bibnamefont {Tabak}},\ and\ \bibinfo
  {author} {\bibfnamefont {A.}~\bibnamefont {Langdon}},\ }\bibfield  {title}
  {\bibinfo {title} {Absorption of ultra-intense laser pulses},\ }\href@noop {}
  {\bibfield  {journal} {\bibinfo  {journal} {Physical Review Letters}\
  }\textbf {\bibinfo {volume} {69}},\ \bibinfo {pages} {1383} (\bibinfo {year}
  {1992})}\BibitemShut {NoStop}%
\bibitem [{\citenamefont {Malka}\ and\ \citenamefont
  {Miquel}(1996)}]{malka1996experimental}%
  \BibitemOpen
  \bibfield  {author} {\bibinfo {author} {\bibfnamefont {G.}~\bibnamefont
  {Malka}}\ and\ \bibinfo {author} {\bibfnamefont {J.}~\bibnamefont {Miquel}},\
  }\bibfield  {title} {\bibinfo {title} {Experimental confirmation of
  ponderomotive-force electrons produced by an ultrarelativistic laser pulse on
  a solid target},\ }\href@noop {} {\bibfield  {journal} {\bibinfo  {journal}
  {Physical Review Letters}\ }\textbf {\bibinfo {volume} {77}},\ \bibinfo
  {pages} {75} (\bibinfo {year} {1996})}\BibitemShut {NoStop}%
\bibitem [{\citenamefont {Fasso}\ \emph {et~al.}(2005)\citenamefont {Fasso},
  \citenamefont {Ferrari}, \citenamefont {Ranft},\ and\ \citenamefont
  {Sala}}]{fasso2005fluka}%
  \BibitemOpen
  \bibfield  {author} {\bibinfo {author} {\bibfnamefont {A.}~\bibnamefont
  {Fasso}}, \bibinfo {author} {\bibfnamefont {A.}~\bibnamefont {Ferrari}},
  \bibinfo {author} {\bibfnamefont {J.}~\bibnamefont {Ranft}},\ and\ \bibinfo
  {author} {\bibfnamefont {P.~R.}\ \bibnamefont {Sala}},\ }\href@noop {} {\emph
  {\bibinfo {title} {FLUKA: a multi-particle transport code}}},\ \bibinfo
  {type} {Tech. Rep.}\ (\bibinfo  {institution} {CERN-2005-10},\ \bibinfo
  {year} {2005})\BibitemShut {NoStop}%
\bibitem [{\citenamefont {Battistoni}\ \emph {et~al.}(2007)\citenamefont
  {Battistoni}, \citenamefont {Cerutti}, \citenamefont {Fasso}, \citenamefont
  {Ferrari}, \citenamefont {Muraro}, \citenamefont {Ranft}, \citenamefont
  {Roesler},\ and\ \citenamefont {Sala}}]{battistoni2007fluka}%
  \BibitemOpen
  \bibfield  {author} {\bibinfo {author} {\bibfnamefont {G.}~\bibnamefont
  {Battistoni}}, \bibinfo {author} {\bibfnamefont {F.}~\bibnamefont {Cerutti}},
  \bibinfo {author} {\bibfnamefont {A.}~\bibnamefont {Fasso}}, \bibinfo
  {author} {\bibfnamefont {A.}~\bibnamefont {Ferrari}}, \bibinfo {author}
  {\bibfnamefont {S.}~\bibnamefont {Muraro}}, \bibinfo {author} {\bibfnamefont
  {J.}~\bibnamefont {Ranft}}, \bibinfo {author} {\bibfnamefont
  {S.}~\bibnamefont {Roesler}},\ and\ \bibinfo {author} {\bibfnamefont
  {P.}~\bibnamefont {Sala}},\ }\bibfield  {title} {\bibinfo {title} {The fluka
  code: Description and benchmarking},\ }\href@noop {} {\bibfield  {journal}
  {\bibinfo  {journal} {AIP Conference proceedings}\ }\textbf {\bibinfo
  {volume} {896}},\ \bibinfo {pages} {31} (\bibinfo {year} {2007})}\BibitemShut
  {NoStop}%
\bibitem [{\citenamefont {B{\"o}hlen}\ \emph {et~al.}(2014)\citenamefont
  {B{\"o}hlen}, \citenamefont {Cerutti}, \citenamefont {Chin}, \citenamefont
  {Fass{\`o}}, \citenamefont {Ferrari}, \citenamefont {Ortega}, \citenamefont
  {Mairani}, \citenamefont {Sala}, \citenamefont {Smirnov},\ and\ \citenamefont
  {Vlachoudis}}]{bohlen2014fluka}%
  \BibitemOpen
  \bibfield  {author} {\bibinfo {author} {\bibfnamefont {T.}~\bibnamefont
  {B{\"o}hlen}}, \bibinfo {author} {\bibfnamefont {F.}~\bibnamefont {Cerutti}},
  \bibinfo {author} {\bibfnamefont {M.}~\bibnamefont {Chin}}, \bibinfo {author}
  {\bibfnamefont {A.}~\bibnamefont {Fass{\`o}}}, \bibinfo {author}
  {\bibfnamefont {A.}~\bibnamefont {Ferrari}}, \bibinfo {author} {\bibfnamefont
  {P.~G.}\ \bibnamefont {Ortega}}, \bibinfo {author} {\bibfnamefont
  {A.}~\bibnamefont {Mairani}}, \bibinfo {author} {\bibfnamefont {P.~R.}\
  \bibnamefont {Sala}}, \bibinfo {author} {\bibfnamefont {G.}~\bibnamefont
  {Smirnov}},\ and\ \bibinfo {author} {\bibfnamefont {V.}~\bibnamefont
  {Vlachoudis}},\ }\bibfield  {title} {\bibinfo {title} {The fluka code:
  developments and challenges for high energy and medical applications},\
  }\href@noop {} {\bibfield  {journal} {\bibinfo  {journal} {Nuclear Data
  Sheets}\ }\textbf {\bibinfo {volume} {120}},\ \bibinfo {pages} {211}
  (\bibinfo {year} {2014})}\BibitemShut {NoStop}%
\bibitem [{\citenamefont {{Kluge}}\ \emph {et~al.}(2011)\citenamefont
  {{Kluge}}, \citenamefont {{Cowan}}, \citenamefont {{Debus}}, \citenamefont
  {{Schramm}}, \citenamefont {{Zeil}},\ and\ \citenamefont
  {{Bussmann}}}]{kluge2011electron}%
  \BibitemOpen
  \bibfield  {author} {\bibinfo {author} {\bibfnamefont {T.}~\bibnamefont
  {{Kluge}}}, \bibinfo {author} {\bibfnamefont {T.}~\bibnamefont {{Cowan}}},
  \bibinfo {author} {\bibfnamefont {A.}~\bibnamefont {{Debus}}}, \bibinfo
  {author} {\bibfnamefont {U.}~\bibnamefont {{Schramm}}}, \bibinfo {author}
  {\bibfnamefont {K.}~\bibnamefont {{Zeil}}},\ and\ \bibinfo {author}
  {\bibfnamefont {M.}~\bibnamefont {{Bussmann}}},\ }\bibfield  {title}
  {\bibinfo {title} {{Electron Temperature Scaling in Laser Interaction with
  Solids}},\ }\href {https://doi.org/10.1103/PhysRevLett.107.205003} {\bibfield
   {journal} {\bibinfo  {journal} {\prl}\ }\textbf {\bibinfo {volume} {107}},\
  \bibinfo {eid} {205003} (\bibinfo {year} {2011})}\BibitemShut {NoStop}%
\bibitem [{\citenamefont {Boutoux}\ \emph {et~al.}(2015)\citenamefont
  {Boutoux}, \citenamefont {Rabhi}, \citenamefont {Batani}, \citenamefont
  {Binet}, \citenamefont {Ducret}, \citenamefont {Jakubowska}, \citenamefont
  {N{\`e}gre}, \citenamefont {Reverdin},\ and\ \citenamefont
  {Thfoin}}]{boutoux2015study}%
  \BibitemOpen
  \bibfield  {author} {\bibinfo {author} {\bibfnamefont {G.}~\bibnamefont
  {Boutoux}}, \bibinfo {author} {\bibfnamefont {N.}~\bibnamefont {Rabhi}},
  \bibinfo {author} {\bibfnamefont {D.}~\bibnamefont {Batani}}, \bibinfo
  {author} {\bibfnamefont {A.}~\bibnamefont {Binet}}, \bibinfo {author}
  {\bibfnamefont {J.-E.}\ \bibnamefont {Ducret}}, \bibinfo {author}
  {\bibfnamefont {K.}~\bibnamefont {Jakubowska}}, \bibinfo {author}
  {\bibfnamefont {J.-P.}\ \bibnamefont {N{\`e}gre}}, \bibinfo {author}
  {\bibfnamefont {C.}~\bibnamefont {Reverdin}},\ and\ \bibinfo {author}
  {\bibfnamefont {I.}~\bibnamefont {Thfoin}},\ }\bibfield  {title} {\bibinfo
  {title} {Study of imaging plate detector sensitivity to 5-18 mev electrons},\
  }\href@noop {} {\bibfield  {journal} {\bibinfo  {journal} {Review of
  Scientific Instruments}\ }\textbf {\bibinfo {volume} {86}},\ \bibinfo {pages}
  {113304} (\bibinfo {year} {2015})}\BibitemShut {NoStop}%
\bibitem [{\citenamefont {{Antici}}\ \emph {et~al.}(2008)\citenamefont
  {{Antici}}, \citenamefont {{Fuchs}}, \citenamefont {{Borghesi}},
  \citenamefont {{Gremillet}}, \citenamefont {{Grismayer}}, \citenamefont
  {{Sentoku}}, \citenamefont {{D'Humi{\`e}res}}, \citenamefont {{Cecchetti}},
  \citenamefont {{Man{\v{c}}i{\'c}}}, \citenamefont {{Pipahl}}, \citenamefont
  {{Toncian}}, \citenamefont {{Willi}}, \citenamefont {{Mora}},\ and\
  \citenamefont {{Audebert}}}]{antici2008hot}%
  \BibitemOpen
  \bibfield  {author} {\bibinfo {author} {\bibfnamefont {P.}~\bibnamefont
  {{Antici}}}, \bibinfo {author} {\bibfnamefont {J.}~\bibnamefont {{Fuchs}}},
  \bibinfo {author} {\bibfnamefont {M.}~\bibnamefont {{Borghesi}}}, \bibinfo
  {author} {\bibfnamefont {L.}~\bibnamefont {{Gremillet}}}, \bibinfo {author}
  {\bibfnamefont {T.}~\bibnamefont {{Grismayer}}}, \bibinfo {author}
  {\bibfnamefont {Y.}~\bibnamefont {{Sentoku}}}, \bibinfo {author}
  {\bibfnamefont {E.}~\bibnamefont {{D'Humi{\`e}res}}}, \bibinfo {author}
  {\bibfnamefont {C.~A.}\ \bibnamefont {{Cecchetti}}}, \bibinfo {author}
  {\bibfnamefont {A.}~\bibnamefont {{Man{\v{c}}i{\'c}}}}, \bibinfo {author}
  {\bibfnamefont {A.~C.}\ \bibnamefont {{Pipahl}}}, \bibinfo {author}
  {\bibfnamefont {T.}~\bibnamefont {{Toncian}}}, \bibinfo {author}
  {\bibfnamefont {O.}~\bibnamefont {{Willi}}}, \bibinfo {author} {\bibfnamefont
  {P.}~\bibnamefont {{Mora}}},\ and\ \bibinfo {author} {\bibfnamefont
  {P.}~\bibnamefont {{Audebert}}},\ }\bibfield  {title} {\bibinfo {title} {{Hot
  and Cold Electron Dynamics Following High-Intensity Laser Matter
  Interaction}},\ }\href {https://doi.org/10.1103/PhysRevLett.101.105004}
  {\bibfield  {journal} {\bibinfo  {journal} {Physical Review Letters}\
  }\textbf {\bibinfo {volume} {101}},\ \bibinfo {eid} {105004} (\bibinfo {year}
  {2008})}\BibitemShut {NoStop}%
\bibitem [{\citenamefont {Cowan}\ \emph {et~al.}(2004)\citenamefont {Cowan},
  \citenamefont {Fuchs}, \citenamefont {Ruhl}, \citenamefont {Kemp},
  \citenamefont {Audebert}, \citenamefont {Roth}, \citenamefont {Stephens},
  \citenamefont {Barton}, \citenamefont {Blazevic}, \citenamefont {Brambrink}
  \emph {et~al.}}]{cowan2004ultralow}%
  \BibitemOpen
  \bibfield  {author} {\bibinfo {author} {\bibfnamefont {T.}~\bibnamefont
  {Cowan}}, \bibinfo {author} {\bibfnamefont {J.}~\bibnamefont {Fuchs}},
  \bibinfo {author} {\bibfnamefont {H.}~\bibnamefont {Ruhl}}, \bibinfo {author}
  {\bibfnamefont {A.}~\bibnamefont {Kemp}}, \bibinfo {author} {\bibfnamefont
  {P.}~\bibnamefont {Audebert}}, \bibinfo {author} {\bibfnamefont
  {M.}~\bibnamefont {Roth}}, \bibinfo {author} {\bibfnamefont {R.}~\bibnamefont
  {Stephens}}, \bibinfo {author} {\bibfnamefont {I.}~\bibnamefont {Barton}},
  \bibinfo {author} {\bibfnamefont {A.}~\bibnamefont {Blazevic}}, \bibinfo
  {author} {\bibfnamefont {E.}~\bibnamefont {Brambrink}}, \emph {et~al.},\
  }\bibfield  {title} {\bibinfo {title} {Ultralow emittance, multi-mev proton
  beams from a laser virtual-cathode plasma accelerator},\ }\href@noop {}
  {\bibfield  {journal} {\bibinfo  {journal} {Physical Review Letters}\
  }\textbf {\bibinfo {volume} {92}},\ \bibinfo {pages} {204801} (\bibinfo
  {year} {2004})}\BibitemShut {NoStop}%
\bibitem [{\citenamefont {Snavely}\ \emph {et~al.}(2000)\citenamefont
  {Snavely}, \citenamefont {Key}, \citenamefont {Hatchett}, \citenamefont
  {Cowan}, \citenamefont {Roth}, \citenamefont {Phillips}, \citenamefont
  {Stoyer}, \citenamefont {Henry}, \citenamefont {Sangster}, \citenamefont
  {Singh} \emph {et~al.}}]{snavely2000intense}%
  \BibitemOpen
  \bibfield  {author} {\bibinfo {author} {\bibfnamefont {R.}~\bibnamefont
  {Snavely}}, \bibinfo {author} {\bibfnamefont {M.}~\bibnamefont {Key}},
  \bibinfo {author} {\bibfnamefont {S.}~\bibnamefont {Hatchett}}, \bibinfo
  {author} {\bibfnamefont {T.}~\bibnamefont {Cowan}}, \bibinfo {author}
  {\bibfnamefont {M.}~\bibnamefont {Roth}}, \bibinfo {author} {\bibfnamefont
  {T.}~\bibnamefont {Phillips}}, \bibinfo {author} {\bibfnamefont
  {M.}~\bibnamefont {Stoyer}}, \bibinfo {author} {\bibfnamefont
  {E.}~\bibnamefont {Henry}}, \bibinfo {author} {\bibfnamefont
  {T.}~\bibnamefont {Sangster}}, \bibinfo {author} {\bibfnamefont
  {M.}~\bibnamefont {Singh}}, \emph {et~al.},\ }\bibfield  {title} {\bibinfo
  {title} {Intense high-energy proton beams from petawatt-laser irradiation of
  solids},\ }\href@noop {} {\bibfield  {journal} {\bibinfo  {journal} {Physical
  Review Letters}\ }\textbf {\bibinfo {volume} {85}},\ \bibinfo {pages} {2945}
  (\bibinfo {year} {2000})}\BibitemShut {NoStop}%
\bibitem [{\citenamefont {Mora}(2005)}]{mora2005thin}%
  \BibitemOpen
  \bibfield  {author} {\bibinfo {author} {\bibfnamefont {P.}~\bibnamefont
  {Mora}},\ }\bibfield  {title} {\bibinfo {title} {Thin-foil expansion into a
  vacuum},\ }\href@noop {} {\bibfield  {journal} {\bibinfo  {journal} {Physical
  Review E}\ }\textbf {\bibinfo {volume} {72}},\ \bibinfo {pages} {056401}
  (\bibinfo {year} {2005})}\BibitemShut {NoStop}%
\bibitem [{\citenamefont {Adam}\ \emph {et~al.}(2006)\citenamefont {Adam},
  \citenamefont {H{\'e}ron},\ and\ \citenamefont {Laval}}]{adam2006dispersion}%
  \BibitemOpen
  \bibfield  {author} {\bibinfo {author} {\bibfnamefont {J.}~\bibnamefont
  {Adam}}, \bibinfo {author} {\bibfnamefont {A.}~\bibnamefont {H{\'e}ron}},\
  and\ \bibinfo {author} {\bibfnamefont {G.}~\bibnamefont {Laval}},\ }\bibfield
   {title} {\bibinfo {title} {Dispersion and transport of energetic particles
  due to the interaction of intense laser pulses with overdense plasmas},\
  }\href@noop {} {\bibfield  {journal} {\bibinfo  {journal} {Physical Review
  Letters}\ }\textbf {\bibinfo {volume} {97}},\ \bibinfo {pages} {205006}
  (\bibinfo {year} {2006})}\BibitemShut {NoStop}%
\bibitem [{\citenamefont {{Derouillat}}\ \emph {et~al.}(2018)\citenamefont
  {{Derouillat}}, \citenamefont {{Beck}}, \citenamefont {{P{\'e}rez}},
  \citenamefont {{Vinci}}, \citenamefont {{Chiaramello}}, \citenamefont
  {{Grassi}}, \citenamefont {{Fl{\'e}}}, \citenamefont {{Bouchard}},
  \citenamefont {{Plotnikov}}, \citenamefont {{Aunai}}, \citenamefont
  {{Dargent}}, \citenamefont {{Riconda}},\ and\ \citenamefont
  {{Grech}}}]{derouillat2018smilei}%
  \BibitemOpen
  \bibfield  {author} {\bibinfo {author} {\bibfnamefont {J.}~\bibnamefont
  {{Derouillat}}}, \bibinfo {author} {\bibfnamefont {A.}~\bibnamefont
  {{Beck}}}, \bibinfo {author} {\bibfnamefont {F.}~\bibnamefont {{P{\'e}rez}}},
  \bibinfo {author} {\bibfnamefont {T.}~\bibnamefont {{Vinci}}}, \bibinfo
  {author} {\bibfnamefont {M.}~\bibnamefont {{Chiaramello}}}, \bibinfo {author}
  {\bibfnamefont {A.}~\bibnamefont {{Grassi}}}, \bibinfo {author}
  {\bibfnamefont {M.}~\bibnamefont {{Fl{\'e}}}}, \bibinfo {author}
  {\bibfnamefont {G.}~\bibnamefont {{Bouchard}}}, \bibinfo {author}
  {\bibfnamefont {I.}~\bibnamefont {{Plotnikov}}}, \bibinfo {author}
  {\bibfnamefont {N.}~\bibnamefont {{Aunai}}}, \bibinfo {author} {\bibfnamefont
  {J.}~\bibnamefont {{Dargent}}}, \bibinfo {author} {\bibfnamefont
  {C.}~\bibnamefont {{Riconda}}},\ and\ \bibinfo {author} {\bibfnamefont
  {M.}~\bibnamefont {{Grech}}},\ }\bibfield  {title} {\bibinfo {title} {{SMILEI
  : A collaborative, open-source, multi-purpose particle-in-cell code for
  plasma simulation}},\ }\href {https://doi.org/10.1016/j.cpc.2017.09.024}
  {\bibfield  {journal} {\bibinfo  {journal} {Computer Physics Communications}\
  }\textbf {\bibinfo {volume} {222}},\ \bibinfo {pages} {351} (\bibinfo {year}
  {2018})}\BibitemShut {NoStop}%
\bibitem [{\citenamefont {{P{\'e}rez}}\ \emph {et~al.}(2013)\citenamefont
  {{P{\'e}rez}}, \citenamefont {{Kemp}}, \citenamefont {{Divol}}, \citenamefont
  {{Chen}},\ and\ \citenamefont {{Patel}}}]{perez2013deflection}%
  \BibitemOpen
  \bibfield  {author} {\bibinfo {author} {\bibfnamefont {F.}~\bibnamefont
  {{P{\'e}rez}}}, \bibinfo {author} {\bibfnamefont {A.~J.}\ \bibnamefont
  {{Kemp}}}, \bibinfo {author} {\bibfnamefont {L.}~\bibnamefont {{Divol}}},
  \bibinfo {author} {\bibfnamefont {C.~D.}\ \bibnamefont {{Chen}}},\ and\
  \bibinfo {author} {\bibfnamefont {P.~K.}\ \bibnamefont {{Patel}}},\
  }\bibfield  {title} {\bibinfo {title} {{Deflection of MeV Electrons by
  Self-Generated Magnetic Fields in Intense Laser-Solid Interactions}},\ }\href
  {https://doi.org/10.1103/PhysRevLett.111.245001} {\bibfield  {journal}
  {\bibinfo  {journal} {Physical Review Letters}\ }\textbf {\bibinfo {volume}
  {111}},\ \bibinfo {eid} {245001} (\bibinfo {year} {2013})}\BibitemShut
  {NoStop}%
\bibitem [{\citenamefont {Xu}\ \emph {et~al.}(2016)\citenamefont {Xu},
  \citenamefont {Qiao}, \citenamefont {Chang}, \citenamefont {Yao},
  \citenamefont {Wu}, \citenamefont {Yan}, \citenamefont {Zhou}, \citenamefont
  {Wang},\ and\ \citenamefont {He}}]{xu2016characterization}%
  \BibitemOpen
  \bibfield  {author} {\bibinfo {author} {\bibfnamefont {Z.}~\bibnamefont
  {Xu}}, \bibinfo {author} {\bibfnamefont {B.}~\bibnamefont {Qiao}}, \bibinfo
  {author} {\bibfnamefont {H.}~\bibnamefont {Chang}}, \bibinfo {author}
  {\bibfnamefont {W.}~\bibnamefont {Yao}}, \bibinfo {author} {\bibfnamefont
  {S.}~\bibnamefont {Wu}}, \bibinfo {author} {\bibfnamefont {X.}~\bibnamefont
  {Yan}}, \bibinfo {author} {\bibfnamefont {C.}~\bibnamefont {Zhou}}, \bibinfo
  {author} {\bibfnamefont {X.}~\bibnamefont {Wang}},\ and\ \bibinfo {author}
  {\bibfnamefont {X.}~\bibnamefont {He}},\ }\bibfield  {title} {\bibinfo
  {title} {Characterization of magnetic reconnection in the high-energy-density
  regime},\ }\href@noop {} {\bibfield  {journal} {\bibinfo  {journal} {Physical
  Review E}\ }\textbf {\bibinfo {volume} {93}},\ \bibinfo {pages} {033206}
  (\bibinfo {year} {2016})}\BibitemShut {NoStop}%
\bibitem [{\citenamefont {{Kemp}}\ and\ \citenamefont
  {{Wilks}}(2020)}]{kemp2020direct}%
  \BibitemOpen
  \bibfield  {author} {\bibinfo {author} {\bibfnamefont {A.~J.}\ \bibnamefont
  {{Kemp}}}\ and\ \bibinfo {author} {\bibfnamefont {S.~C.}\ \bibnamefont
  {{Wilks}}},\ }\bibfield  {title} {\bibinfo {title} {{Direct electron
  acceleration in multi-kilojoule, multi-picosecond laser pulses}},\ }\href
  {https://doi.org/10.1063/5.0007159} {\bibfield  {journal} {\bibinfo
  {journal} {Physics of Plasmas}\ }\textbf {\bibinfo {volume} {27}},\ \bibinfo
  {eid} {103106} (\bibinfo {year} {2020})}\BibitemShut {NoStop}%
\bibitem [{\citenamefont {{P{\'e}rez}}\ \emph {et~al.}(2012)\citenamefont
  {{P{\'e}rez}}, \citenamefont {{Gremillet}}, \citenamefont {{Decoster}},
  \citenamefont {{Drouin}},\ and\ \citenamefont
  {{Lefebvre}}}]{perez2012improved}%
  \BibitemOpen
  \bibfield  {author} {\bibinfo {author} {\bibfnamefont {F.}~\bibnamefont
  {{P{\'e}rez}}}, \bibinfo {author} {\bibfnamefont {L.}~\bibnamefont
  {{Gremillet}}}, \bibinfo {author} {\bibfnamefont {A.}~\bibnamefont
  {{Decoster}}}, \bibinfo {author} {\bibfnamefont {M.}~\bibnamefont
  {{Drouin}}},\ and\ \bibinfo {author} {\bibfnamefont {E.}~\bibnamefont
  {{Lefebvre}}},\ }\bibfield  {title} {\bibinfo {title} {{Improved modeling of
  relativistic collisions and collisional ionization in particle-in-cell
  codes}},\ }\href {https://doi.org/10.1063/1.4742167} {\bibfield  {journal}
  {\bibinfo  {journal} {Physics of Plasmas}\ }\textbf {\bibinfo {volume}
  {19}},\ \bibinfo {eid} {083104} (\bibinfo {year} {2012})}\BibitemShut
  {NoStop}%
\bibitem [{\citenamefont {Higginson}\ \emph {et~al.}(2020)\citenamefont
  {Higginson}, \citenamefont {Holod},\ and\ \citenamefont
  {Link}}]{higginson2020corrected}%
  \BibitemOpen
  \bibfield  {author} {\bibinfo {author} {\bibfnamefont {D.~P.}\ \bibnamefont
  {Higginson}}, \bibinfo {author} {\bibfnamefont {I.}~\bibnamefont {Holod}},\
  and\ \bibinfo {author} {\bibfnamefont {A.}~\bibnamefont {Link}},\ }\bibfield
  {title} {\bibinfo {title} {A corrected method for coulomb scattering in
  arbitrarily weighted particle-in-cell plasma simulations},\ }\href@noop {}
  {\bibfield  {journal} {\bibinfo  {journal} {Journal of Computational
  Physics}\ }\textbf {\bibinfo {volume} {413}},\ \bibinfo {pages} {109450}
  (\bibinfo {year} {2020})}\BibitemShut {NoStop}%
\bibitem [{\citenamefont {{Nuter}}\ \emph {et~al.}(2011)\citenamefont
  {{Nuter}}, \citenamefont {{Gremillet}}, \citenamefont {{Lefebvre}},
  \citenamefont {{L{\'e}vy}}, \citenamefont {{Ceccotti}},\ and\ \citenamefont
  {{Martin}}}]{nuter2011field}%
  \BibitemOpen
  \bibfield  {author} {\bibinfo {author} {\bibfnamefont {R.}~\bibnamefont
  {{Nuter}}}, \bibinfo {author} {\bibfnamefont {L.}~\bibnamefont
  {{Gremillet}}}, \bibinfo {author} {\bibfnamefont {E.}~\bibnamefont
  {{Lefebvre}}}, \bibinfo {author} {\bibfnamefont {A.}~\bibnamefont
  {{L{\'e}vy}}}, \bibinfo {author} {\bibfnamefont {T.}~\bibnamefont
  {{Ceccotti}}},\ and\ \bibinfo {author} {\bibfnamefont {P.}~\bibnamefont
  {{Martin}}},\ }\bibfield  {title} {\bibinfo {title} {{Field ionization model
  implemented in Particle In Cell code and applied to laser-accelerated carbon
  ions}},\ }\href {https://doi.org/10.1063/1.3559494} {\bibfield  {journal}
  {\bibinfo  {journal} {Physics of Plasmas}\ }\textbf {\bibinfo {volume}
  {18}},\ \bibinfo {eid} {033107} (\bibinfo {year} {2011})}\BibitemShut
  {NoStop}%
\bibitem [{\citenamefont {{Greenwood}}\ \emph {et~al.}(2004)\citenamefont
  {{Greenwood}}, \citenamefont {{Cartwright}}, \citenamefont {{Luginsland}},\
  and\ \citenamefont {{Baca}}}]{greenwood2004elimination}%
  \BibitemOpen
  \bibfield  {author} {\bibinfo {author} {\bibfnamefont {A.~D.}\ \bibnamefont
  {{Greenwood}}}, \bibinfo {author} {\bibfnamefont {K.~L.}\ \bibnamefont
  {{Cartwright}}}, \bibinfo {author} {\bibfnamefont {J.~W.}\ \bibnamefont
  {{Luginsland}}},\ and\ \bibinfo {author} {\bibfnamefont {E.~A.}\ \bibnamefont
  {{Baca}}},\ }\bibfield  {title} {\bibinfo {title} {{On the elimination of
  numerical Cerenkov radiation in PIC simulations}},\ }\href
  {https://doi.org/10.1016/j.jcp.2004.06.021} {\bibfield  {journal} {\bibinfo
  {journal} {Journal of Computational Physics}\ }\textbf {\bibinfo {volume}
  {201}},\ \bibinfo {pages} {665} (\bibinfo {year} {2004})}\BibitemShut
  {NoStop}%
\bibitem [{\citenamefont {{Vay}}\ \emph {et~al.}(2011)\citenamefont {{Vay}},
  \citenamefont {{Geddes}}, \citenamefont {{Cormier-Michel}},\ and\
  \citenamefont {{Grote}}}]{vay2011numerical}%
  \BibitemOpen
  \bibfield  {author} {\bibinfo {author} {\bibfnamefont {J.~L.}\ \bibnamefont
  {{Vay}}}, \bibinfo {author} {\bibfnamefont {C.~G.~R.}\ \bibnamefont
  {{Geddes}}}, \bibinfo {author} {\bibfnamefont {E.}~\bibnamefont
  {{Cormier-Michel}}},\ and\ \bibinfo {author} {\bibfnamefont {D.~P.}\
  \bibnamefont {{Grote}}},\ }\bibfield  {title} {\bibinfo {title} {{Numerical
  methods for instability mitigation in the modeling of laser wakefield
  accelerators in a Lorentz-boosted frame}},\ }\href
  {https://doi.org/10.1016/j.jcp.2011.04.003} {\bibfield  {journal} {\bibinfo
  {journal} {Journal of Computational Physics}\ }\textbf {\bibinfo {volume}
  {230}},\ \bibinfo {pages} {5908} (\bibinfo {year} {2011})}\BibitemShut
  {NoStop}%
\bibitem [{\citenamefont {{Davies}}\ \emph {et~al.}(1997)\citenamefont
  {{Davies}}, \citenamefont {{Bell}}, \citenamefont {{Haines}},\ and\
  \citenamefont {{Gu{\'e}rin}}}]{Davies_1997}%
  \BibitemOpen
  \bibfield  {author} {\bibinfo {author} {\bibfnamefont {J.~R.}\ \bibnamefont
  {{Davies}}}, \bibinfo {author} {\bibfnamefont {A.~R.}\ \bibnamefont
  {{Bell}}}, \bibinfo {author} {\bibfnamefont {M.~G.}\ \bibnamefont
  {{Haines}}},\ and\ \bibinfo {author} {\bibfnamefont {S.~M.}\ \bibnamefont
  {{Gu{\'e}rin}}},\ }\bibfield  {title} {\bibinfo {title} {{Short-pulse
  high-intensity laser-generated fast electron transport into thick solid
  targets}},\ }\href {https://doi.org/10.1103/PhysRevE.56.7193} {\bibfield
  {journal} {\bibinfo  {journal} {\pre}\ }\textbf {\bibinfo {volume} {56}},\
  \bibinfo {pages} {7193} (\bibinfo {year} {1997})}\BibitemShut {NoStop}%
\bibitem [{\citenamefont {Bell}\ and\ \citenamefont
  {Kingham}(2003)}]{bell2003resistive}%
  \BibitemOpen
  \bibfield  {author} {\bibinfo {author} {\bibfnamefont {A.}~\bibnamefont
  {Bell}}\ and\ \bibinfo {author} {\bibfnamefont {R.}~\bibnamefont {Kingham}},\
  }\bibfield  {title} {\bibinfo {title} {Resistive collimation of electron
  beams in laser-produced plasmas},\ }\href@noop {} {\bibfield  {journal}
  {\bibinfo  {journal} {Physical Review Letters}\ }\textbf {\bibinfo {volume}
  {91}},\ \bibinfo {pages} {035003} (\bibinfo {year} {2003})}\BibitemShut
  {NoStop}%
\bibitem [{\citenamefont {{Evans}}(2006)}]{Evans_2006}%
  \BibitemOpen
  \bibfield  {author} {\bibinfo {author} {\bibfnamefont {R.~G.}\ \bibnamefont
  {{Evans}}},\ }\bibfield  {title} {\bibinfo {title} {{Modelling short pulse,
  high intensity laser plasma interactions}},\ }\href
  {https://doi.org/10.1016/j.hedp.2006.02.002} {\bibfield  {journal} {\bibinfo
  {journal} {High Energy Density Physics}\ }\textbf {\bibinfo {volume} {2}},\
  \bibinfo {pages} {35} (\bibinfo {year} {2006})}\BibitemShut {NoStop}%
\bibitem [{\citenamefont {{Debayle}}\ \emph {et~al.}(2010)\citenamefont
  {{Debayle}}, \citenamefont {{Honrubia}}, \citenamefont {{D'Humi{\`e}res}},\
  and\ \citenamefont {{Tikhonchuk}}}]{debayle2010divergence}%
  \BibitemOpen
  \bibfield  {author} {\bibinfo {author} {\bibfnamefont {A.}~\bibnamefont
  {{Debayle}}}, \bibinfo {author} {\bibfnamefont {J.~J.}\ \bibnamefont
  {{Honrubia}}}, \bibinfo {author} {\bibfnamefont {E.}~\bibnamefont
  {{D'Humi{\`e}res}}},\ and\ \bibinfo {author} {\bibfnamefont {V.~T.}\
  \bibnamefont {{Tikhonchuk}}},\ }\bibfield  {title} {\bibinfo {title}
  {{Divergence of laser-driven relativistic electron beams}},\ }\href
  {https://doi.org/10.1103/PhysRevE.82.036405} {\bibfield  {journal} {\bibinfo
  {journal} {Phys. Rev. E}\ }\textbf {\bibinfo {volume} {82}},\ \bibinfo {eid}
  {036405} (\bibinfo {year} {2010})}\BibitemShut {NoStop}%
\bibitem [{\citenamefont {{Robinson}}\ \emph {et~al.}(2014)\citenamefont
  {{Robinson}}, \citenamefont {{Strozzi}}, \citenamefont {{Davies}},
  \citenamefont {{Gremillet}}, \citenamefont {{Honrubia}}, \citenamefont
  {{Johzaki}}, \citenamefont {{Kingham}}, \citenamefont {{Sherlock}},\ and\
  \citenamefont {{Solodov}}}]{Robinson_2014}%
  \BibitemOpen
  \bibfield  {author} {\bibinfo {author} {\bibfnamefont {A.~P.~L.}\
  \bibnamefont {{Robinson}}}, \bibinfo {author} {\bibfnamefont {D.~J.}\
  \bibnamefont {{Strozzi}}}, \bibinfo {author} {\bibfnamefont {J.~R.}\
  \bibnamefont {{Davies}}}, \bibinfo {author} {\bibfnamefont {L.}~\bibnamefont
  {{Gremillet}}}, \bibinfo {author} {\bibfnamefont {J.~J.}\ \bibnamefont
  {{Honrubia}}}, \bibinfo {author} {\bibfnamefont {T.}~\bibnamefont
  {{Johzaki}}}, \bibinfo {author} {\bibfnamefont {R.~J.}\ \bibnamefont
  {{Kingham}}}, \bibinfo {author} {\bibfnamefont {M.}~\bibnamefont
  {{Sherlock}}},\ and\ \bibinfo {author} {\bibfnamefont {A.~A.}\ \bibnamefont
  {{Solodov}}},\ }\bibfield  {title} {\bibinfo {title} {{Theory of fast
  electron transport for fast ignition}},\ }\href
  {https://doi.org/10.1088/0029-5515/54/5/054003} {\bibfield  {journal}
  {\bibinfo  {journal} {Nuclear Fusion}\ }\textbf {\bibinfo {volume} {54}},\
  \bibinfo {eid} {054003} (\bibinfo {year} {2014})}\BibitemShut {NoStop}%
\bibitem [{\citenamefont {{Sentoku}}\ \emph {et~al.}(2011)\citenamefont
  {{Sentoku}}, \citenamefont {{D'Humi{\`e}res}}, \citenamefont {{Romagnani}},
  \citenamefont {{Audebert}},\ and\ \citenamefont
  {{Fuchs}}}]{sentoku2011dynamic}%
  \BibitemOpen
  \bibfield  {author} {\bibinfo {author} {\bibfnamefont {Y.}~\bibnamefont
  {{Sentoku}}}, \bibinfo {author} {\bibfnamefont {E.}~\bibnamefont
  {{D'Humi{\`e}res}}}, \bibinfo {author} {\bibfnamefont {L.}~\bibnamefont
  {{Romagnani}}}, \bibinfo {author} {\bibfnamefont {P.}~\bibnamefont
  {{Audebert}}},\ and\ \bibinfo {author} {\bibfnamefont {J.}~\bibnamefont
  {{Fuchs}}},\ }\bibfield  {title} {\bibinfo {title} {{Dynamic Control over
  Mega-Ampere Electron Currents in Metals Using Ionization-Driven Resistive
  Magnetic Fields}},\ }\href {https://doi.org/10.1103/PhysRevLett.107.135005}
  {\bibfield  {journal} {\bibinfo  {journal} {\prl}\ }\textbf {\bibinfo
  {volume} {107}},\ \bibinfo {eid} {135005} (\bibinfo {year}
  {2011})}\BibitemShut {NoStop}%
\bibitem [{\citenamefont {{Gremillet}}\ \emph {et~al.}(2002)\citenamefont
  {{Gremillet}}, \citenamefont {{Bonnaud}},\ and\ \citenamefont
  {{Amiranoff}}}]{Gremillet_2002}%
  \BibitemOpen
  \bibfield  {author} {\bibinfo {author} {\bibfnamefont {L.}~\bibnamefont
  {{Gremillet}}}, \bibinfo {author} {\bibfnamefont {G.}~\bibnamefont
  {{Bonnaud}}},\ and\ \bibinfo {author} {\bibfnamefont {F.}~\bibnamefont
  {{Amiranoff}}},\ }\bibfield  {title} {\bibinfo {title} {{Filamented transport
  of laser-generated relativistic electrons penetrating a solid target}},\
  }\href {https://doi.org/10.1063/1.1432994} {\bibfield  {journal} {\bibinfo
  {journal} {Physics of Plasmas}\ }\textbf {\bibinfo {volume} {9}},\ \bibinfo
  {pages} {941} (\bibinfo {year} {2002})}\BibitemShut {NoStop}%
\end{thebibliography}%

\end{document}